\documentclass[useAMS,usenatbib,fleqn]{mnras}
\usepackage{amsmath}
\usepackage{graphicx}
\usepackage{color}
\usepackage{natbib}
\usepackage{amssymb}

\newcommand{\eqa}[1]{\begin{eqnarray}   #1 \end{eqnarray}}

\newcommand{\br}[1]{\left( #1 \right)}
\newcommand{\bc}[1]{\left\{ #1 \right\}}
\newcommand{\bb}[1]{\left[ #1 \right]}
\newcommand{\ba}[1]{\left\langle #1 \right\rangle}

\newcommand{\nn}{\nonumber}

\newcommand{\dd}{{\rm d}}

\bibpunct{(}{)}{;}{a}{}{,}

\title[KiDS+GAMA: combining cosmological probes]{KiDS+GAMA: Cosmology constraints from a joint analysis of cosmic shear, galaxy-galaxy lensing and angular clustering}

\author[van Uitert, Joachimi et al.]
{Edo van Uitert$^1$\thanks{vuitert@ucl.ac.uk}, Benjamin Joachimi$^1$\thanks{b.joachimi@ucl.ac.uk}, Shahab Joudaki$^{2,3,4}$, Alexandra Amon$^5$, \and Catherine Heymans$^5$, Fabian  K{\"o}hlinger$^{6,7}$, Marika Asgari$^5$, Chris Blake$^2$, Ami Choi$^8$, \and Thomas Erben$^9$, Daniel J. Farrow$^{10}$, Joachim Harnois-D{\'e}raps$^5$, Hendrik Hildebrandt$^9$, \and Henk Hoekstra$^6$, Thomas D. Kitching$^{11}$, Dominik Klaes$^9$, Konrad Kuijken$^6$, \and Julian Merten$^4$, Lance Miller$^4$, Reiko Nakajima$^9$,  Peter Schneider$^9$, Edwin Valentijn$^{12}$, \and Massimo Viola$^6$\\ 
\\
$^1$ Department of Physics and Astronomy, University College London, Gower Street, London WC1E 6BT, UK \\ 
$^2$ Centre for Astrophysics \& Supercomputing, Swinburne University of Technology, PO Box 218, Hawthorn, VIC 3122, Australia \\ 
$^3$ ARC Centre of Excellence for All-sky Astrophysics (CAASTRO) \\
$^4$ Department of Physics, University of Oxford, Keble Road, Oxford OX1 3RH, UK \\
$^5$ Institute for Astronomy, University of Edinburgh, Blackford Hill, Edinburgh, EH9 3HJ, UK \\
$^6$ Leiden Observatory, Leiden University, Niels Bohrweg 2, NL-2333 CA Leiden, The Netherlands \\
$^{7}$ Kavli Institute for the Physics and Mathematics of the Universe (Kavli IPMU, WPI), The University of Tokyo Institutes for Advanced \\ \hspace{0.25cm}Study, The University of Tokyo, Kashiwa, Chiba 277-8583, Japan \\
$^8$ Center for Cosmology and AstroParticle Physics, The Ohio State University, 191 West Woodruff Avenue, Columbus, OH 43210, USA \\
$^9$ Argelander-Institut f\"ur Astronomie, Auf dem H\"ugel 71, 53121 Bonn, Germany \\ 
$^{10}$ Max-Planck-Institut f\"ur extraterrestrische Physik, Postfach 1312 Giessenbachstrasse, 85741 Garching, Germany  \\ 
$^{11}$ Mullard Space Science Laboratory, University College London, Holmbury St Mary, Dorking, Surrey RH5 6NT, UK \\
$^{12}$ Kapteyn Institute, University of Groningen, PO Box 800, NL 9700 AV Groningen \\
}

\pubyear{2018}

\begin{document}

\maketitle

\begin{abstract}
We present cosmological parameter constraints from a joint analysis of three cosmological probes: the tomographic cosmic shear signal in $\sim$450 deg$^2$ of data from the Kilo Degree Survey (KiDS), the galaxy-matter cross-correlation signal of galaxies from the Galaxies And Mass Assembly (GAMA) survey determined with KiDS weak lensing, and the angular correlation function of the same GAMA galaxies. We use fast power spectrum estimators that are based on simple integrals over the real-space correlation functions, and show that they are practically unbiased over relevant angular frequency ranges. We test our full pipeline on numerical simulations that are tailored to KiDS and retrieve the input cosmology. By fitting different combinations of power spectra, we demonstrate that the three probes are internally consistent. For all probes combined, we obtain $S_8\equiv \sigma_8 \sqrt{\Omega_{\rm m}/0.3}=0.800_{-0.027}^{+0.029}$, consistent with Planck and the fiducial KiDS-450 cosmic shear correlation function results. Marginalising over wide priors on the mean of the tomographic redshift distributions yields consistent results for $S_8$ with an increase of $28\%$ in the error. The combination of probes results in a 26\% reduction in uncertainties of $S_8$ over using the cosmic shear power spectra alone. The main gain from these additional probes comes through their constraining power on nuisance parameters, such as the galaxy intrinsic alignment amplitude or potential shifts in the redshift distributions, which are up to a factor of two better constrained compared to using cosmic shear alone, demonstrating the value of large-scale structure probe combination.
\end{abstract}

\begin{keywords}
large-scale structure of Universe - methods: statistical - methods: data analysis \vspace{-2.2cm}
\end{keywords}

\clearpage


\section{Introduction}  
\indent The total mass-energy content of the Universe is dominated by two components, dark matter and dark energy, whose unknown nature constitutes one of the largest scientific mysteries of our time. Our knowledge of these components will increase dramatically in the coming decade, due to dedicated large-scale imaging and spectroscopic surveys such as Euclid\footnote{http://euclid-ec.org} \citep{Laureijs11}, the Large Synoptic Survey Telescope\footnote{https://www.lsst.org/} \citep[LSST;][]{LSST09} and the Wide-Field Infrared Survey Telescope\footnote{https://wfirst.gsfc.nasa.gov/} \citep[WFIRST;][]{Spergel15}, which will increase the mapped volume of the Universe by more than an order of magnitude. The two main cosmological probes from these surveys are the clustering of galaxies and weak gravitational lensing. Combined, they provide a particularly powerful framework for constraining properties of dark energy \citep{Albrecht06}. \\
\indent Weak gravitational lensing measures correlations in the distortion of galaxy shapes caused by the gravitational field of the large-scale structure in the foreground \citep{Bartelmann01} and is sensitive to the geometry of the Universe and the growth rate. These distortions can be extracted by correlating the positions of galaxies in the foreground (which trace the large-scale structure) with the shapes of the galaxies in the background, which is the galaxy-matter cross-correlation (often referred to as galaxy-galaxy lensing), or by correlating the observed shapes of galaxies, which is commonly referred to as cosmic shear \citep[for a review, see][]{Kilbinger15}. \\
\indent Most cosmic shear studies to date used the shear correlation functions \citep[e.g.][]{Heymans13,Jee13,Abbott16,Hildebrandt17} or the shear power spectrum \citep[e.g.][]{Brown03,Heymans05,Kitching07,Lin12,Kitching14,Kohlinger16,Abbott16,Alsing17,Kohlinger17} to constrain cosmological parameters. An intriguing finding of the fiducial cosmic shear analyses of the Canada-France-Hawaii Lensing Survey \citep[CFHTLenS;][]{Heymans13} and the Kilo Degree Survey \citep[KiDS;][]{Hildebrandt17}, two of the most constraining surveys to date, is that they prefer a cosmological model that is in mild tension with the best-fitting cosmological model from \citet{Planck15}. The first cosmological results from the Dark Energy Survey (DES) are consistent with Planck, but their uncertainties are considerably larger. Also, the result from the Deep Lens Survey \citep[DLS;][]{Jee16} agrees with Planck. Further investigation of this tension is warranted, because if it is real and not due to systematics, the implications would be far-reaching \citep[see e.g.][]{Battye14,MacCrann15,Kitching16b,Joudaki17kids}. \\
\indent To tighten the constraints, we combine the cosmic shear measurements from KiDS with two other large-scale structure probes that are sensitive to cosmological parameters: the galaxy-matter cross-correlation function and the two-point clustering auto-correlation function of galaxies. These probes have been used to constrain cosmological parameters \citep[e.g.][]{Cacciato13,Mandelbaum13,More15,Kwan17,Nicola17}. Instead of combining the different cosmological probes at the likelihood level, which is what is usually done, we follow a more optimal `self-calibration' approach by modelling them within a single framework, as this enables a coherent treatment of systematic effects and a lifting of parameter degeneracies \citep{Nicola16}. \\
\indent In this work, we adopt a formalism from \citet{Schneider02} to estimate power spectra by performing simple integrals over the real-space correlation functions using appropriate weight functions. \citet{Schneider02} demonstrate that this method works using analytical predictions of cosmic shear measurements. \citet{Brown03} applied this formalism to data to measure shear power spectra, while \citet{Hoekstra02} used it to constrain aperture masses. We extend the formalism to the galaxy-matter power spectrum and the angular power spectrum, and apply these power spectrum estimators for the first time to data. Although this approach is formally only unbiased if the correlation function measurements were available from zero lag to infinity, we show that it produces unbiased band power estimates over a considerable range of angular multipoles. This method is much faster than established methods for estimating power spectra. Furthermore, these cosmic shear power spectra are insensitive to the survey masks. Modelling the power spectra instead of the real-space correlation functions enables us to cleanly separate scales and to separate the cosmic shear signal in E-modes and B-modes, with the latter serving as a test for systematics, although it should be noted that this advantage is not exclusive to power spectra, as COSEBIs \citep{Schneider10}, for example, also split the signal in E- and B-modes. Finally, it puts the different probes on the same angular-frequency scale, which could help with identifying certain types of systematics that affect particular angular frequency ranges. \\
\indent We use the most recent shape measurement catalogues from the KiDS survey, the KiDS-450 catalogues \citep{Hildebrandt17}, to measure the weak lensing signals, and the foreground galaxies from the GAMA survey \citep{Driver09,Driver11,Liske15} from the three equatorial patches that are completely covered by KiDS, to determine the galaxy-matter cross-correlation as well as the projected clustering signal. A parallel KiDS analysis that is similar in nature, in which KiDS-450 cosmic shear measurements are combined with galaxy-galaxy lensing and redshift space distortions from BOSS \citep{Dawson13} and the 2dFLenS survey \citep{Blake16}, will be released imminently in \citet{Joudaki17kids2df}.  \\ 
\indent The outline of the paper is as follows. We introduce the three power spectrum estimators in Sect. \ref{sec_theory}. The data and the measurements are presented in Sect. \ref{sec_meas}, which is followed by the results in Sect. \ref{sec_res}. We conclude in Sect. \ref{sec_conc}. We validate our power spectrum estimators in Appendix \ref{app_val_ps}, and the entire fitting pipeline using $N$-body simulations tailored to KiDS in Appendix \ref{app_val}. In Appendix \ref{app_qe} we compare our cosmic shear power spectra to those estimated with a quadratic estimator, and in Appendix \ref{app_iter} we present our iterative scheme for determining the analytical covariance matrix. The full posterior of all fit parameters is shown in Appendix \ref{app_fullpost}. Finally, in Appendix \ref{app_flat} we check the impact of the flat-sky approximation on our power spectrum estimators, and in Appendix \ref{app_covarmat} we discuss the effect of cross-survey covariance when probes from surveys with different footprints on the sky are combined.


\section{Power spectrum estimators}\label{sec_theory}
Computing power spectra directly from the data, for example using a quadratic estimator \citep{Hu01}, is usually a complicated and CPU-intensive task \citep[e.g.][]{Kohlinger16}. This is particularly challenging for cosmic shear studies as the high signal-to-noise regime of the cosmological measurements is on relatively small scales, thus requiring high resolution measurements. Alternatively, pseudo-$C_\ell$ methods can be used \citep{Hikage11,Asgari16}, but they are sensitive to the details of the survey mask. Here, we adopt a much simpler and faster approach: we integrate over the corresponding real-space correlation functions, which can be readily measured with existing public code. We will demonstrate that this method accurately recovers the power spectra over a relevant range of $\ell$. This {\it ansatz} is very similar to the `Spice/PolSpice' methods \citep[e.g.][]{Chon04,Becker16}, except that we calculate correlation functions via direct galaxy pair counts instead of passing through map-making and pseudo-$C_\ell$ estimation steps first. 


\subsection{Cosmic shear power spectrum}

The weak lensing convergence power spectrum can be obtained from the 3-D matter power spectrum $P_\delta$ via
\begin{equation}\label{eq_Pkappa}
P_{\kappa}(\ell)=\left(\frac{3H_0^2 \Omega_{\rm m}}{2c^2} \right)^2 \int_0^{\chi_{\rm H}} {\rm d}\chi \; \frac{g^2(\chi)}{a^2(\chi)} P_\delta \left(\frac{\ell+1/2}{f_K(\chi)},\chi \right) \;,
\end{equation}
with $H_0$ the Hubble constant, $\Omega_{\rm m}$ the present-day matter density parameter, $c$ the speed of light, $\chi$ the comoving distance, $a(\chi)$ the scale-factor, $f_K(\chi)$ the comoving angular diameter distance, $\chi_{\rm H}$ the comoving horizon distance, and $g(\chi)$ a geometric weight factor, which depends on the source redshift distribution $p_z(z){\rm d}z=p_\chi(\chi){\rm d}\chi$:
\begin{eqnarray} \label{eq_lenseff}
g(\chi)=\int_\chi^{\chi_{\rm H}} {\rm d}\chi' \; p_\chi(\chi') \frac{f_K(\chi'-\chi)}{f_K({\chi'})} \;.
\end{eqnarray}
Hence for a given theoretical matter power spectrum $P_\delta$, we can predict the observed convergence power spectrum once the source redshift distribution is specified. \\
\indent As in Eq. (\ref{eq_Pkappa}), we assume the Limber and flat-sky approximations throughout in our power spectrum estimator. We validate the latter explicitly in Appendix \ref{app_flat}. A number of recent papers have demonstrated for the case of cosmic shear that these approximations are very good on the scales that we consider \citep{Kitching16,Lemos17,Kilbinger17}. For all signals we employ the hybrid approximation proposed by \citet{Loverde08}, which uses $\ell+1/2$ in the argument of the matter power spectrum but no additional prefactors. Limber's approximation is more accurate the more extended along the line of sight the kernel of the signal under consideration is \citep[see e.g.][]{Giannantonio12}. We will therefore assess the validity of our galaxy clustering estimator and model more carefully in Sect. \ref{sec_pgg}. \\
\indent The convergence power spectrum can be converted into the shear correlation functions:
\begin{eqnarray}\label{eq_xipm}
\xi_+(\theta)=\int_0^\infty \frac{{\rm d}\ell \; \ell}{2\pi} J_0(\ell\theta)P_\kappa(\ell) \; ; \nonumber \\
\xi_-(\theta)=\int_0^\infty \frac{{\rm d}\ell \; \ell}{2\pi} J_4(\ell\theta)P_\kappa(\ell) \;,
\end{eqnarray}
where $J_n(x)$ are the $n$-th order Bessel functions of the first kind. The use of shear correlation functions is popular in observational studies \citep{Kilbinger15} because they can be readily measured from the data using \mbox{$\widehat{\xi_\pm}=\widehat{\xi_{\rm tt}} \pm \widehat{\xi_{\times\times}}$}, with 
\begin{eqnarray}\label{eq_xipm2}
\widehat{\xi_{\rm tt}}(\theta)=\frac{\sum w_i w_j \epsilon_{{\rm t},i} \epsilon_{{\rm t},j}}{\sum w_i w_j} \;\; ; \;\; \widehat{\xi_{\times\times}}(\theta)=\frac{\sum w_i w_j \epsilon_{\times,i} \epsilon_{\times,j}}{\sum w_i w_j} \;, 
\end{eqnarray}
with $\epsilon_{\rm t}$ and $\epsilon_\times$ the tangential and cross component of the ellipticities of galaxies $i$ and $j$, measured with respect to their separation vector, and $w$ the inverse variance weight of the shape measurements, which comes from our shape measurement method \emph{lens}fit \citep{Miller13,Fenech16}. The sum runs over all galaxy pairs whose projected separation on the sky falls inside a radial bin centred at $\theta$ and with a width $\Delta\theta$. \\
\indent Although the shear correlation functions are easy to measure, power spectrum estimators have a number of advantages \citep{Kohlinger16}. Firstly, they enable a clean separation of different $\ell$-modes, while $\xi_\pm$ averages over them; if systematics are present that affect only certain $\ell$-modes, they are more easily identified in the power spectra. Furthermore, the covariance matrix of the power spectra is more diagonal than its real-space counterpart, also leading to a cleaner separation of scales, that is easier to model. Finally, the power spectrum estimators can be readily modified to extract the $B$-mode part of the signal, which should be consistent with zero if systematics are absent and hence serves as a systematic check. \\
\indent We estimate $\ell^2 P_\kappa(\ell)$ in a band with an upper and lower $\ell$-limit of $\ell_{i{\rm u}}$ and $\ell_{i{\rm l}}$ directly from the observed shear correlation functions using the estimator from \citet{Schneider02}:
\begin{eqnarray} \label{eq_pe}
&&P^{\rm E}_{{\rm band},i} = \frac{1}{\Delta_i} \int_{\ell_{i{\rm l}}}^{\ell_{i{\rm u}}}{\rm d}\ell \; \ell \; P_\kappa(\ell) \nonumber \\
&&\hspace{6mm}= \frac{2\pi}{\Delta_i} \int_{\ell_{i{\rm l}}}^{\ell_{i{\rm u}}}{\rm d}\ell \; \ell \nonumber \\
&& \times \label{eq_pe2} \int_{\theta_{\rm min}}^{\theta_{\rm max}}{\rm d}\theta \; \theta [K_+\xi_+(\theta)J_0(\ell\theta)+(1-K_+)\xi_-(\theta)J_4(\ell\theta)] \hspace{5mm} \\
&&\hspace{6mm}= \frac{2\pi}{\Delta_i} \int_{\theta_{\rm min}}^{\theta_{\rm max}} \frac{{\rm d}\theta}{\theta} \lbrace K_+ \xi_+(\theta) \left [ g_+(\ell_{i{\rm u}}\theta)- g_+(\ell_{i{\rm l}}\theta)\right ] +  \nonumber \\
&&(1-K_+) \xi_-(\theta) \left [ g_-(\ell_{i{\rm u}}\theta)- g_-(\ell_{i{\rm l}}\theta)\right ] \rbrace \;,
\end{eqnarray} 
with $\theta_{\rm min}$ and $\theta_{\rm max}$ the minimum and maximum angular scale that can be used, $\Delta_i = \ln(\ell_{i{\rm u}}/\ell_{i{\rm l}})$, and
\begin{eqnarray}
g_+(x)=xJ_1(x) \; \; ; \; \; g_-(x)=\left (x-\frac{8}{x} \right ) J_1(x)-8J_2(x) \;.
\end{eqnarray}
To ensure a clean E/B mode separation, the scalar $K_+$ should be fixed to 0.5. This can be seen by expressing $\xi_{+/-}$ as a function of the E/B mode power spectra \citep[see e.g. Eq. 9 in][]{Joachimi08} and inserting that into Eq. (\ref{eq_pe2}). \\
\indent This estimator is only unbiased if $\theta_{\rm min}=0$ and $\theta_{\rm max}=\infty$. However, even if we restrict the range of the integral to what can be realistically measured in our data, we can retrieve unbiased estimates of $P^E_{{\rm band},i}$ over a large $\ell$ range, as is shown in Appendix \ref{app_val_ps}, because most of the information of a given $\ell$ mode comes from a finite angular range of the shear correlation functions. The lowest $\ell$ bins we adopt may have a small remaining bias, for which we derive an integral bias correction (IBC), as detailed in Appendix \ref{app_val_ps}. To compute the IBC, we need to adopt a cosmology, which makes the correction cosmology-dependent. However, since the correction is smaller than the statistical errors, a small bias in the IBC due to adopting the wrong cosmology does not impact our results, and we will demonstrate that not applying the correction at all does not affect our results.  \\
\indent The B-mode part of the signal is measured by:
\begin{eqnarray} \label{eq_pb}
P^{\rm B}_{{\rm band},i} := \frac{\pi}{\Delta_i} \int_{\theta_{\rm min}}^{\theta_{\rm max}} \frac{{\rm d}\theta}{\theta} \lbrace \xi_+(\theta) \left [ g_+(\ell_{i{\rm u}}\theta)- g_+(\ell_{i{\rm l}}\theta)\right ] - \nonumber \\
\xi_-(\theta) \left [ g_-(\ell_{i{\rm u}}\theta)- g_-(\ell_{i{\rm l}}\theta)\right ] \rbrace \;,
\end{eqnarray}
which we measure simultaneously in the data to test for the presence of systematics. \\
\indent A similar power spectrum estimator has been proposed in \citet{Becker16P} and applied to data in \citet{Becker16}, specifically designed to minimize E-mode/B-mode mixing. However, how this estimator performs when $\xi_\pm$ have been measured in limited angular ranges, has not yet been explored. Although our estimator has some E-mode/B-mode mixing, we demonstrate that it is negligible for all but the lowest ell bin, and we derive a robust correction scheme for it.


\subsection{Galaxy-matter power spectrum}

The projected galaxy-matter power spectrum is related to the matter power spectrum via: 
\begin{eqnarray}
&&\hspace*{-1.5cm}P^{\rm gm}(\ell)= b \left(\frac{3H_0^2 \Omega_{\rm m}}{2c^2} \right) \nn \\
&&\times \int_0^{\chi_{\rm H}} {\rm d}\chi \; \frac{p_{\rm F}(\chi)g(\chi) }{a(\chi)f_K(\chi)} P_\delta \left( \frac{\ell+1/2}{f_K(\chi)};\chi\right) \;,
\end{eqnarray}
with $p_{\rm F}(\chi)$ the redshift distribution of the foreground sample. We assume that the galaxy bias is linear and deterministic\footnote{In other words, the cross-correlation coefficient $r$ \citep[e.g.][]{Pen98,Dekel99} is fixed to unity.} such that $b$ is the effective bias of the lens sample. We will motivate this choice in Sect. \ref{sec_fit}. \\
\indent In analogy with Eq. (\ref{eq_xipm2}) and (\ref{eq_pe}), we estimate the projected galaxy-matter power spectrum as:
\begin{eqnarray} \label{eq_pgm}
P^{\rm gm}(\ell) =2\pi \int_0^{\infty} {\rm d}\theta \hspace{0.5mm} \theta \hspace{0.5mm} \gamma_{\rm t}(\theta) J_2(\ell \theta) \;,
\end{eqnarray}
with $\gamma_{\rm t}(\theta)$ the tangential shear around foreground galaxies. The band galaxy-matter power spectrum estimator then follows from:
\begin{eqnarray} \label{eq_pgm2}
\hspace*{-0.5cm}P^{\rm gm}_{{\rm band},i} &:= & \frac{1}{\Delta_i} \int_{\ell_{i{\rm l}}}^{\ell_{i{\rm u}}} {\rm d}\ell \hspace{0.5mm} \ell \hspace{0.5mm} P^{\rm gm}(\ell) \nonumber \\
\hspace*{-0.5cm}&=&\frac{2\pi}{\Delta_i}\int_{\theta_{\rm min}}^{\theta_{\rm max}} \frac{{\rm d}\theta}{\theta} \gamma_{\rm t}(\theta)\left [ h(\ell_{i{\rm u}}\theta) - h(\ell_{i{\rm l}}\theta) \right ] \;,
\end{eqnarray}
with 
\begin{eqnarray}
h(x)=-xJ_1(x)-2J_0(x).
\end{eqnarray}
The final result is derived by inserting Eq. (\ref{eq_pgm}) into the first line of Eq. (\ref{eq_pgm2}), changing the order of the integrals, renaming the variables and making use of the derivative identity of Bessel functions. The analogy for the B-mode part of the signal is obtained by replacing $\gamma_{\rm t}$ with the cross shear part, $\gamma_\times$:
\begin{eqnarray}
P^{\rm g\times}_{{\rm band},i} :=  \frac{2\pi}{\Delta_i}\int_{\theta_{\rm min}}^{\theta_{\rm max}} \frac{{\rm d}\theta}{\theta} \gamma_{\times}
(\theta)\left [ h(\ell_{i{\rm u}}\theta) - h(\ell_{i{\rm l}}\theta) \right ].
\label{eq_pxm}
\end{eqnarray}
The tangential shear and cross shear are measured with the following estimators:
\begin{eqnarray}
\widehat{\gamma_{\rm t}}(\theta) = \frac{\sum_i \epsilon_{{\rm t},i}  w_i}{\sum w_i}  \; \; ; \; \; \widehat{\gamma_\times}(\theta) = \frac{\sum_i \epsilon_{\times,i}  w_i}{\sum w_i}.
\end{eqnarray}
\indent In practise, we also measured the tangential shear and cross shear signals around random points and subtracted that from the measurements around galaxies, as discussed in Sect. \ref{sec_measmeas}. As for the cosmic shear power spectra, we verify that our galaxy-matter power spectrum estimator is unbiased using analytical correlation functions and $N$-body simulations tailored to KiDS (see Appendix \ref{app_val_ps} and \ref{app_val}). We also derive and apply the IBC, which is negligible for all but the lowest $\ell$ bin, and for the first $\ell$ bin it is smaller than the measurements errors.


\subsection{Angular power spectrum} \label{sec_pgg}

The angular power spectrum can be determined from the matter power spectrum via:
\begin{eqnarray} \label{eq_pgg}
P^{\rm gg}(\ell)= b^2  \int_0^{\chi_{\rm H}} {\rm d}\chi \; \frac{p^2_{\rm F}(\chi)}{f^2_K(\chi)}P_\delta \left( \frac{\ell+1/2}{f_K(\chi)};\chi\right) \;,
\end{eqnarray}
where, as above, $b$ corresponds to the effective bias of the sample (as motivated in Sect. \ref{sec_fit}). \\
\indent The 0$^{th}$ order Limber approximation for the angular correlation function is accurate to less than a percent at scales $\ell > 5 \chi(z_0)/\sigma_\chi$, with $\chi(z_0)$ the comoving distance of the mean redshift of the foreground sample and $\sigma_\chi$ the standard deviation of the galaxies' comoving distances around the mean \citep[see Sect. IV-B of][]{Loverde08}. For our low- and high-redshift foreground samples (defined in Sect. \ref{sec_meas}), we obtain scales of $\ell\gtrsim15$ and $\ell\gtrsim25$, respectively. Since the minimum $\ell$ scale entering the analysis is 150, the Limber approximation is valid here. \\
\indent Analogous to the cosmic shear and the projected galaxy-matter power spectra, we derive an estimator for the angular power spectrum:
\begin{eqnarray}
P^{\rm gg}(\ell) =2\pi \int_0^{\infty} {\rm d}\theta \hspace{0.5mm} \theta \hspace{0.5mm} w(\theta) J_0(\ell \theta),
\end{eqnarray}
with $w(\theta)$ the angular correlation function. We estimate the galaxy-galaxy band powers using:
\begin{eqnarray}\label{eq_pgg2}
\hspace*{-0.5cm}P^{\rm gg}_{{\rm band},i} &:= & \frac{1}{\Delta_i} \int_{\ell_{i{\rm l}}}^{\ell_{i{\rm u}}} {\rm d}\ell \hspace{0.5mm} \ell \hspace{0.5mm} P^{\rm gg}(\ell) \nonumber \\
\hspace*{-0.5cm}&=&\frac{2\pi}{\Delta_i}\int_{\theta_{\rm min}}^{\theta_{\rm max}} \frac{{\rm d}\theta}{\theta} w(\theta)\left [ f(\ell_{i{\rm u}}\theta) - f(\ell_{i{\rm l}}\theta) \right ] \;,
\end{eqnarray}
with 
\begin{eqnarray}
f(x)=xJ_1(x).
\end{eqnarray}
The angular correlation function is estimated from the data using the standard LS estimator \citep{Landy93}:
\begin{eqnarray}
\widehat{w}(\theta)=\frac{D D - 2 D R + R R}{R R} \;,
\end{eqnarray}
with $D D$ the number of galaxy pairs, $D R$ the number of galaxy - random point pairs, and $R R$ the number of random point pairs. The counts with random points are scaled with the ratio of the total number of galaxies and the total number of random points. \\
\indent As for the cosmic shear and galaxy-matter power spectra, we verify that our angular power spectrum estimator is unbiased using analytical correlation functions and $N$-body simulations tailored to KiDS (see Appendix \ref{app_val_ps} and \ref{app_val}). For completeness, we also apply the IBC, but the impact on the power spectra is negligible. Note that in the remainder of this paper, we omit the subscript `${\rm band},i$' from the band power estimates for convenience, which we do not expect to cause any confusion.


\section{Data analysis}\label{sec_meas}

\subsection{Data}
The Kilo Degree Survey \citep[KiDS;][]{DeJong13} is an optical imaging survey that aims to span 1500 deg$^2$ of the sky in four optical bands, $u$, $g$, $r$ and $i$, complemented with observations in five infrared bands from the VISTA Kilo-degree Infrared Galaxy (VIKING) survey \citep{Edge13}. The exceptional imaging quality particularly suits the main science objective of the survey, which is constraining cosmology using weak gravitational lensing. \\
\indent In this study, we use data from the most recent public data release, the KiDS-450 catalogues \citep{Hildebrandt17,DeJong17}, which contains the shape measurement and photometric redshifts of 450 deg$^2$ of data, split over five different patches on the sky, which include the three equatorial patches that completely overlap with GAMA. Below, we give an overview of the main characteristics of this data set. \\
\begin{figure}
   \centering
   \includegraphics[width=1.0\linewidth]{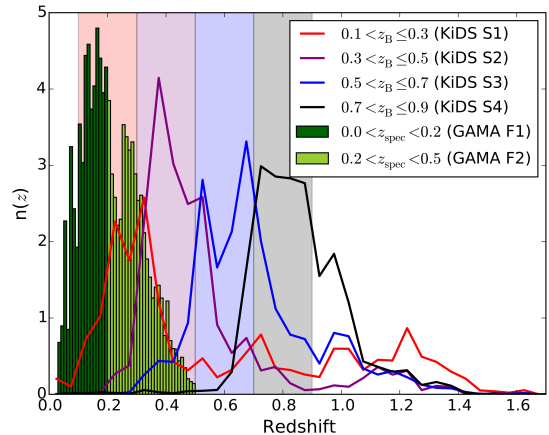}
   \caption{Normalised redshift distribution of the four tomographic source bins of KiDS (solid lines), used to measure the weak gravitational lensing signal, and the normalised redshift distribution of the two spectroscopic samples of GAMA galaxies (histograms), that serve as the foreground sample in the galaxy-galaxy lensing analysis and that are used to determine the angular correlation function. For plotting purposes, the redshift distribution of GAMA galaxies has been multiplied by a factor 0.5. The shaded regions indicate the photometric redshift ($z_{\rm B}$) selection of the tomographic source bins.}
   \label{plot_zdistr}
\end{figure}
\begin{figure*}
   \centering
   \includegraphics[width=1.0\linewidth]{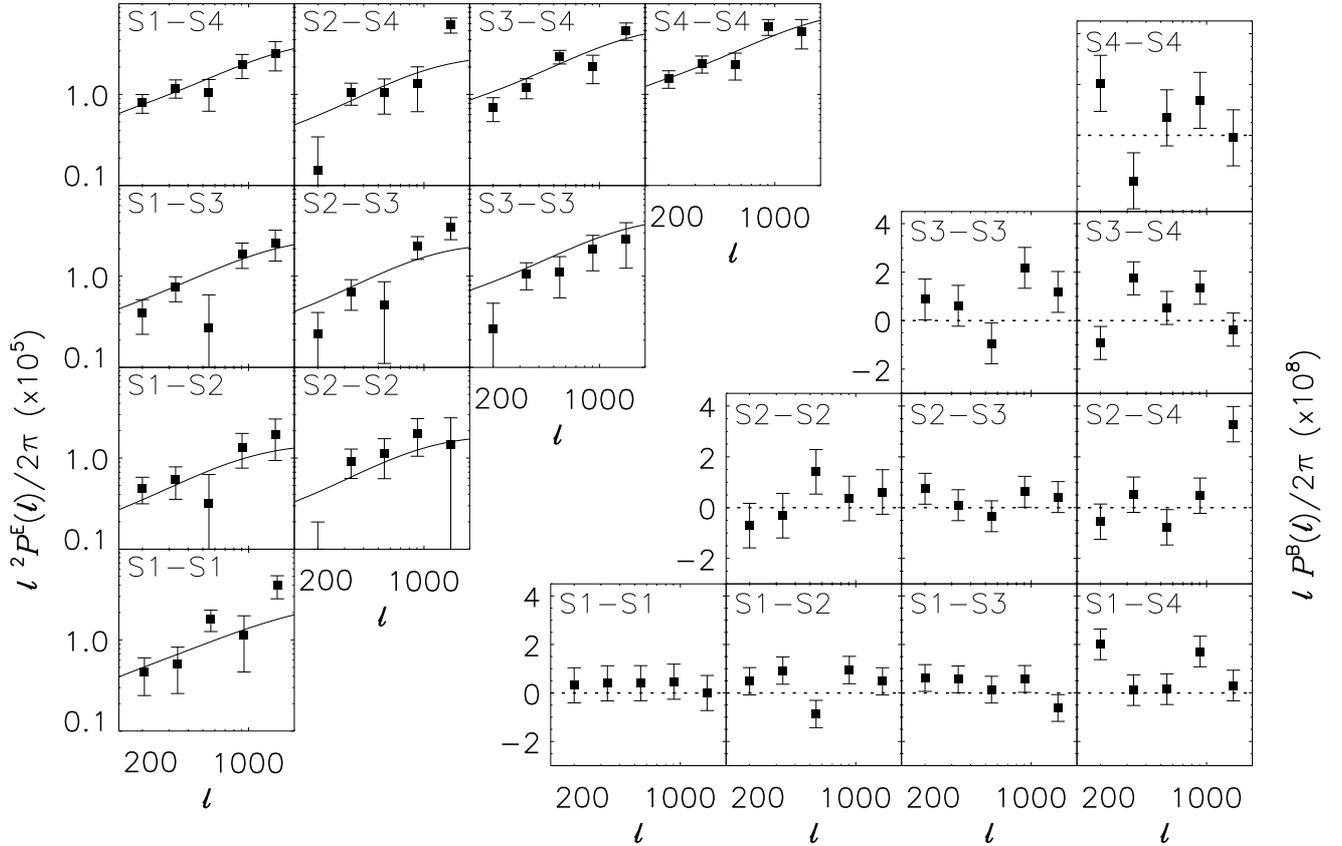}
   \caption{Cosmic shear power spectra for KiDS-450, derived with our power spectrum estimator that integrates the shear correlation functions in the range $0.06<\theta<120$ arcmin. The numbers in each panel indicate which shape (S) samples are correlated, with the numbers defined in the legend of Fig. \ref{plot_zdistr}. The panels on the left show the E-modes, and the ones on the right the B-modes. Error bars have been computed analytically. The B-modes have been multiplied with $\ell$ instead of $\ell^2$ for improved visibility of the error bars. Solid lines correspond to the best-fitting model, for our combined fit to $P^{\rm E}$, $P^{\rm gm}$ and $P^{\rm gg}$. There is one $\ell$ bin whose B-mode deviates from zero by more than 3$\sigma$, the highest $\ell$ of the S2--S4 cross-correlation; the corresponding E-mode is high as well. We have verified that excluding this bin from the analysis does not change our results.}
   \label{plot_pnom}
\end{figure*}
\indent The redshift distribution of the source galaxies was determined using four different methods in KiDS-450. The most robust is the weighted direct calibration method (hereafter refered to as DIR), which is based on the work of \citet{Lima08}. In this method, catalogues from deep spectroscopic surveys are weighted in such a way as to remove incompleteness caused by their spectroscopic selection functions \citep[see][for details]{Hildebrandt17}. The true redshift distribution for a  sample of KiDS galaxies selected using their bayesian photometric redshifts from BPZ \citep{Benitez00} can then be determined by matching to these weighted spectroscopic catalogues. The resulting redshift distribution is well-calibrated in the range 0.1 $<z_{\rm B}\leq$ 0.9, with $z_{\rm B}$ the peak of the posterior photometric redshift distribution from BPZ. In this work, we use the same four tomographic source redshift bins as adopted in \citet{Hildebrandt17} by selecting galaxies with 0.1 $<z_{\rm B}\leq$ 0.3, 0.3 $<z_{\rm B}\leq$ 0.5, 0.5 $<z_{\rm B}\leq$ 0.7 and 0.7 $<z_{\rm B}\leq$ 0.9. The redshift distribution of the four source samples from the DIR method is shown in Fig. \ref{plot_zdistr}. The main properties of the source samples, such as their average redshift, number density and ellipticity dispersion, can be found in Table 1 of \citet{Hildebrandt17}. \\
\indent The galaxy shapes were measured from the $r$-band data using an updated version of the \emph{lens}fit method \citep{Miller13}, carefully calibrated to a large suite of image simulations tailored to KiDS \citep{Fenech16}. The resulting multiplicative bias is of the order of a percent with a statistical uncertainty of less than 0.3 percent, and is determined in each tomographic bin separately. The additive shape measurement bias is determined separately in each patch on the sky and in each tomographic redshift bin as the weighted average galaxy ellipticity per ellipticity component, and has typical values of $\sim10^{-3}$. We corrected the additive bias at the catalogue level, while the multiplicative bias was accounted for during the correlation function estimation. \\
\indent To avoid confirmation bias, the fiducial cosmological analysis of KiDS \citep{Hildebrandt17} was blinded: three different shape catalogues were analysed, the original and two copies in which the galaxy ellipticities were modified such that the resulting cosmological constraints would differ. Only after the analysis was written up, an external blinder revealed which catalogue was the correct one. Since the lead authors of this paper were already unblinded elsewhere, the current analysis could no longer be performed blindly. However, since the shear catalogues were not changed after unblinding, we still partly benefit from the original blinding exercise.\\
\indent We used the KiDS galaxies to measure the cosmic shear correlation functions, and to measure the tangential shear around the foreground galaxies from the Galaxy And Mass Assembly (GAMA) survey \citep{Driver09,Driver11,Liske15}. GAMA is a highly complete spectroscopic survey up to a Petrosian $r$-band magnitude of 19.8. In total, it targeted $\sim$240$\,$000 galaxies. We use a subset of $\sim$180$\,$000 galaxies that reside in the three patches of 60 deg$^2$ each near the celestial equator, G09, G12 and G15, as those patches fully overlap with KiDS. The tangential shear measurements in these three patches are combined with equal weighting. Due to the flux limit of the survey, GAMA galaxies have redshifts between 0 and 0.5. We select two GAMA samples, a low redshift sample with $z_{\rm spec}<0.2$, and a high redshift sample with $0.2<z_{\rm spec}<0.5$. Their redshift distributions are also shown in Fig. \ref{plot_zdistr}.  \\
\indent We also use the same subset of GAMA galaxies to determine the angular correlation function, and thus the corresponding angular power spectrum. To determine the clustering, we make use of the GAMA random catalogue version 0.3, which closely resembles the random catalogue that was used in \citet{Farrow15} to measure the angular correlation function of GAMA galaxies. We sample the random catalogue such that we have ten times more random points than real GAMA galaxies.


\subsection{Measurements}\label{sec_measmeas}
We use the shape measurement catalogues of KiDS-450 to measure the cosmic shear correlation functions, $\xi_+$ and $\xi_-$, and the tangential shear around GAMA galaxies. All projected real-space correlation functions in this work are measured with {\sc TreeCorr}\footnote{https://github.com/rmjarvis/TreeCorr}\citep{Jarvis04}. Since the $\xi_+$ and $\xi_-$ measurements have already been presented in \citet{Hildebrandt17}, we will not show them here. The $\ell$-range in which we can obtain unbiased estimates of the power spectra depends on the angular range where we trust the correlation functions. For $\xi_+$ and $\xi_-$, we use an upper limit of $\theta<120$ arcmin, as the measurements on larger scales become increasingly sensitive to residual uncertainties on the additive bias correction. The lower limit is 0.06 arcmin, but our power spectrum estimator is insensitive to any signal below 1 arcmin. The $P^{\rm E}$ band powers are nearly unbiased in the range $\ell>150$ (see Appendix \ref{app_val_ps}). We measure $\xi_+$ and $\xi_-$ in 600 logarithmically-spaced bins between 0.06 and 600 arcmin, to account for the rapid oscillations of the window functions used to convert the shear correlation functions to the power spectra, but we only use scales $0.06<\theta<120$ arcmin in the integral. \\
\indent To test the sensitivity of our estimator to a residual additive shear bias, we also measured the power spectra without applying the additive  bias correction. This only affected the lowest $\ell$ bins by shifting them with a typical amount of 0.5$\sigma$; the impact on other bins was negligible. Since the error on the additive bias correction is smaller than the correction itself, its impact on the power spectra is even smaller and can therefore be safely ignored. \\
\begin{figure*}
   \centering
   \includegraphics[width=1\linewidth]{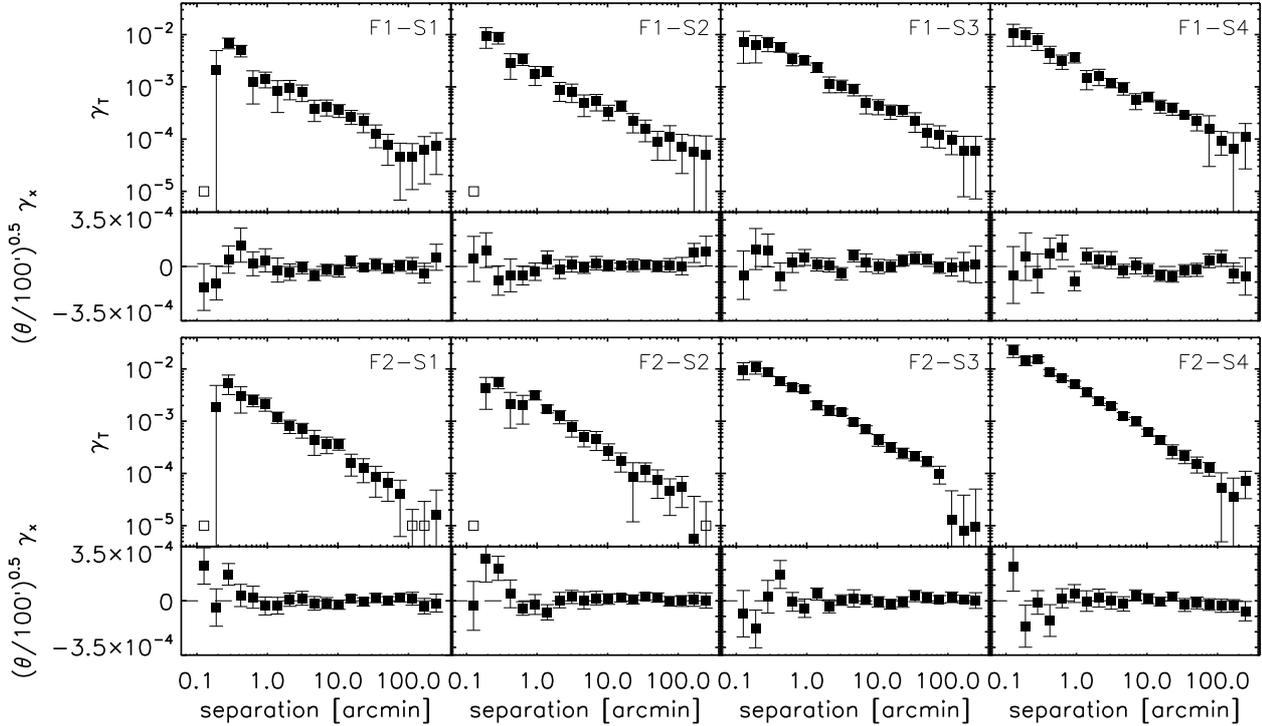}
   \caption{Tangential shear and cross shear around GAMA galaxies measured with KiDS sources in tomographic bins, as indicated in the panels. The cross shear measurements have been multiplied with a factor $(\theta/100)^{0.5}$ to ensure that the error bars are visible over the plotted angular range. Open squares show negative points of $\gamma_{\rm t}$ with unaltered error bars. The lensing signal measured around random points has been subtracted, which is consistent with zero on the scales of interest for all but the third tomographic source bin, where it is small but positive on scales $>$20 arcmin. Furthermore, the signal has been corrected for the contamination of source galaxies that are physically associated with the lenses. The errors are derived from jackknifing over 2.5$\times$3 degree non-overlapping patches. They are only used to assess on which scales the signal is consistent with not being affected by systematics; when we fit models to our power spectra we use analytical errors throughout.}
   \label{plot_gt}
\end{figure*}
\begin{figure*}
   \centering
   \includegraphics[width=1\linewidth]{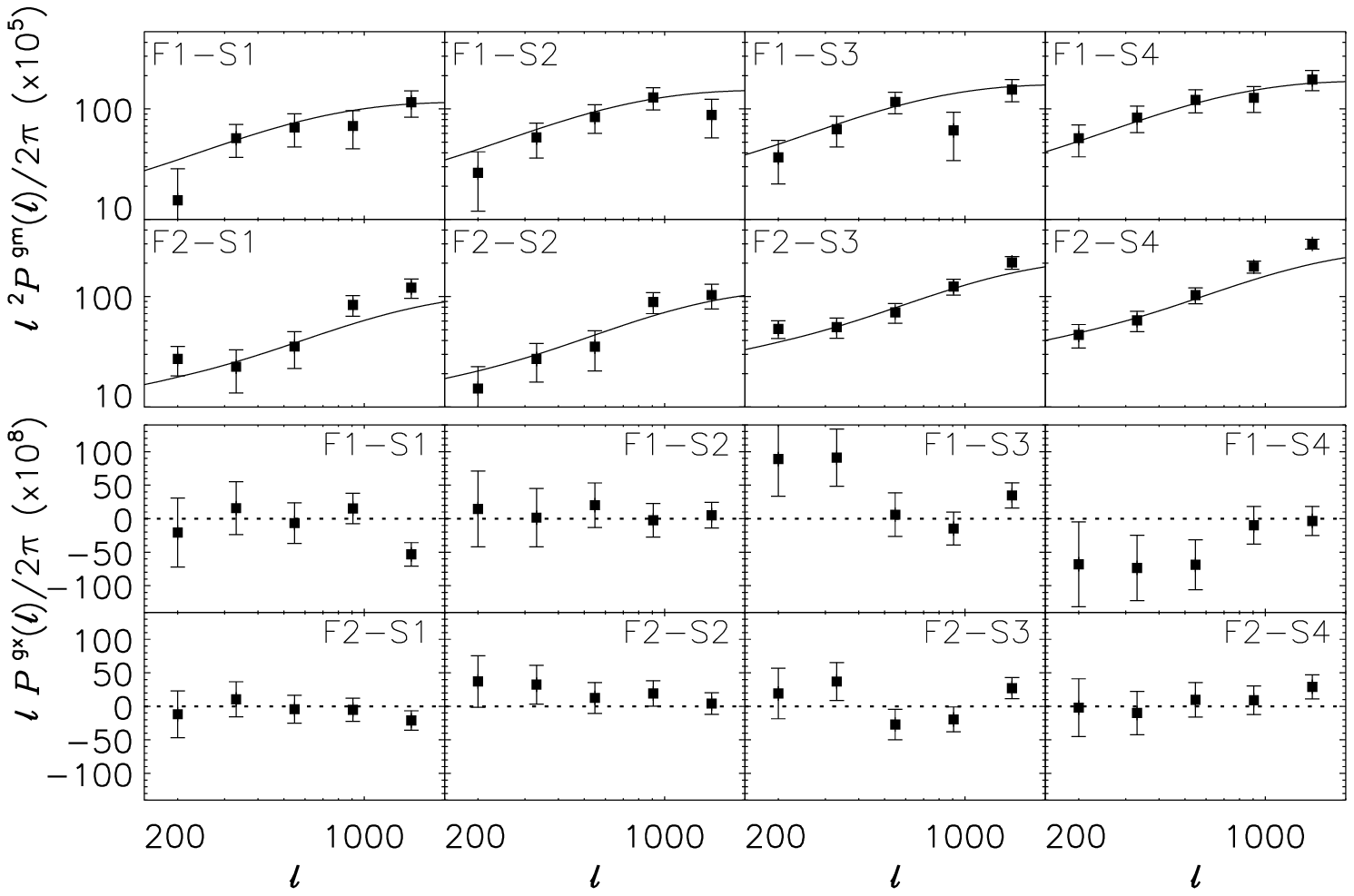}
   \caption{Galaxy-matter power spectrum (top) and galaxy-cross shear power spectrum (bottom) around GAMA galaxies in two lens redshift bins, measured with KiDS sources using four tomographic source bins. The numbers in each panel indicate the foreground (F) sample - shape (S) sample combination, as defined in Fig. \ref{plot_zdistr}. The errors are computed analytically and correspond to the 68\% confidence interval. $P^{\rm g\times}$ has been multiplied with $\ell$ instead of $\ell^2$ for improved visibility of the error bars. Solid lines correspond to the best-fitting model, for our combined fit to $P^{\rm E}$, $P^{\rm gm}$ and $P^{\rm gg}$. The $P^{\rm g\times}$ in the bottom rows serves as a systematic test, and it is consistent with zero.}
   \label{plot_pgdm}
\end{figure*}
\indent Since $P^{\rm E}$ does not vary rapidly with $\ell$, we only need a small number of $\ell$-bins to capture most of the cosmological information. We use five logarithmically-spaced bins, whose logarithmic means range from $\ell=200$ to $\ell=1500$; the $\ell$-ranges they cover can be read off from Fig. \ref{plot_P_scaletst}. Truncating the integral to $\theta<120$ arcmin leads to a small negative additive bias of the order $10^{-6}$ in the lowest $\ell$ bin (smaller than the statistical errors). We derive an integral bias correction (IBC) for this in Appendix \ref{app_val_ps} and apply it to all power spectra, although not applying this correction leads to negligible changes of our results. The resulting E-modes and B-modes are shown in Fig. \ref{plot_pnom}. \\
\indent We obtain a clear detection for $P^{\rm E}$ in each tomographic bin combination. The signal increases with redshift, which is expected as the impact of more structures is imprinted on the galaxy ellipticities if their light traversed more large-scale structure and because of the geometric scaling of the lensing signal (see Eq. \ref{eq_lenseff}). \\
\indent Fig. \ref{plot_pnom} also shows $P^{\rm B}$, the B-modes that serve as a systematic test. Note that the IBC has also been applied to the B-modes. There are a number of $\ell$ bins which appear to be affected by B-modes; the most prominent feature is the highest $\ell$ bin for the cross-correlation between the second and fourth tomographic bins. To quantify this, we determined the reduced $\chi^2$ value of the null hypothesis for all bins combined, which has a value of 1.96. This corresponds to a $p$-value of 0.0001.  This number is driven by this single $\ell$ bin; excluding this bin alone lowers the reduced $\chi^2$ to 1.55 (and a $p$-value of 0.0082), which is still a tentative sign of residual B-modes. Not applying the IBC slightly improves the overall reduced $\chi^2$ to 1.87 (1.45 after removing the suspicious $\ell$ bin). The origin of the B-modes in KiDS is under active investigation and will be presented in Asgari et al. (in prep). To test how it may affect our cosmological results, we repeat the test of \citet{Hildebrandt17}, subtract the B-modes from the E-modes, and run the cosmological inference. This B-mode correction shifts our main cosmological result by less than 0.5$\sigma$, thus demonstrating that if the source of the B-modes also generates E-modes in equal amounts, our results are not significantly biased if we do not account for that. More details of this test are provided in Appendix \ref{app_qe}. \\
\indent The large amplitude of $P^{\rm B}$ of this suspicious $\ell$ bin suggests that the corresponding $P^{\rm E}$ measurement might not be trustworthy, and indeed, it appears high. We have tested that removing this single $\ell$-bin from the analysis does not affect the cosmological inference except for the goodness of fit. Another apparent feature is that the $P^{\rm E}$ of the first $\ell$ bins of the cross-correlation between the second tomographic bin and the second, third and fourth tomographic bins are $\sim$2$\sigma$ below the best-fitting model. However, the first $\ell$ bins of the various tomographic bin combinations are fairly correlated (see e.g. Fig. \ref{plot_errcompshape} in Appendix \ref{app_val_err}), so this feature is less significant than it appears. Furthermore, in Sect. \ref{sec_res} we will show that excluding the lowest $\ell$ bins from the fit does not impact our results. \\
\indent We have also compared our power spectrum estimates with those derived using the quadratic estimator from \citet{Kohlinger17}. A detailed comparison is presented in Appendix \ref{app_qe}. Overall, we find good agreement between the E-modes, although for one tomographic bin combination we find a noticeable difference at high $\ell$. A possible explanation is the presence of some B-modes in the cosmic shear correlation functions \citep[as reported in ][]{Hildebrandt17}. This is further supported by the fact that we detect B-modes at a higher significance than \citet{Kohlinger17}, where they are found to be consistent with zero. It is still unclear if or how this affects the cosmological inference, although the B-mode correction test we did in Appendix \ref{app_qe} suggests that the impact is small. \\
\indent Next, we determined the galaxy-matter power spectrum, for which we needed to measure the tangential shear signal around GAMA galaxies first. This lensing signal is shown in Fig. \ref{plot_gt}. We also measured the signal with an independent code, and the results agreed very well. For illustrative purposes, we used 20 logarithmically-spaced bins between 0.1 and 300 arcmin. To compute the power spectra, we need a much finer sampling, as the window functions used to convert the correlation functions to power spectra oscillate rapidly. Hence we measured the signal in 600 logarithmically-spaced bins in the range $0.06<\theta<600$ arcmin, but only used the measurements on scales $\theta<120$ arcmin to compute the power spectrum. \\
\indent Some of the galaxies from the source sample are physically associated with the lenses. They are not lensed and bias the tangential shear measurements. As demonstrated in \citet{Mandelbaum05}, this bias can easily be corrected by multiplying the lensing signal with a boost factor, which contains the overdensity of source galaxies as a function of projected radial distance to the lens. The boost factor generally increases towards smaller separations, but decreases very close to the lens, due to problems with the background estimation caused by the lens light \citep[see e.g.][]{Dvornik17}. The boost factor can be made smaller by applying redshift cuts to the source sample; here, we do not apply such cuts because we want to use the exact same sources as in the cosmic shear measurements. In our case, the impact of the boost correction is negligible, as our estimator is insensitive to scales $\theta<2$ arcmin (see Appendix \ref{app_val_ps}). At 2 arcmin, the boost factor is 7\% at most for the F2-S2 bin, and decreases quickly with radius. For all other bins, the correction is much smaller. We have checked that not applying the boost correction does not significantly affect the power spectra\footnote{The boost correction implicitly assumes  that satellite galaxies are not intrinsically aligned with the foreground galaxies, although our model can account for such alignments. Most dedicated studies of this type of alignments show that it is consistent with zero \citep[see e.g.][and references therein]{Sifon15}. If it is not, this could incur a small bias in the boost correction. We will address this in a future work.}.\\
\indent The impact of magnification on the boost factor is negligible in this radial range and can safely be ignored. Furthermore, we measured the tangential shear around random points from the GAMA random catalogue, and subtracted that from the real signal. Apart from removing potential additive systematics in the shape measurement catalogues, this procedure also suppresses sampling variance errors \citep{Singh16}.  \\
\indent To obtain the errors on our galaxy-galaxy lensing measurements, we split the survey into 24 non-overlapping patches of 2.5$\times$3 degrees, and used those for a `delete one jackknife' error analysis. These errors should give a fair representation of the true errors, and thus be sufficient to assess at which scales we consider the measurements robust. Note that we used jackknife errors instead of analytical errors on these real-space measurements for convenience; we stress that in the cosmological inference, we used an analytical covariance matrix for all power spectra.\\
\indent Figure \ref{plot_gt} also shows the cross shear, the projection of source ellipticities at an angle of 45 degrees with respect to the lens-source separation vector. Galaxy-galaxy lensing does not produce a parity violating cross shear once the signal is azimuthally averaged, and hence it serves as a standard test for the presence of systematics. The cross shear is consistent with zero on most scales, although some deviations are visible, e.g. at scales of half a degree for the F1-S4 bin. The cross shear at small separations for the F2-S1 and F2-S2 bins is not worrisome, as our estimator is not sensitive to the galaxy-galaxy lensing signal on those scales. For consistency with the cosmic shear power spectrum, we only use the galaxy-galaxy lensing measurements in the range $<120$ arcmin. As demonstrated in Appendix \ref{app_val_ps}, we can obtain unbiased estimates on $P^{\rm gm}$ from $\gamma_{\rm t}$ in the range $\ell\geq150$. \\
\indent We estimate $P^{\rm gm}$ using the same $\ell$-range as for $P^{\rm E/B}$. The measurements are shown in Fig. \ref{plot_pgdm}. We apply the IBC, which on average causes a 6\% change in the lowest $\ell$ bin, and much smaller changes for the higher $\ell$ bins. We obtain significant detections for all lens-source bin combinations. The error bars have been computed analytically as discussed in Sect. \ref{sec_covar}. The amplitude of the power spectrum increases for higher source redshift bins as expected, because of the geometric scaling of the lensing signal. We also show $P^{\rm g\times}$, the power spectrum computed using the cross shear, which serves as a systematic test. There are a few neighbouring $\ell$-bins that are systematically offset, for example the low-$\ell$ bins of F1-S3 and F1-S4. We already pointed out the presence of some cross shear in Fig. \ref{plot_gt} on the scale of half a degree for those bins, which translates into those $P^{\rm g\times}$ bins. On average, however, the amplitude of $P^{\rm g\times}$ is not worrisome as the reduced $\chi^2$ of the null hypothesis has a value of 1.13. The corresponding $p$-value is 0.27. \\
\indent Finally, to determine $P^{\rm gg}$, we first measure the angular correlation function of the two foreground galaxy samples from GAMA. We show the signal in Fig. \ref{plot_clus}. Errors come from jackknifing over 2.5$\times$3 deg patches and only serve as an illustration; in the cosmological inference, we use analytical errors for $P^{\rm gg}$. The angular correlation function is robustly measured on all scales depicted. Therefore, we use an upper limit of 240 arcmin in the integral to determine $P^{\rm gg}$. We adopt the same $\ell$ ranges as for $P^{\rm E}$ and $P^{\rm gm}$ and show the band powers of $P^{\rm gg}$ in Fig. \ref{plot_pgg}. The angular power spectrum of the F2 sample is lower than that of the F1 sample because the redshift range of F2 is wider. Note that the angular correlation function $w(\theta)$ has an additive contribution due to the fact that the mean galaxy density is estimated from the same dataset. This integral constraint only contributes to the $\ell=0$ mode in $P^{\rm gg}$ and therefore does not have to be considered further in our modelling.
\begin{figure}
   \centering
   \includegraphics[width=1\linewidth]{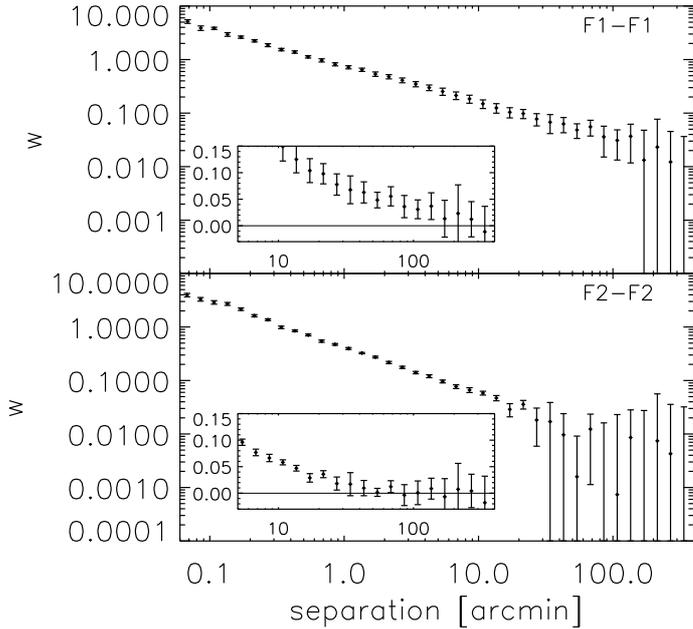}
   \caption{Angular correlation function of the two foreground galaxy samples from GAMA. The inset in each panel shows the signal on large scales with a linear vertical axis. The errors are derived from jackknifing over 2.5$\times$3 degree non-overlapping patches and serve for illustration. When we fit models to our power spectra we used analytical errors throughout.}
   \label{plot_clus}
\end{figure}


\subsection{Covariance matrix}\label{sec_covar}

\indent We determine the covariance matrix of the combined set of power spectra analytically, following a similar formalism as in \citet{Hildebrandt17}. The covariance matrix includes the cross-covariance between the different probes. One particular advantage of this approach is that it properly accounts for super-sample covariance, which are the cosmic variance modes that are larger than the survey window and couple to smaller modes within. This term is typically underestimated when the covariance matrix is estimated from the data itself, for example through jackknifing, or when it is estimated from numerical simulations. Another advantage is that it is free of simulation sampling noise, which could otherwise pose a significant hindrance for joint probe analyses with large data vectors. \\
\indent The analytical covariance matrix consists of three terms: (i) a Gaussian term that combines the Gaussian contribution to sample variance, shape noise, and a mixed noise-sample variance term, estimated following \citet{Joachimi08}, (ii) an in-survey non-Gaussian term from the connected matter trispectrum, and (iii) a super-sample covariance term. To compute the latter two terms, we closely follow the formalism outlined in \citet{Takada13}, which can be readily expanded to galaxy-galaxy lensing and clustering measurements \citep[e.g.][]{Krause16}. \\
\begin{figure}
   \centering
   \includegraphics[width=1\linewidth]{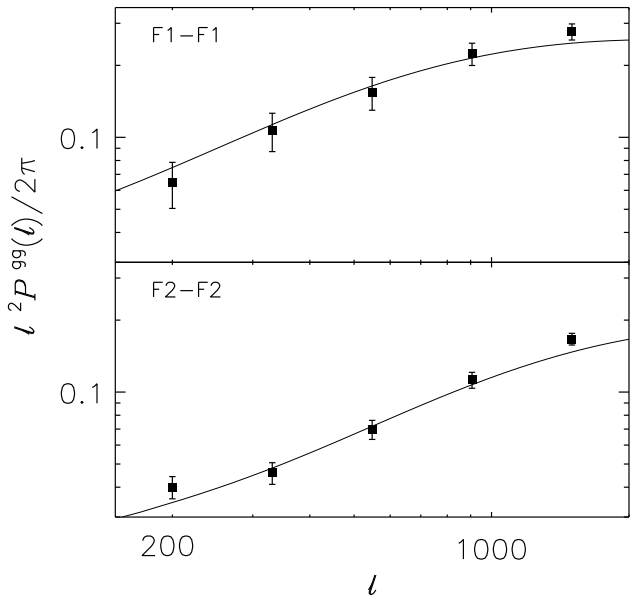}
   \caption{Angular power spectrum of the two foreground galaxy samples from GAMA. The depicted errors are determined analytically. Solid lines correspond to the best-fitting model, for our combined fit to $P^{\rm E}$, $P^{\rm gm}$ and $P^{\rm gg}$.}
   \label{plot_pgg}
\end{figure}
\indent By subtracting the signal around random points from the galaxy-matter cross-correlation, we effectively normalise fluctuations in the galaxy distribution with respect to the mean galaxy density in the survey area instead of the global mean density. This substantially reduces the response to super-survey modes \citep{Takada13} and diminishes error bars \citep{Singh16}, and we do account for this effect in our covariance model.\\
\indent One further complication is that the KiDS survey area is larger than GAMA. While the galaxy-matter power spectrum and the angular power spectrum are measured in the 180 deg$^2$ of the three GAMA patches near the equator that are fully covered by KiDS, the cosmic shear power spectrum is measured on the full 450 deg$^2$ of KiDS-450. This partial sky overlap of the different probes affects the cross-correlation and is accounted for (see Appendix \ref{app_covarmat}). \\
\indent In order to compute the covariance matrix, we need to adopt an initial fiducial cosmology as well as values for the effective galaxy bias. For the fiducial cosmology, we use the best-fit parameters from \citet{Planck15}, and for the effective galaxy biases we assume values of unity for both bins. If our data prefers different values for these parameters, the size of our posteriors could be affected \citep[as illustrated in][for the case of cosmic shear only]{Eifler09}. Therefore, after the initial cosmological inference, the analytical covariance matrix is updated with the parameter values of the best-fitting model. This is turned into an iterative approach, as detailed in Appendix \ref{app_iter}. It is made possible by the use of an analytical covariance matrix, which is relatively fast and easy to compute. Since the parameter constraints do not change significantly at the second iteration, we adopt the resulting analytical covariance matrix for all cosmological inferences in this paper.  \\
\indent The analytical covariance matrix for $\xi_+$ and $\xi_-$ has been validated against mocks in \citet{Hildebrandt17}. We repeat that exercise for the three power spectra in Appendix \ref{app_val}. The analytical covariance matrix agrees well with the one estimated from the $N$-body simulations. Our choice of power spectrum estimator is not guaranteed to reach the expected errors that we calculate analytically, but the comparison with the simulations did not reveal any evidence for significant excess noise. We did not include intrinsic alignments or baryonic feedback in the covariance modelling, but since all our measurements are dominated by the cosmological signals, the impact of the astrophysical nuisances on sample variance is small\footnote{By far the most strongly affected bin combination is F2--S1 whose redshift distributions have substantial overlap. For $A_{\rm IA}=1$, the galaxy position-intrinsic shape correlation contributes at most 17\% to the total signal, with little dependence on angular scale.}. We have checked that a potential error on the additive bias correction has a negligible contribution to the covariance matrix.


\subsection{Model fitting}\label{sec_fit}

To constrain the cosmological parameters, we used {\sc cosmoMC}\footnote{http://cosmologist.info/cosmomc/} \citep{Lewis02}, which is a fast Markov Chain Monte Carlo code for cosmological parameter estimation. The version we use is based on \citet{Joudaki17}\footnote{https://github.com/sjoudaki/CosmoLSS}, which includes prescriptions to deal with intrinsic alignment, the effect of baryons on the non-linear power spectrum, and systematic errors in the redshift distribution. This framework has been further developed to simultaneously model the tangential shear signal of a sample of foreground galaxies and redshift space distortions \citep{Joudaki17kids2df}. We extended it by modelling the angular correlation function of the same foreground sample. Furthermore, we modified the code in order to fit the power spectra instead of the correlation functions. Since the conversion from power spectra to correlation functions could be skipped, the runtime decreased by a factor of two. We computed the power spectra at the logarithmic mean of the band instead of integrating over the band width, as the difference between the two was found to be at the percent level and therefore ignored. We checked that the impact of this simplification on our cosmological parameter constraints was less than 0.3$\sigma$ for our fiducial data vector.\\ 
\indent The effect of non-linear structure formation and baryonic feedback are modelled in {\sc cosmoMC} using a module called {\sc hmcode}, which is based on the results of \citet{Mead15}. Baryonic effects are accounted for by modifying the parameters that describe the shape of dark matter haloes. AGN and supernova feedback, for example, blow material out of the haloes, making them less concentrated. This is incorporated in {\sc hmcode} by choosing the following form for the mass-concentration relation,
\begin{eqnarray}
c(M,z)=B\frac{1+z_{\rm f}}{1+z} \;,
\end{eqnarray}
with $z_{\rm f}$ the formation redshift of a halo, which depends on halo mass. The free parameter in the fit, $B$, modulates the amplitude of this mass-concentration relation. It also sets the amplitude of a `halo bloating' parameter $\eta_0$ which changes the halo profile in a mass dependent way \citep[see equation 26 of][]{Mead15}, where we follow the recommendation of \citet{Mead15} by fixing $\eta_0 = 1.03 -0.11 B$. Setting $B=3.13$ corresponds to a dark-matter-only model. The resulting model is verified with power spectra measured on large hydrodynamical simulations, and found to be accurate to 5\% for $k\leq10 h/$Mpc. This is a relative uncertainty, not an absolute one (the absolute accuracy of any theoretical matter power spectrum prediction is not well established), and indicates the relative accuracy of their halo model fits with respect to hydrodynamical simulations, which are uncertain themselves. In addition, as Fig. 2 of \citet{Mead15} shows, this accuracy is strongly $k$-dependent, and at small $k$ ($k<0.05\, h/{\rm Mpc}$), the agreement is much better than 5\%. Therefore, putting a meaningful prior on the accuracy of the theory predictions is currently out of reach. However, the main source of theoretical uncertainty is caused by baryonic feedback, which mainly affects the small scales (high $k$). By marginalizing over $B$, we account for this main source of uncertainty. \\
\indent Intrinsic alignments affect both the cosmic shear power spectrum and the galaxy-matter power spectrum. For the cosmic shear power spectrum, there are two contributions, the intrinsic-intrinsic (II) and the shear-intrinsic (GI) terms \citep[see Eq. 5 and 6 of][]{Joudaki17}. The galaxy-matter power spectrum has a galaxy-intrinsic contribution \citep[e.g.][]{Joachimi10}. These three terms can be computed once the intrinsic alignment power spectrum is specified, which is assumed to follow the non-linear modification of the linear alignment model \citep{Catelan01,Hirata04,Bridle07,Hirata10}:
\begin{eqnarray}
P_{\delta{\rm I}}(k,z) = - A_{\rm IA} C_1 \rho_{\rm crit} \frac{\Omega_{\rm m}}{D(z)} P_\delta(k,z) \;,
\label{eq_pi}
\end{eqnarray}
with $P_\delta(k,z)$ the full non-linear matter power spectrum, $D(z)$ the growth factor, normalised to unity at $z=0$, $\rho_{\rm crit}$ the critical density, $C_1=5\times10^{-14}h^{-2}M_\odot^{-1}$Mpc$^3$ a normalization constant, and $A_{\rm IA}$ the overall amplitude, which is a free parameter in our model. Our intrinsic alignment model is minimally flexible with a single, global amplitude parameter. Since the mean luminosities of the different tomographic bins are similar, there is no need to account for a luminosity dependence in the model; in addition, there currently does not exist observational evidence for a significant redshift dependence \citep[see e.g.][]{Joudaki17kids,Joudaki17kids2df}. Adding flexibility to the intrinsic alignment model is therefore currently not warranted by the data.\\
\indent To model $P^{\rm gm}$ and $P^{\rm gg}$, we assume that the galaxy bias is constant and scale-independent. Since we include non-linear scales in our fit, this bias should be interpreted as an effective bias. It is fitted separately for the low-redshift and high-redshift foreground sample. The scale dependence of the bias has been constrained in observations by combining galaxy-galaxy lensing and galaxy clustering measurements for various flux-limited samples and was found to be small \citep[e.g.][]{Hoekstra02,Simon07,Jullo12,Cacciato12}. In a recent study on data from the Dark Energy Survey, \citet{Crocce16} constrained the scale dependence of the bias using the clustering signal of flux-limited samples, selected with $i<22.5$, modelling the signal with a non-linear power spectrum from \citet{Takahashi12} with a fixed, linear bias as fit parameter. They report that their linear bias model reproduces their measurements down to a minimum angle of 3 arcmin for their low-redshift samples (although the caveat should be added that our foreground sample is selected with a different apparent magnitude cut). While the aforementioned studies report little scale dependence of the bias in real space, our assumption of a scale-independent bias is made in Fourier space. The largest $\ell$ bin is centred at 1500, which uses information from $\xi_{+/-}$ down to scales of less than an arcminute (see Appendix \ref{app_val_ps}). Hence a strong scale dependence of the bias on scales less than 3 arcminutes could violate our assumption. However, if the bias is strongly scale-dependent on scales of $\ell<1500$, this will show up in our measurements as a systematic offset between data and model for the highest $\ell$ bin of $P^{\rm gg}$ (and, to a lesser extent, $P^{\rm gm}$). Also, on small scales, the cross-correlation coefficient $r$ might differ from one, which would lead to discrepancies between $P^{\rm gm}$ and $P^{\rm gg}$. However, as Fig. \ref{plot_pgdm} and \ref{plot_pgg} show, there is no clear evidence for such a systematic difference, which serves as further evidence that our approach is robust. Also, when we exclude the highest $\ell$ bin of $P^{\rm gm}$ and $P^{\rm gg}$ from our analysis, our results do not change significantly (see Sect. \ref{sec_cosmo}).  \\
\indent We validated the $P^{\rm gg}$ model predictions using an independent code that was internally available to us. The signal agreed to within 3\% in the range $150<\ell<2000$, with a mean difference of 2\%. The small remaining difference is caused by different redshift interpolation schemes of the galaxy number density; in our code, we used a spline interpolation, while a linear interpolation was used in the independent code. When we adopted a spline interpolation in the independent code, the model signal agreed to within 1.5\%, with a mean difference of $\sim$1\%. Since it is not a priori clear which interpolation scheme is better, we decided to keep using the spline interpolation scheme. The model prediction of $P^{\rm E}$ has been compared to independent code in \citet{Hildebrandt17} and was found to agree well. We have not explicitly compared the predictions of $P^{\rm gm}$ with an independent code, but since that model is built of components used in the computation of $P^{\rm gg}$ and $P^{\rm E}$, we expect a similar level of accuracy.\\
\begin{table}
  \centering
  \caption{Priors on the fit parameters. Rows 1--6 contain the priors on cosmological parameters, rows 7--10 the priors on astrophysical `nuisance' parameters. All priors are flat within their ranges.}
  \begin{tabular}{c c c} 
  \hline
Parameter & Description &  Prior range \\
  \hline\hline
$100\theta_{\rm MC}$ & 100 $\times$ angular size of sound horizon & $[0.5,10]$ \\
$\Omega_{\rm c}h^2$ & Cold dark matter density & $[0.01,0.99]$ \\
$\Omega_{\rm b}h^2$ & Baryon density & $[0.019,0.026]$ \\
$\ln(10^{10}A_{\rm s})$ & Scalar spectrum amplitude & $[1.7,5.0]$ \\
$n_{\rm s}$ & Scalar spectral index & $[0.7,1.3]$ \\
$h$ & Dimensionless Hubble parameter & $[0.64,0.82]$ \\
$A_{\rm IA}$ & Intrinsic alignment amplitude & $[-6,6]$ \\
$B$ & Baryonic feedback amplitude & $[2,4]$ \\
$b_{\rm z1}$ & Galaxy bias of low-$z$ lens sample & $[0.1,5]$ \\
$b_{\rm z2}$ & Galaxy bias of high-$z$ lens sample & $[0.1,5]$ \\
  \hline
  \end{tabular}
  \label{tab_prior}
\end{table} 
\indent We marginalize over the systematic uncertainty of the redshift distribution of our source bins following the same methodology adopted in \citet{Hildebrandt17} and \citet{Kohlinger17}, that is by drawing a random realization of the redshift distribution in each step of the MCMC. This approach fully propagates the statistical uncertainties included in the redshift probability distributions, but does not account for sample variance in the spectroscopic calibration data. We investigated the robustness of this method by also fitting models in which we allowed for a constant shift in the redshift distributions. This procedure basically marginalizes over the first moment of the redshift distribution, which is, to first order, what the weak lensing signal is sensitive to \citep{Amara07}. We discuss the result of this test in Sect. \ref{sec_zshift}. We do not account for the uncertainty of the multiplicative shear calibration correction, as \citet{Hildebrandt17} showed that it has a negligible impact on correlation function measurements. \\
\indent To obtain a crude estimate of how much cosmic variance in the source redshift distribution affects our cosmological results, we performed the following test. We used the DIR method separately on the different spectroscopic fields. The variation between the resulting redshift distributions suggests that cosmic variance and Poisson noise contribute roughly equally to the total uncertainty. To estimate the potential impact on our cosmological constraints, we fixed the redshift distribution to the mean from the DIR method, but allowed for a shift in the mean redshift of each tomographic bin, using a Gaussian prior with a width that equals the error on the mean redshift \citep[from Table 1 of][]{Hildebrandt17}. Using this set-up, we recovered practically identical errors on the cosmological parameters compared to our fiducial approach. Next, we increased the width of the Gaussian prior by a generous factor of 1.5, to roughly include the impact of cosmic variance. This increased the error on our cosmology results by 5\%. Note that this is a conservative upper limit, as the cosmic variance between the separate spectroscopic fields is larger than the cosmic variance of all the fields combined. Hence we conclude that cosmic variance of the source redshift distribution affects our cosmological constraints by a few percent at most. We do not adopt this as our fiducial approach, however, since our current method of estimating the impact is not sufficiently accurate. \\
\indent We adopt top-hat priors on the cosmological parameters, as well as the physical `nuisance' parameters discussed earlier in this section. The prior ranges are listed in Table \ref{tab_prior}. Furthermore, we fix $k_{\rm pivot}$, the pivot scale where the scalar spectrum has an amplitude of $A_s$, to 0.05/Mpc. Even though the sum of the neutrino masses is known to be non-zero, we adopt the same prior as \citet{Hildebrandt17} and fix it to zero. We have tested that adopting 0.06 eV instead leads to a negligible change in our results. Note that the priors and fiducial values we adopted are the same as in \citet{Hildebrandt17}, which makes a comparison of the results easier. As a test, we also fitted our joint data vector adopting the broader priors on $H_0$ and $\Omega_{\rm b}$ from \citet{Joudaki17kids} and found negligible changes to our results, showing that we are not sensitive to the adopted prior ranges of these parameters.  \\
\indent A number of the assumptions we made could affect the measured or theoretical power spectra, and thus our cosmological constraints, at the per cent level. We have decided to ignore the assumptions whose impact is of the order 1 percent or less. This includes the Limber approximation, the flat-sky approximation, and the uncertainty on the multiplicative bias correction. Other effects whose impact is either uncertain or expected to be larger are addressed in the text.\\
\indent We ran {\sc cosmoMC} with twelve independent chains. To assess whether the chains have converged, we used a Gelman-Rubin test \citep{Gelman92} with the criterion that the ratio between the variance of any of the fit parameters in a single chain and the variance of that parameter in all chains combined is smaller than 1.03. Furthermore, we have checked that the chains are stable against further exploration. When analysing the chains, we removed the first 30\% of the chains as the burn-in phase. Before fitting the measured power spectra from the data, we ran {\sc cosmoMC} on our mock results, and verified that we retrieved the input cosmology. Details of this test can be found in Appendix \ref{app_val_cosm}.  


\section{Results}\label{sec_res}
\begin{figure}
   \centering
   \includegraphics[width=\linewidth]{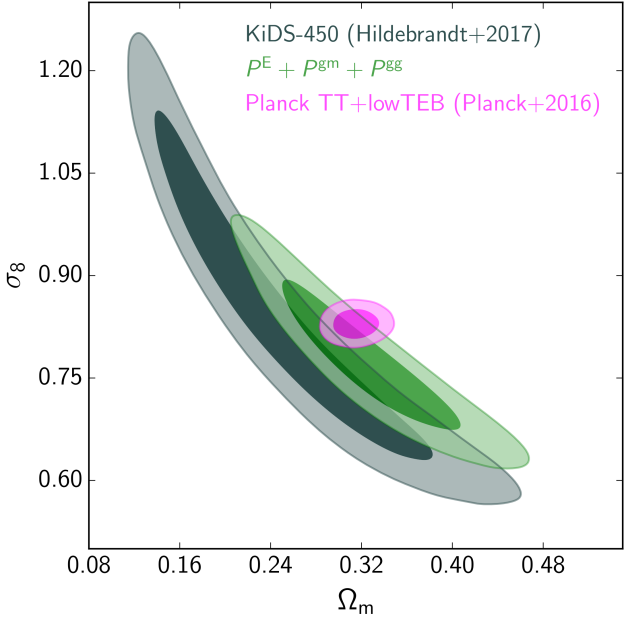}
   \caption{Constraints on $\Omega_{\rm m}$ and $\sigma_8$ from this work, from the fiducial KiDS-450 analysis \citep{Hildebrandt17} and from \citet{Planck15}. Our combined-probe constraints lie between those from the fiducial KiDS-450 analysis and those from Planck, and are consistent with both.}
   \label{plot_cosm_data}
\end{figure}
We fitted all power spectra simultaneously and show the best-fit model as solid lines in Figs. \ref{plot_pnom}, \ref{plot_pgdm} and \ref{plot_pgg}. Overall, the model describes the trends in the data well. The reduced $\chi^2$ of the best-fitting model has a value of 1.29 (115.9/ [100 data points - 10 fit parameters]) and the $p$-value is 0.034. Hence our model provides a fair fit. If we exclude the highest $\ell$ bin of the S2--S4 correlation of $P^{\rm E}$, whose corresponding B-mode is high, the best-fitting reduced $\chi^2$ becomes 1.19 without affecting any of the results (a shift of 0.1$\sigma$ in $S_8$). We do include this particular $\ell$ bin in all our results below to avoid a posteriori selection. \\

\subsection{Cosmological inference}\label{sec_cosmo}

\begin{figure}
   \centering
   \includegraphics[width=\linewidth]{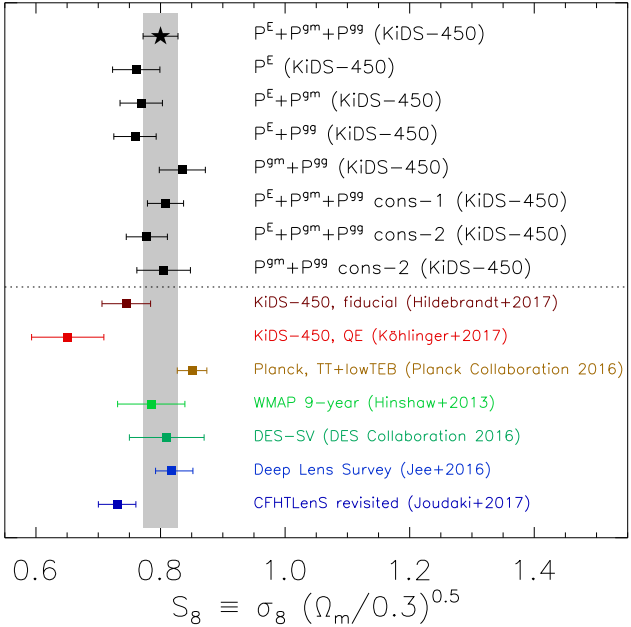}
   \caption{Comparison of our constraints on $S_8$ with a number of recent results from the literature. We show the results for different combinations of power spectra on top with black squares, as well as the results from our conservative runs where we excluded the lowest $\ell$ bin of all power spectra (`cons-1') and the highest $\ell$ bin of $P^{\rm gm}$ and $P^{\rm gg}$ (`cons-2') in the fit. In general, our results agree well with those from the literature, including those from Planck.}
   \label{plot_S8}
\end{figure}
\indent The main result of this work is the constraint on $\Omega_{\rm m} - \sigma_8$, which is shown in Fig. \ref{plot_cosm_data}. It is this combination of cosmological parameters to which weak lensing is most sensitive. We recover the familiar `banana-shape' degeneracy between these two parameters, which is expected as gravitational lensing roughly scales as $\sigma_8^2 \Omega_{\rm m}$ \citep{Jain97}. Also shown are the main fiducial results of KiDS-450 \citep{Hildebrandt17} and the constraints from \citet{Planck15}. Our confidence regions are somewhat displaced with respect to those of \citet{Hildebrandt17} and our error on $S_8$ is 28\% smaller. Interestingly, our results lie somewhat closer to those of \citet{Planck15}, showing better consistency with Planck than KiDS-450 cosmic shear alone. As discussed below, our cosmic shear-only results are fully consistent with the results from \citet{Hildebrandt17}, although not identical, because our power spectra weight the angular scales differently than the correlation functions. Hence this shift towards Planck must either be caused by $P^{\rm gm}$ or $P^{\rm gg}$ or a combination of the two.  \\
\indent We computed the marginalised constraint on \mbox{$S_8\equiv \sigma_8 \sqrt{\Omega_{\rm m}/0.3}$} and show the results in Fig. \ref{plot_S8}. The joint constraints for our fiducial setup is $S_8=0.800_{-0.027}^{+0.029}$. The fiducial result from KiDS-450 is $S_8=0.745\pm0.039$ \citep{Hildebrandt17}, whilst those of \citet{Planck15} is $S_8=0.851\pm0.024$. \\
\indent Compared to the results from \citet{Hildebrandt17}, our posteriors have considerably shrunk along the degeneracy direction. Since we applied the same priors, this improvement is purely due to the gain in information from the additional probes. Hence the real improvement becomes clear when we compare the constraints on $\Omega_{\rm m}$ and $\sigma_8$, for which we find $\Omega_{\rm m}=0.326_{-0.057}^{+0.048}$ and $\sigma_8=0.776_{-0.081}^{+0.064}$, while \citet{Hildebrandt17} report $\Omega_{\rm m}=0.250_{-0.103}^{+0.053}$ and $\sigma_8=0.849_{-0.204}^{+0.120}$. Hence our constraint on $\sigma_8$ has improved by roughly a factor of two compared to \citet{Hildebrandt17}\footnote{The improvement compared to the $P^{\rm E}$ only results that are discussed below is $\sim44$\%.}. \\
\begin{figure}
   \centering
   \includegraphics[width=\linewidth]{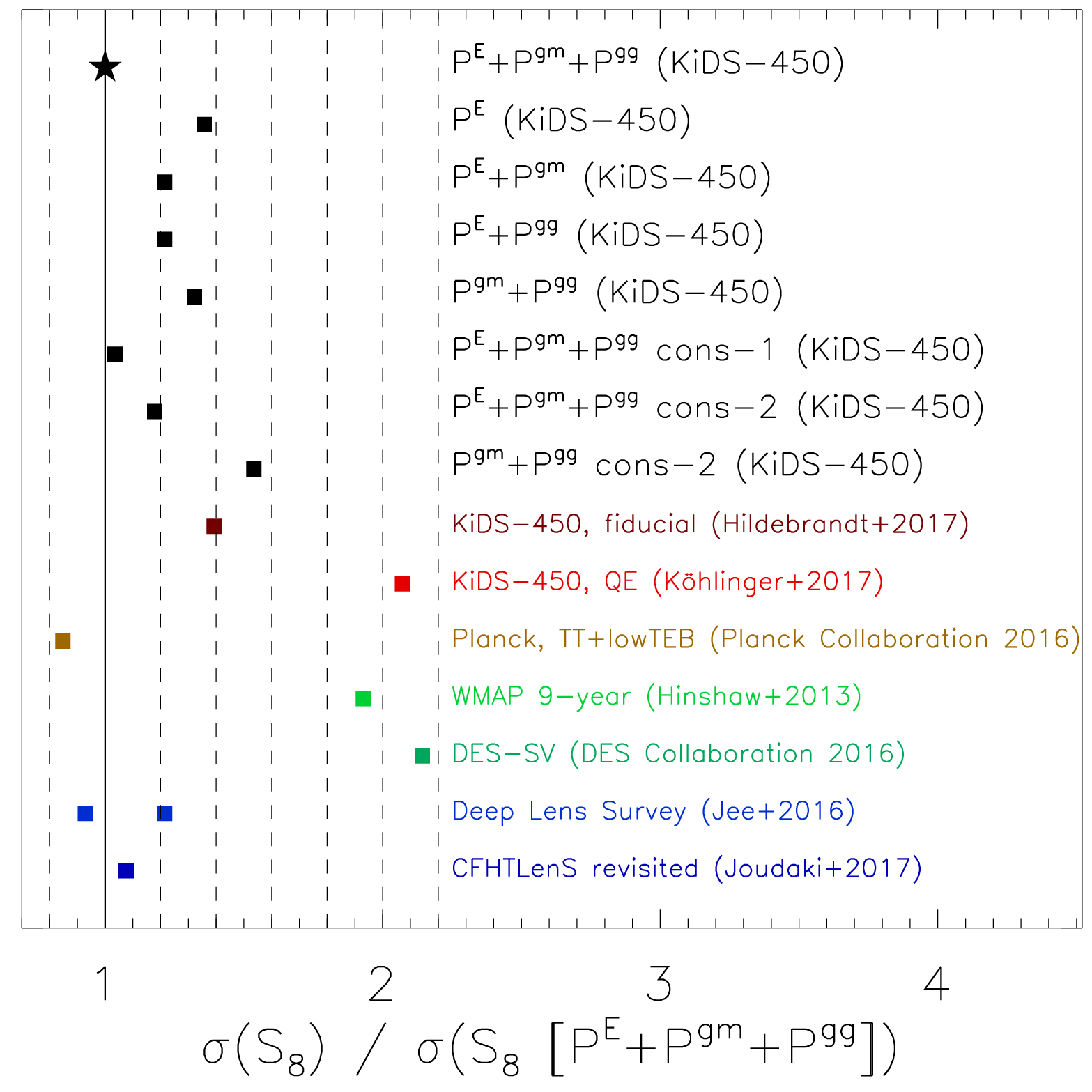}
   \caption{Ratio of the error bar on $S_8$ for various combinations of our data vector and for results from the literature, relative to our fiducial results ($P^{\rm E}+P^{\rm gm}+P^{\rm gg}$). The solid vertical line indicates a ratio of unity, while the dashed lines are displaced by relative shifts of 0.2. Our error bar is 28\% smaller than the one from \citet{Hildebrandt17}, while the error bar from \citet{Planck15} is 18\% smaller than ours. The two points shown for \citet{Jee16} are for the quoted lower and upper limit on $S_8$. }
   \label{plot_S8err}
\end{figure}
\begin{figure*}
\begin{minipage}[t]{0.48\linewidth}
   \includegraphics[width=\linewidth]{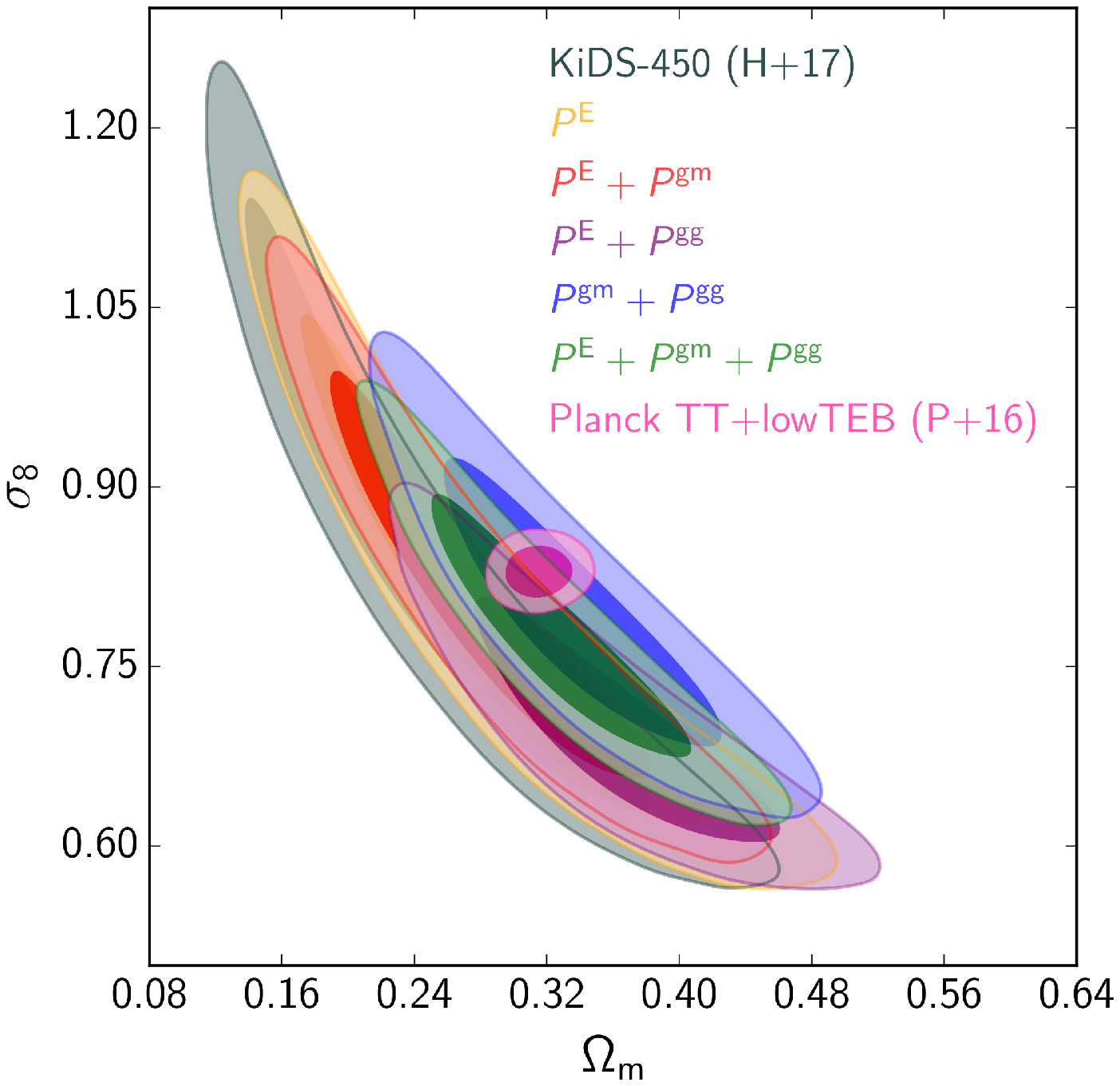}
\end{minipage}
\begin{minipage}[t]{0.48\linewidth}
   \includegraphics[width=\linewidth]{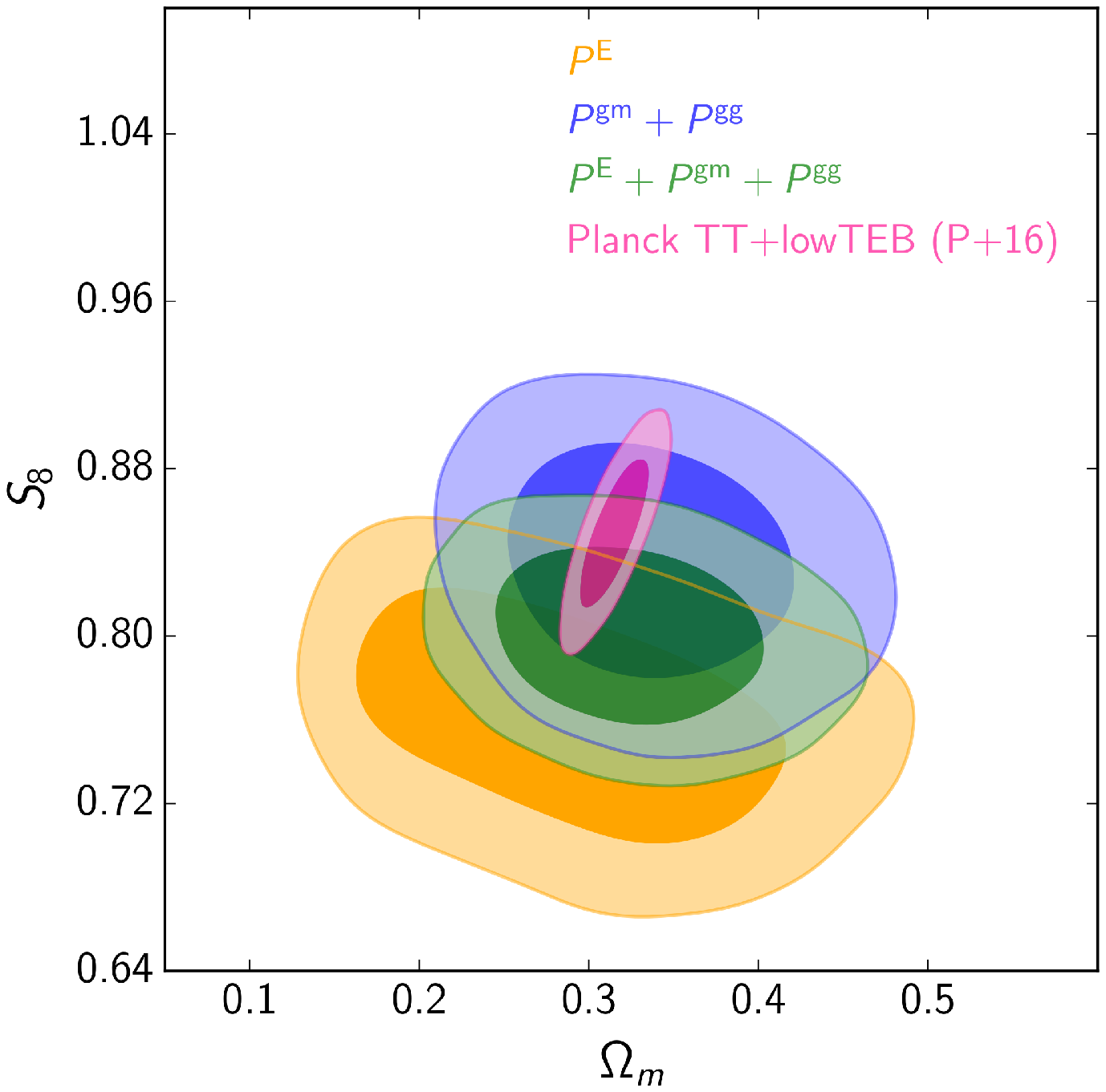}
\end{minipage}
   \caption{Constraints on $\Omega_{\rm m}$ - $\sigma_8$ and $\Omega_{\rm m}$ - $S_8$ from this work for different combinations of power spectra. Also shown are the fiducial results for KiDS-450 \citep[H+17;][]{Hildebrandt17} and Planck \citep[P+16;][]{Planck15}.}
   \label{plot_cosm_PE}
\end{figure*}
\indent To understand where the difference between our results and \citet{Hildebrandt17} comes from, and to learn how much $P^{\rm gm}$ and $P^{\rm gg}$ help with constraining cosmological parameters, we also ran our cosmological inference on all pairs of power spectra, as well as on $P^{\rm E}$ alone. The resulting constraints are shown in Fig. \ref{plot_S8}. Fig. \ref{plot_S8err} shows the relative difference of the size of the error bars, while Fig. \ref{plot_cosm_PE} shows the marginalized posterior of $\Omega_{\rm m} - \sigma_8$ and $\Omega_{\rm m} - S_8$. Interestingly, the constraints from $P^{\rm E}$ and $P^{\rm gm}+P^{\rm gg}$ are somewhat offset, with the latter preferring larger values. The constraint on $S_8$ from $P^{\rm E}$ alone is $0.761\pm0.038$, hence close to the results from \citet{Hildebrandt17}, while for $P^{\rm gm}+P^{\rm gg}$ we obtain $S_8=0.835\pm0.037$. $P^{\rm E}$ is only weakly correlated with $P^{\rm gg}$ and $P^{\rm gm}$ (see e.g. Fig. \ref{plot_errcompshape}), and if we ignore this correlation (it is fully accounted for in all our fits), the constraints on $S_8$ from $P^{\rm E}$ and $P^{\rm gm}+P^{\rm gg}$ differ by 1.4$\sigma$. Since the reduced $\chi^2$ is not much worse for the joint fit, our data does not point at a strong tension between the probes, and they can be safely combined. \\
\indent Combining $P^{\rm E}$ with $P^{\rm gm}$ or $P^{\rm gg}$ results in a relatively minor decrease of the errors of $S_8$ of 11\%. Also, the mean value of $S_8$ does not change much. The reason is that the amplitude of $P^{\rm gm}$ and $P^{\rm gg}$, which contains most of the cosmological information, is degenerate with the effective galaxy bias, and as a result, $P^{\rm E}$ drives the cosmological constraints. When both $P^{\rm gm}$ and $P^{\rm gg}$  are included in the fit, this degeneracy is broken. Fitting all probes jointly leads therefore to a larger decrease of 26\% compared to fitting only $P^{\rm E}$ (see Fig. \ref{plot_S8err}), although this could partly be driven by the displacement of the posteriors in the $\Omega_{\rm m} - \sigma_8$ plane between $P^{\rm E}$ and $P^{\rm gm}+P^{\rm gg}$. Finally, it is interesting to note that $P^{\rm E}$ and $P^{\rm gm}+P^{\rm gg}$ have similar statistical power, even though the latter is measured on less than half the survey area \citep[see also][]{Seljak05,Mandelbaum13}. Using the full 3D information content of $P^{\rm gg}$ instead of the projected quantities that we used here, will improve the cosmological constraining power of this probe even further. \\
\begin{figure*}
   \centering
   \includegraphics[width=\linewidth]{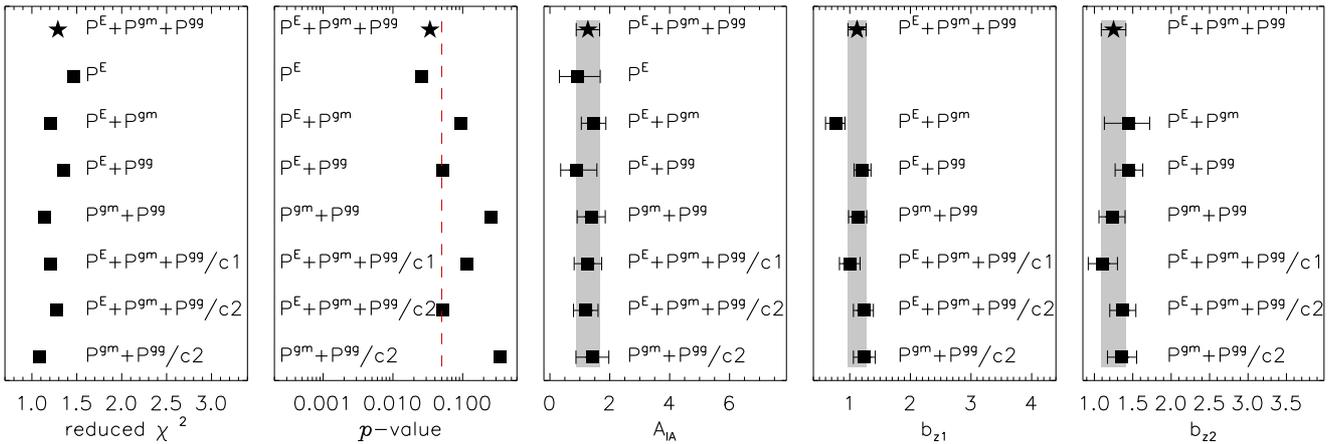}
   \caption{Reduced $\chi^2$ values of the best-fitting models, corresponding $p$-values of the fit, and constraints on the amplitude of the intrinsic alignment model $A_{\rm IA}$ and effective biases of the two foreground samples, $b_{\rm z1}$ and $b_{\rm z2}$, for the different combinations of power spectra. The lower points show the results of the conservative run, where we excluded the lowest $\ell$ bin from $P^{\rm E}$ (c1) and the highest $\ell$ bin from $P^{\rm gm }$ and $P^{\rm gg}$ (c2) in the fit. The red, vertical dashed line in the second panel indicates a $p$-value of 0.05, the 2$\sigma$ discrepancy line.}
   \label{plot_nuisance}
\end{figure*}
\indent We also performed two conservative runs to test the robustness of our results. In the first run, we excluded the lowest $\ell$ bins of all power spectra, as it has the largest IBC and our results might be biased if the correction is cosmology-dependent. In the second conservative run, we only removed the highest $\ell$ bins of $P^{\rm gg}$ and $P^{\rm gm}$, as these bins are potentially most biased if the effective galaxy bias (which we assumed to be constant) has some scale dependency, which would affect the small scales (largest $\ell$ bin) most. The constraints on $S_8$ are also shown in Fig. \ref{plot_S8}, and the relative increase in errors is shown in Fig. \ref{plot_S8err}. We find fully consistent results. The errors on $S_8$ increase by 4\% and 11\% for the first and second conservative run, compared to the fiducial results. As a final test, we fitted $P^{\rm gm}+P^{\rm gg}$ only excluding the highest $\ell$ bins. The difference between the constraint on $S_8$ from this run and the fit of $P^{\rm E}$  has decreased to 1.0$\sigma$, because of the increase of the error bars and because the results from $P^{\rm gm}+P^{\rm gg}$ are shifted to a slightly lower value. \\
\indent Figure \ref{plot_S8} shows that our results agree fairly well with a number of recent results from the literature. There is a mild discrepancy with the results from \citet{Kohlinger17}, which is noteworthy as they also used the KiDS-450 data set to estimate power spectra, but with a quadratic estimator. The difference is likely caused by a conspiracy of several effects. First of all, \citet{Kohlinger17} employed a different redshift binning and fitted the signal up to lower values of $\ell$, that is in the range $76<\ell<1310$; they report in their work that the signal on large scales prefers somewhat smaller values of $S_8$. In Appendix \ref{app_qe}, we directly compare the power spectrum estimators for the same redshift and $\ell$ bins. For the highest tomographic bins, the quadratic estimator band powers are lower than our $P^{\rm E}$ estimates at high $\ell$. This is accommodated by the model fit of \citet{Kohlinger17} with a large, negative intrinsic alignment amplitude of $A_{\rm IA}=-1.72$. Since $A_{\rm IA}$ and $S_8$ are correlated (e.g. see Fig. \ref{plot_zbias}), this pushes the $S_8$ from \citet{Kohlinger17} down relative to our results. Note that a thorough internal consistency check of KiDS-450 data, including a comparison of the information content from large and small scales, is currently underway (K\"ohlinger et al., in prep.). A more in-depth discussion of the difference is presented in Appendix \ref{app_qe}.  \\
\indent The constraints from the other works from the literature that we compare to are consistent with ours (i.e. differences are less than 2$\sigma$), as is shown in Fig. \ref{plot_S8}. This includes the results from \citet{Joudaki17}, who re-analysed the shear correlation functions from CFHTLenS \citep{Heymans13} using the extended version of {\sc CosmoMC} that we used here as well. \citet{Jee16} presented results based on a tomographic cosmic shear analysis of the Deep Lens Survey, a deep 20 deg$^2$ survey with a median source redshift of 1.2. Furthermore, we show the first constraints from the DES \citep{Abbott16}, who used 139 deg$^2$ of Science Verification data for a tomographic cosmic shear analysis, and finally, we show the results for {\it WMAP}9 \citep{Hinshaw13}. We caution that the above works have been analysed with different models and assumptions, which complicates a detailed comparison of the results. \\

\subsection{Constraints on astrophysical nuisance parameters} \label{sec_nuisance}

\indent Our analysis constrains a number of physical `nuisance' parameters, which are interesting in themselves. Their 1-D marginalized posterior means and 68\% confidence intervals are shown in Fig. \ref{plot_nuisance}, together with the reduced $\chi^2$ of the best-fitting model, for all combinations of power spectra as well as for the conservative runs. Overall, we find a fair agreement between the constraints between probes. Interestingly, the fit of $P^{\rm E}$ alone has the worst reduced $\chi^2$ of 1.46. However, that fit is relatively more affected by the highest $\ell$ bin of the S2--S4 cross-correlation, compared to the joint fit; excluding that bin from the fit leads to a reduced $\chi^2$ of 1.28, more in line with the $\chi^2$ values of the other fits. \\ 
\indent The amplitude of the intrinsic alignment model is well constrained in the combined fit, with $A_{\rm IA}=1.27\pm0.39$. Most of the constraining power on $A_{\rm IA}$ comes from $P^{\rm gm}$, as the redshift distributions of the foreground samples and the shape samples partly overlap; fitting only $P^{\rm E}$, $A_{\rm IA}=0.92_{-0.60}^{+0.76}$ and is therefore only inconclusively detected. In an analysis of cosmic shear data from CFHTLenS combined with WMAP7 results, \citet{Heymans13} reported $A_{\rm IA}=-1.18_{-1.17}^{+0.96}$. \citet{Joudaki17} analysed CFHTLenS data and found $A_{\rm IA}=-3.6\pm1.6$, while the correlation function analysis of KiDS \citep{Hildebrandt17} reported $A_{\rm IA}=1.10\pm0.64$. Hence, similar to \citet{Hildebrandt17}, our results prefer a positive intrinsic alignment amplitude, but we detect it with a larger significance. The preference for negative values in CFHTLenS but positive values in KiDS suggests that $A_{\rm IA}$ is not simply a measure of the amount of intrinsic alignments of galaxies, but that in fact it accounts for systematic effects that might differ between surveys. Further evidence for this scenario is that the amplitude we obtain is larger than what is expected based on results from previous dedicated intrinsic alignment studies; although intrinsic alignments have been detected for luminous red galaxies \citep[e.g.][]{Joachimi11,Singh15}, the constraints for less luminous red galaxies and blue galaxies are consistent with zero \citep{Mandelbaum06,Hirata07,Mandelbaum11}. We provide evidence that $A_{\rm IA}$ effectively accounts for uncertainty in the redshift distributions in Sect. \ref{sec_zshift}. \\
\indent The effective biases of the foreground samples are constrained to $b_{\rm z1}=1.12\pm0.15$ and $b_{\rm z2}=1.25\pm0.16$ in the combined fit. The most remarkable difference is the lower value for $b_{\rm z1}=0.78_{-0.18}^{+0.14}$ for $P^{\rm E}+P^{\rm gm}$, compared to $b_{\rm z1}=1.21\pm0.14$ for $P^{\rm E}+P^{\rm gg}$, which is a 2.1$\sigma$ difference. The constraint on the bias, however, is dominated by the angular correlation functions, which is expected as it scales quadratically with the effective bias while the galaxy-matter power spectrum only linearly. A direct comparison of our bias constraints with results from other work is complicated, since most studies focus on volume-limited rather than flux-limited samples, and because the fitting methodology is different. However, values a bit larger than unity are typical for samples selected in luminosity or stellar mass bins close to the mean of our sample \citep[e.g.][]{Zehavi11,Ying15,Crocce16}. Furthermore, we note that our cosmological results are not sensitive to the actual values of the biases, as the bias is degenerate with the $\Omega_{\rm m} - \sigma_8$ degeneracy, as illustrated in Fig. \ref{plot_cosm_mock_bias} in Sect. \ref{app_val_cosm}. \\
\indent The last physical nuisance parameter we fit is the baryonic feedback parameter $B$. Even when we include $P^{\rm gm}$ and $P^{\rm gg}$ in the fit, it is rather poorly constrained at $B=2.97_{-0.69}^{+0.56}$. Fitting $P^{\rm E}$ only, we obtain $B=3.26_{-0.22}^{+0.74}$, while \citet{Hildebrandt17} reported $B=2.88_{-0.88}^{+0.30}$. All results are consistent with $B=3.13$, a pure dark-matter-only model, but the errors are still large and do not rule out that baryonic feedback has some impact on the matter power spectrum \citep[which is supported by observational results on the scaling relation between baryonic properties of haloes and their dark matter content, see e.g.][]{Viola15}. \\
\indent The full marginalized 2-D posteriors of all fit parameter pairs is shown in Appendix \ref{app_fullpost}, and the mean and 68\% credible regions of the marginalized 1-D posteriors are listed in Table \ref{tab_posterior}.\\
\begin{figure*}
   \centering
   \includegraphics[width=\linewidth]{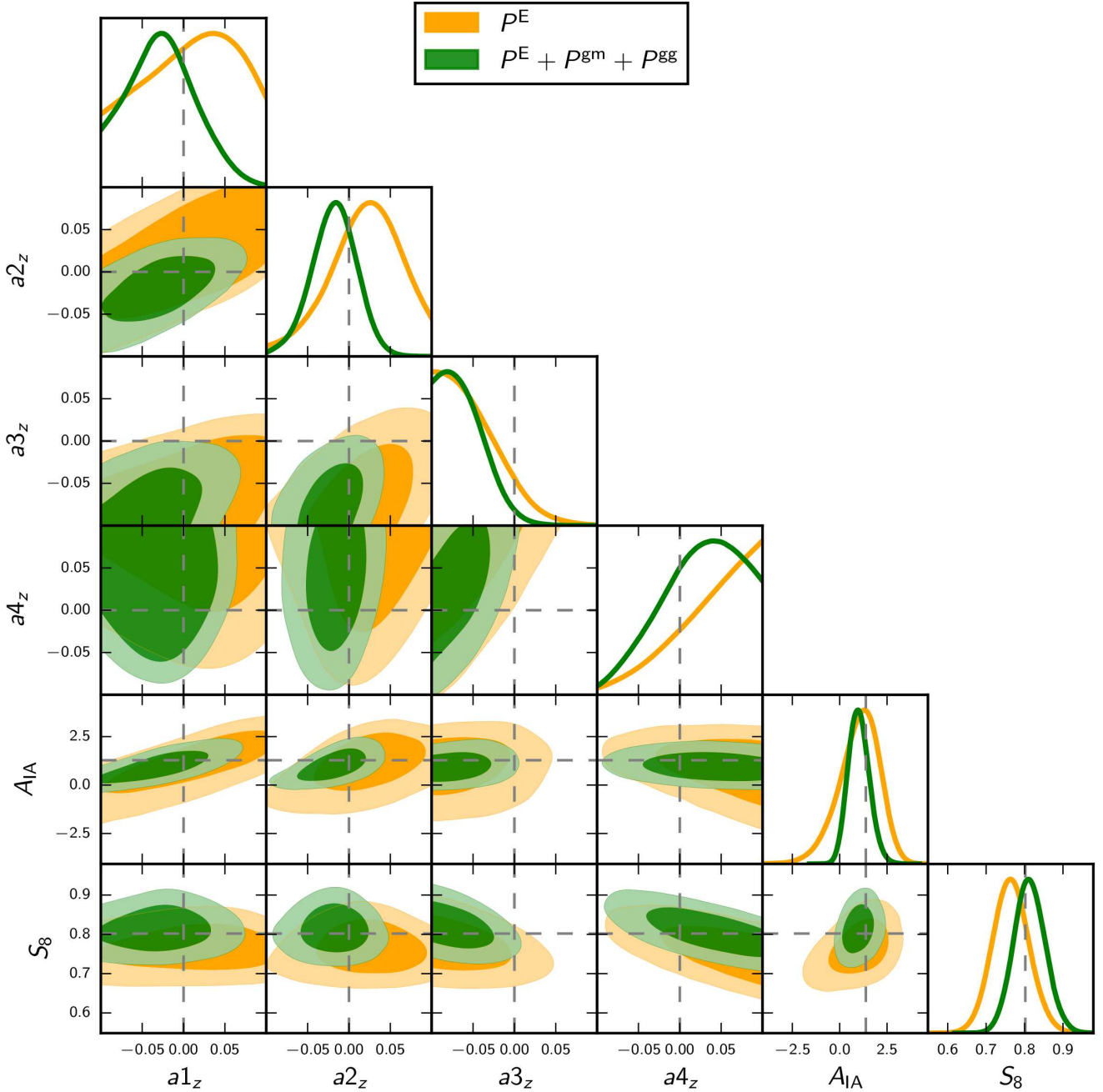}
   \caption{Posteriors on the shifts of the redshift distributions of our four tomographic source redshift bins $a1_z$ to $a4_z$. Grey dashed lines correspond to our fiducial results where the shifts are fixed to zero.}
   \label{plot_zbias}
\end{figure*}

\subsection{Redshift distribution uncertainty} \label{sec_zshift}

To investigate uncertainty in the redshift distribution, we performed an analysis where we allowed a constant shift in the redshift distributions of our source samples, independently for each tomographic source bin. These shifts $a[x]_z$ are defined as $n^{\rm shift}(z)=n^{\rm orig}(z+a[x]_z)$, with $n^{\rm orig}$ and $n^{\rm shift}$ the original and shifted source redshift distribution of tomographic bin $[x]$. We adopted priors in the range $[-0.1,0.1]$, as larger shifts are extremely unlikely, given the differences between the various photometric redshift methods tested in \citet{Hildebrandt17}. The purpose of this test is two-fold: it enables us to roughly estimate the impact of unknown systematic uncertainties in the redshift distributions on our results, and it also tests whether our data point to systematic biases in the redshift distributions. The actual redshift bias may be more complicated than a simple shift of the distribution, and future work could explore more complicated redshift bias models, such as changes to the tails of the distribution. \\
\indent We fit this extended model to our fiducial data set of $P^{\rm E}+P^{\rm gm}+P^{\rm gg}$ and to $P^{\rm E}$ only, to test whether they are internally consistent and to assess how much additional constraining power $P^{\rm gm}+P^{\rm gg}$ brings. The 2-D marginalized posteriors of these four shift parameters, together with those on $S_8$ and the intrinsic alignment amplitude, are shown in Fig. \ref{plot_zbias}. The constraints on the shifts are listed in Table \ref{tab_zbias}.  Both data sets clearly prefer a negative offset for the third tomographic bin of $\sim-0.05$. The joint analysis disfavours a zero shift in this bin at $\sim2\sigma$. \\
\indent It has been suggested that the redshift distribution obtained by the DIR method (our fiducial one) and the CC method \citep[a cross-correlation method based on the work of][]{Schmidt13,Menard13} are discrepant for this tomographic bin (Efstathiou \& Efstathiou, {\it priv. comm.}), and Fig. 2 of \citet{Hildebrandt17} indeed indicates that the redshift distribution of the CC method is shifted by roughly this amount towards lower values, relative to the DIR method. A similar shift between the mean of the redshift distributions of DIR and CC is reported in Table 1 of \citet{Morrison17}, although this shift is not significant there given that the error on the mean for the CC method is large. Further evidence is presented in Appendix A of \citet{Joudaki17kids}, who fit for an unknown constant offset of the KiDS-450 shear correlation functions per tomographic bin, and report a weak preference for a negative shift of the third tomographic bin,  and finally in \citet{Johnson17}, who cross-correlated galaxies from the 2dFLenS survey with KiDS galaxies using the same photometric redshift bins. For the other tomographic bins, the shifts are consistent with zero. Furthermore, it is interesting to note that including $P^{\rm gm}+P^{\rm gg}$ leads to tighter constraints on the shift for the first and second tomographic redshift bin, due to the overlap with the foreground sample. \\
\indent Allowing for a shift does not significantly change our constraints on $S_8$, as Fig. \ref{plot_zbias} shows. For the combined data set, we obtain $S_8=0.808_{-0.035}^{+0.036}$, entirely consistent with our fiducial $0.800_{-0.027}^{+0.029}$. The uncertainty in $S_8$ is 29\% larger when we include the redshift shifts in the fit. \\
\indent We also show the constraints on $A_{\rm IA}$ in Fig. \ref{plot_zbias}. As already alluded to in Sect. \ref{sec_nuisance}, $A_{\rm IA}$ may effectively serve as a genuine nuisance parameter, rather than a parameter which corresponds to the actual intrinsic alignment amplitude of galaxies. Allowing for the shifts already leads to weaker constraints centred at lower values, that is $A_{\rm IA}=0.89_{-0.58}^{+0.48}$ (for the combined fit), but even more interesting is the degeneracy with the shift of the first and second tomographic redshift bin. If the shifts of these two bins are negative, the intrinsic alignment amplitude becomes smaller, which shows that in our fiducial runs, where the shifts are fixed to zero, $A_{\rm IA}$ is at least partly serving as a nuisance parameter that absorbs potential biases in the redshift distributions.  

\begin{table}
  \centering
  \caption{Mean and 68\% credible intervals of the shifts of the four tomographic redshift bins, $S_8$ and $A_{\rm IA}$.}
  \begin{tabular}{c c c } 
  \hline
Bin shift parameter &  $P^{\rm E}+P^{\rm gm}+P^{\rm gg}$ & $P^{\rm E}$ \\
  \hline\hline
\\
$a1_z$ & $-0.033_{-0.050}^{+0.030} $ & $0.008_{-0.041}^{+0.077}$ \\
$a2_z$ & $-0.022_{-0.027}^{+0.030} $ & $0.018_{-0.036}^{+0.050}$ \\
$a3_z$ & $-0.057_{-0.042}^{+0.012}$ & $-0.051_{-0.049}^{+0.013}$ \\
$a4_z$ & $0.032_{-0.019}^{+0.068}$ & $0.039_{-0.017}^{+0.061}$ \\
$S_8$ & $0.808_{-0.035}^{+0.036}$ & $0.765\pm0.045$\\
$A_{\rm IA}$ & $0.89_{-0.58}^{+0.48}$ & $1.01_{-0.90}^{+1.18}$ \\
  \hline
  \end{tabular}
  \label{tab_zbias}
\end{table} 


\section{Conclusions}\label{sec_conc}
We constrained parameters of a flat $\Lambda$CDM model by combining three cosmological probes: the cosmic shear measurements from KiDS-450, the galaxy-matter cross correlation from KiDS-450 around two foreground samples of GAMA galaxies, and the angular correlation function of the same foreground galaxies. The analysis employed angular band power estimates determined from integrals over the corresponding two-point correlation functions. This simple formalism provides practically unbiased band powers over a considerable range in $\ell$. In our case, the range was $150<\ell<2000$ (see Appendix \ref{app_val_ps}). \\
\indent We fitted cosmological models to our data using the updated version of the {\sc cosmoMC} pipeline from \citet{Joudaki17}, extended to simultaneously model galaxy-galaxy lensing measurements \citep{Joudaki17kids2df}. The baseline model consists of a flat $\Lambda$CDM model and physically motivated prescriptions for the intrinsic alignment of galaxies and baryonic feedback. We assumed a scale-independent effective galaxy bias for our foreground samples. Fitting this model to the three sets of power spectra simultaneously enabled us to coherently account for the physical nuisance parameters, lifting degeneracies between the fit parameters. We tested our full pipeline on numerical simulations that are tailored to KiDS and recovered the input cosmology. \\
\indent In the model fitting, we used an analytical covariance matrix, accounting for all cross-correlations between power spectra and also for the partial spatial overlap between KiDS-450 and three equatorial GAMA patches. We validated the analytical covariance matrix with numerical simulations and obtained a reasonable level of agreement. Our approach of using an analytical covariance matrix has the advantage that we can accurately account for the effect of super-sample covariance, whose impact is subdominant compared to the other terms but not irrelevant. Furthermore, it enabled us to derive an iterative scheme where we updated the analytical covariance matrix with the best-fit parameters of the previous run (see Appendix \ref{app_iter}). This led to a $\sim1\sigma$ shift of the effective galaxy bias posteriors after one iteration, but the posterior of $S_8$ was not significantly affected (i.e. a shift of less than 0.5$\sigma$ in the mean). \\
\indent We obtained tight constraints on the two cosmological parameters to which weak gravitational lensing is most sensitive, $\Omega_{\rm m}$ and $\sigma_8$. Our results can be summarised with the $S_8$ parameter, for which we obtained $S_8\equiv \sigma_8 \sqrt{\Omega_{\rm m}/0.3}=0.800_{-0.027}^{+0.029}$. We demonstrated that our three probes are internally consistent, and that including $P^{\rm gm}$ and $P^{\rm gg}$ in the fit leads to a 26\% improvement in constraining power on $S_8$. We compared our results to a number of recent studies from the literature and found good overall agreement. The fiducial KiDS-450 cosmic shear correlation function analysis \citep{Hildebrandt17} revealed a value for $S_8$ that is lower than the Planck cosmology, with the tension being 2.3$\sigma$. Interestingly, the results from our combined probe analysis point to a somewhat higher $S_8$ value, in-between the KiDS-450 and Planck results (and consistent with both). Our constraints from cosmic shear alone are fully consistent with the results from \citet{Hildebrandt17} and maintain the same level of discrepancy with Planck. \\
\indent The physical nuisance parameters that we marginalise over are interesting in themselves from an astrophysical perspective. However, they have to be interpreted with care. For example, when taken at face value, our constraints on the intrinsic alignment amplitude, $A_{\rm IA}=1.27\pm0.39$, suggests that galaxies are on average intrinsically aligned with the large-scale density field. However, by allowing for an additional shift in the source redshift distributions in the fit, we demonstrated that $A_{\rm IA}$ could partly work as a nuisance parameter that accounts for such residual biases in the redshift distributions. This test also demonstrated that our data prefers a small, negative shift of the redshift distribution of the third bin with $\Delta z=-0.057_{-0.042}^{+0.012}$, but this does not impact our cosmological results. The nuisance parameters are better constrained when $P^{\rm gm}$ and $P^{\rm gg}$ are included in the fit, which highlights the power of combining these cosmological probes. Marginalising over wide priors on the mean of the tomographic redshift distributions yields consistent results for $S_8$ with an increase of $28\%$ in the error. \\
\indent As in \citet{Hildebrandt17}, we detect B-modes in the cosmic shear signal at a low level. Under the strong assumption that the underlying residual systematic generates the same amount of E- and B-modes, we obtain a $0.5 \sigma$ shift in $S_8$ away from the Planck values. As the cause of the systematic is currently unknown, we caution that its eventual correction could lead to similar changes in the $S_8$ posterior. \\
\indent Another KiDS study of a similar nature has run in parallel to this work and will be presented imminently in \citet{Joudaki17kids2df} In that work, cosmic shear measurements from KiDS-450 are combined with galaxy-galaxy lensing and redshift space distortion measurements for a foreground sample of galaxies from BOSS \citep{Dawson13} and 2dFLenS \citep{Blake16}. Even though the analyses differ in many aspects (e.g. different lens samples, different clustering statistics, different scales used in the fit, different methods to estimate the covariance matrices and different priors in the fit), the combination of probes used in that work lead to a similar $\sim 20\%$ decrease of the error bar of $S_8$, but maintain in tension with Planck. \\
\indent Our work shows that large-scale structure self-calibration methods work on real data. This not only leads to significant improvements in the constraints of cosmological parameters, but also properly accounts for nuisance effects such as galaxy bias, intrinsic alignments, and biases in the source redshift distributions. Future extensions of this work will include other cosmological probes such as redshift space distortions, but also explore extensions of the astrophysical and cosmological models considered. 

\paragraph*{Acknowledgements}
 We thank Elisabeth Krause for a helpful comparison of covariance models, Hiranya Peiris for useful discussions, and Andy Taylor, Niall MacCrann and Gerrit Schellenberger for useful feedback on the draft. We thank our referee, Joe Zuntz, for valuable feedback. EvU and BJ acknowledge support from an STFC Ernest Rutherford Research Grant, grant reference ST/L00285X/1. BJ acknowledges support by an STFC Ernest Rutherford Fellowship, grant reference ST/J004421/1. JHD acknowledges support from the European Commission under a Marie-Sk{l}odwoska-Curie European Fellowship (EU project 656869). CH and MA acknowledge support from the European Research Council under grant number 647112. CB acknowledges the support of the Australian Research Council through the award of a Future Fellowship. Parts of this research were conducted by the Australian Research Council Centre of Excellence for All-sky Astrophysics (CAASTRO), through project number CE110001020. H. Hildebrandt is supported by an Emmy Noether grant (No. Hi 1495/2-1) of the Deutsche Forschungsgemeinschaft. This work was supported by the World Premier International Research Center Initiative (WPI), MEXT, Japan. TDK is supported by a Royal Society URF. This work is supported by the Deutsche Forschungsgemeinschaft in the framework of the TR33 `The Dark Universe'. KK acknowledges support by the Alexander von Humboldt Foundation. The research leading to these results has received funding from the People Programme (Marie Curie Actions) of the European Union's Seventh Framework Programme (FP7/2007-2013) under REA grant agreement number 627288. RN acknowledges support from the German Federal Ministry for Economic Affairs and Energy (BMWi) provided via DLR under project no. 50QE1103. MV acknowledges support from the European Research Council under FP7 grant number 279396 and the Netherlands Organisation for Scientific Research (NWO) through grants 614.001.103. This work is based on data products from observations made with ESO Telescopes at the La Silla Paranal Observatory under programme IDs 177.A-3016, 177.A-3017 and 177.A-3018. GAMA is a joint European-Australasian project based around a spectroscopic campaign using the Anglo-Australian Telescope. The GAMA input catalogue is based on data taken from the Sloan Digital Sky Survey and the UKIRT Infrared Deep Sky Survey. Complementary imaging of the GAMA regions is being obtained by a number of independent survey programs including GALEX MIS, VST KiDS, VISTA VIKING, WISE, Herschel-ATLAS, GMRT and ASKAP providing UV to radio coverage. GAMA is funded by the STFC (UK), the ARC (Australia), the AAO, and the participating institutions. The GAMA website is http://www.gama-survey.org/.
\indent {\it Author Contributions}: All authors contributed to the development and writing of this paper. The authorship list is given in three groups: the lead authors (EvU, BJ, SJ), followed by two alphabetical groups. The first alphabetical group includes those who are key contributors to both the scientific analysis and the data products. The second group covers those who have either made a significant contribution to the data products, or to the scientific analysis.

\bibliographystyle{mnras}

\begin{thebibliography}{}
\makeatletter
\relax
\def\mn@urlcharsother{\let\do\@makeother \do\$\do\&\do\#\do\^\do\_\do\%\do\~}
\def\mn@doi{\begingroup\mn@urlcharsother \@ifnextchar [ {\mn@doi@}
  {\mn@doi@[]}}
\def\mn@doi@[#1]#2{\def\@tempa{#1}\ifx\@tempa\@empty \href
  {http://dx.doi.org/#2} {doi:#2}\else \href {http://dx.doi.org/#2} {#1}\fi
  \endgroup}
\def\mn@eprint#1#2{\mn@eprint@#1:#2::\@nil}
\def\mn@eprint@arXiv#1{\href {http://arxiv.org/abs/#1} {{\tt arXiv:#1}}}
\def\mn@eprint@dblp#1{\href {http://dblp.uni-trier.de/rec/bibtex/#1.xml}
  {dblp:#1}}
\def\mn@eprint@#1:#2:#3:#4\@nil{\def\@tempa {#1}\def\@tempb {#2}\def\@tempc
  {#3}\ifx \@tempc \@empty \let \@tempc \@tempb \let \@tempb \@tempa \fi \ifx
  \@tempb \@empty \def\@tempb {arXiv}\fi \@ifundefined
  {mn@eprint@\@tempb}{\@tempb:\@tempc}{\expandafter \expandafter \csname
  mn@eprint@\@tempb\endcsname \expandafter{\@tempc}}}

\bibitem[\protect\citeauthoryear{{Abbott} et~al.,}{{Abbott}
  et~al.}{2016}]{Abbott16}
{Abbott} T.,  et~al., 2016, \mn@doi [\prd] {10.1103/PhysRevD.94.022001}, \href
  {http://adsabs.harvard.edu/abs/2016PhRvD..94b2001A} {94, 022001}

\bibitem[\protect\citeauthoryear{{Albrecht} et~al.,}{{Albrecht}
  et~al.}{2006}]{Albrecht06}
{Albrecht} A.,  et~al., 2006, preprint, \href
  {http://adsabs.harvard.edu/abs/2006astro.ph..9591A} {} (\mn@eprint {arXiv}
  {0609591})

\bibitem[\protect\citeauthoryear{{Alsing}, {Heavens}  \& {Jaffe}}{{Alsing}
  et~al.}{2017}]{Alsing17}
{Alsing} J.,  {Heavens} A.,   {Jaffe} A.~H.,  2017, \mn@doi [\mnras]
  {10.1093/mnras/stw3161}, \href
  {http://adsabs.harvard.edu/abs/2017MNRAS.466.3272A} {466, 3272}

\bibitem[\protect\citeauthoryear{{Amara} \& {R{\'e}fr{\'e}gier}}{{Amara} \&
  {R{\'e}fr{\'e}gier}}{2007}]{Amara07}
{Amara} A.,  {R{\'e}fr{\'e}gier} A.,  2007, \mn@doi [\mnras]
  {10.1111/j.1365-2966.2007.12271.x}, \href
  {http://adsabs.harvard.edu/abs/2007MNRAS.381.1018A} {381, 1018}

\bibitem[\protect\citeauthoryear{{Asgari}, {Taylor}, {Joachimi}  \&
  {Kitching}}{{Asgari} et~al.}{2016}]{Asgari16}
{Asgari} M.,  {Taylor} A.,  {Joachimi} B.,   {Kitching} T.~D.,  2016, preprint,
  \href {http://adsabs.harvard.edu/abs/2016arXiv161204664A} {} (\mn@eprint
  {arXiv} {1612.04664})

\bibitem[\protect\citeauthoryear{{Bartelmann} \& {Schneider}}{{Bartelmann} \&
  {Schneider}}{2001}]{Bartelmann01}
{Bartelmann} M.,  {Schneider} P.,  2001, \mn@doi [\physrep]
  {10.1016/S0370-1573(00)00082-X}, \href
  {http://adsabs.harvard.edu/abs/2001PhR...340..291B} {340, 291}

\bibitem[\protect\citeauthoryear{{Battye} \& {Moss}}{{Battye} \&
  {Moss}}{2014}]{Battye14}
{Battye} R.~A.,  {Moss} A.,  2014, \mn@doi [Physical Review Letters]
  {10.1103/PhysRevLett.112.051303}, \href
  {http://adsabs.harvard.edu/abs/2014PhRvL.112e1303B} {112, 051303}

\bibitem[\protect\citeauthoryear{{Becker} \& {Rozo}}{{Becker} \&
  {Rozo}}{2016}]{Becker16P}
{Becker} M.~R.,  {Rozo} E.,  2016, \mn@doi [\mnras] {10.1093/mnras/stv3018},
  \href {http://adsabs.harvard.edu/abs/2016MNRAS.457..304B} {457, 304}

\bibitem[\protect\citeauthoryear{{Becker} et~al.,}{{Becker}
  et~al.}{2016}]{Becker16}
{Becker} M.~R.,  et~al., 2016, \mn@doi [\prd] {10.1103/PhysRevD.94.022002},
  \href {http://adsabs.harvard.edu/abs/2016PhRvD..94b2002B} {94, 022002}

\bibitem[\protect\citeauthoryear{{Ben{\'{\i}}tez}}{{Ben{\'{\i}}tez}}{2000}]{Benitez00}
{Ben{\'{\i}}tez} N.,  2000, \mn@doi [\apj] {10.1086/308947}, \href
  {http://adsabs.harvard.edu/abs/2000ApJ...536..571B} {536, 571}

\bibitem[\protect\citeauthoryear{{Blake} et~al.,}{{Blake}
  et~al.}{2016}]{Blake16}
{Blake} C.,  et~al., 2016, \mn@doi [\mnras] {10.1093/mnras/stw1990}, \href
  {http://adsabs.harvard.edu/abs/2016MNRAS.462.4240B} {462, 4240}

\bibitem[\protect\citeauthoryear{{Bridle} \& {King}}{{Bridle} \&
  {King}}{2007}]{Bridle07}
{Bridle} S.,  {King} L.,  2007, \mn@doi [New Journal of Physics]
  {10.1088/1367-2630/9/12/444}, \href
  {http://adsabs.harvard.edu/abs/2007NJPh....9..444B} {9, 444}

\bibitem[\protect\citeauthoryear{{Brown}, {Taylor}, {Bacon}, {Gray}, {Dye},
  {Meisenheimer}  \& {Wolf}}{{Brown} et~al.}{2003}]{Brown03}
{Brown} M.~L.,  {Taylor} A.~N.,  {Bacon} D.~J.,  {Gray} M.~E.,  {Dye} S.,
  {Meisenheimer} K.,   {Wolf} C.,  2003, \mn@doi [\mnras]
  {10.1046/j.1365-8711.2003.06237.x}, \href
  {http://adsabs.harvard.edu/abs/2003MNRAS.341..100B} {341, 100}

\bibitem[\protect\citeauthoryear{{Cacciato}, {Lahav}, {van den Bosch},
  {Hoekstra}  \& {Dekel}}{{Cacciato} et~al.}{2012}]{Cacciato12}
{Cacciato} M.,  {Lahav} O.,  {van den Bosch} F.~C.,  {Hoekstra} H.,   {Dekel}
  A.,  2012, \mn@doi [\mnras] {10.1111/j.1365-2966.2012.21762.x}, \href
  {http://adsabs.harvard.edu/abs/2012MNRAS.426..566C} {426, 566}

\bibitem[\protect\citeauthoryear{{Cacciato}, {van den Bosch}, {More}, {Mo}  \&
  {Yang}}{{Cacciato} et~al.}{2013}]{Cacciato13}
{Cacciato} M.,  {van den Bosch} F.~C.,  {More} S.,  {Mo} H.,   {Yang} X.,
  2013, \mn@doi [\mnras] {10.1093/mnras/sts525}, \href
  {http://adsabs.harvard.edu/abs/2013MNRAS.430..767C} {430, 767}

\bibitem[\protect\citeauthoryear{{Catelan}, {Kamionkowski}  \&
  {Blandford}}{{Catelan} et~al.}{2001}]{Catelan01}
{Catelan} P.,  {Kamionkowski} M.,   {Blandford} R.~D.,  2001, \mn@doi [\mnras]
  {10.1046/j.1365-8711.2001.04105.x}, \href
  {http://adsabs.harvard.edu/abs/2001MNRAS.320L...7C} {320, L7}

\bibitem[\protect\citeauthoryear{{Chon}, {Challinor}, {Prunet}, {Hivon}  \&
  {Szapudi}}{{Chon} et~al.}{2004}]{Chon04}
{Chon} G.,  {Challinor} A.,  {Prunet} S.,  {Hivon} E.,   {Szapudi} I.,  2004,
  \mn@doi [\mnras] {10.1111/j.1365-2966.2004.07737.x}, \href
  {http://adsabs.harvard.edu/abs/2004MNRAS.350..914C} {350, 914}

\bibitem[\protect\citeauthoryear{{Crocce} et~al.,}{{Crocce}
  et~al.}{2016}]{Crocce16}
{Crocce} M.,  et~al., 2016, \mn@doi [\mnras] {10.1093/mnras/stv2590}, \href
  {http://adsabs.harvard.edu/abs/2016MNRAS.455.4301C} {455, 4301}

\bibitem[\protect\citeauthoryear{{Dawson} et~al.,}{{Dawson}
  et~al.}{2013}]{Dawson13}
{Dawson} K.~S.,  et~al., 2013, \mn@doi [\aj] {10.1088/0004-6256/145/1/10},
  \href {http://adsabs.harvard.edu/abs/2013AJ....145...10D} {145, 10}

\bibitem[\protect\citeauthoryear{{Dekel} \& {Lahav}}{{Dekel} \&
  {Lahav}}{1999}]{Dekel99}
{Dekel} A.,  {Lahav} O.,  1999, \mn@doi [\apj] {10.1086/307428}, \href
  {http://adsabs.harvard.edu/abs/1999ApJ...520...24D} {520, 24}

\bibitem[\protect\citeauthoryear{{Driver} et~al.,}{{Driver}
  et~al.}{2009}]{Driver09}
{Driver} S.~P.,  et~al., 2009, \mn@doi [Astronomy and Geophysics]
  {10.1111/j.1468-4004.2009.50512.x}, \href
  {http://adsabs.harvard.edu/abs/2009A%26G....50e..12D} {50, 050000}

\bibitem[\protect\citeauthoryear{{Driver} et~al.,}{{Driver}
  et~al.}{2011}]{Driver11}
{Driver} S.~P.,  et~al., 2011, \mn@doi [\mnras]
  {10.1111/j.1365-2966.2010.18188.x}, \href
  {http://adsabs.harvard.edu/abs/2011MNRAS.413..971D} {413, 971}

\bibitem[\protect\citeauthoryear{{Dvornik} et~al.,}{{Dvornik}
  et~al.}{2017}]{Dvornik17}
{Dvornik} A.,  et~al., 2017, \mn@doi [\mnras] {10.1093/mnras/stx705}, \href
  {http://adsabs.harvard.edu/abs/2017MNRAS.468.3251D} {468, 3251}

\bibitem[\protect\citeauthoryear{{Edge}, {Sutherland}, {Kuijken}, {Driver},
  {McMahon}, {Eales}  \& {Emerson}}{{Edge} et~al.}{2013}]{Edge13}
{Edge} A.,  {Sutherland} W.,  {Kuijken} K.,  {Driver} S.,  {McMahon} R.,
  {Eales} S.,   {Emerson} J.~P.,  2013, The Messenger, \href
  {http://adsabs.harvard.edu/abs/2013Msngr.154...32E} {154, 32}

\bibitem[\protect\citeauthoryear{{Eifler}, {Schneider}  \& {Hartlap}}{{Eifler}
  et~al.}{2009}]{Eifler09}
{Eifler} T.,  {Schneider} P.,   {Hartlap} J.,  2009, \mn@doi [\aap]
  {10.1051/0004-6361/200811276}, \href
  {http://adsabs.harvard.edu/abs/2009A%26A...502..721E} {502, 721}

\bibitem[\protect\citeauthoryear{{Eisenstein} \& {Hu}}{{Eisenstein} \&
  {Hu}}{1998}]{Eisenstein98}
{Eisenstein} D.~J.,  {Hu} W.,  1998, \mn@doi [\apj] {10.1086/305424}, \href
  {http://adsabs.harvard.edu/abs/1998ApJ...496..605E} {496, 605}

\bibitem[\protect\citeauthoryear{{Farrow} et~al.,}{{Farrow}
  et~al.}{2015}]{Farrow15}
{Farrow} D.~J.,  et~al., 2015, \mn@doi [\mnras] {10.1093/mnras/stv2075}, \href
  {http://adsabs.harvard.edu/abs/2015MNRAS.454.2120F} {454, 2120}

\bibitem[\protect\citeauthoryear{{Fenech Conti}, {Herbonnet}, {Hoekstra},
  {Merten}, {Miller}  \& {Viola}}{{Fenech Conti} et~al.}{2017}]{Fenech16}
{Fenech Conti} I.,  {Herbonnet} R.,  {Hoekstra} H.,  {Merten} J.,  {Miller} L.,
    {Viola} M.,  2017, \mn@doi [\mnras] {10.1093/mnras/stx200}, \href
  {http://adsabs.harvard.edu/abs/2017MNRAS.467.1627F} {467, 1627}

\bibitem[\protect\citeauthoryear{{Gelman} \& {Rubin}}{{Gelman} \&
  {Rubin}}{1992}]{Gelman92}
{Gelman} A.,  {Rubin} D.~B.,  1992, Stat. Sci., 7, 457

\bibitem[\protect\citeauthoryear{{Giannantonio}, {Porciani}, {Carron}, {Amara}
  \& {Pillepich}}{{Giannantonio} et~al.}{2012}]{Giannantonio12}
{Giannantonio} T.,  {Porciani} C.,  {Carron} J.,  {Amara} A.,   {Pillepich} A.,
   2012, \mn@doi [\mnras] {10.1111/j.1365-2966.2012.20604.x}, \href
  {http://adsabs.harvard.edu/abs/2012MNRAS.422.2854G} {422, 2854}

\bibitem[\protect\citeauthoryear{{Harnois-D{\'e}raps} \& {van
  Waerbeke}}{{Harnois-D{\'e}raps} \& {van Waerbeke}}{2015}]{HarnoisDeraps15}
{Harnois-D{\'e}raps} J.,  {van Waerbeke} L.,  2015, \mn@doi [\mnras]
  {10.1093/mnras/stv794}, \href
  {http://adsabs.harvard.edu/abs/2015MNRAS.450.2857H} {450, 2857}

\bibitem[\protect\citeauthoryear{{Heymans} et~al.,}{{Heymans}
  et~al.}{2005}]{Heymans05}
{Heymans} C.,  et~al., 2005, \mn@doi [\mnras]
  {10.1111/j.1365-2966.2005.09152.x}, \href
  {http://adsabs.harvard.edu/abs/2005MNRAS.361..160H} {361, 160}

\bibitem[\protect\citeauthoryear{{Heymans} et~al.,}{{Heymans}
  et~al.}{2013}]{Heymans13}
{Heymans} C.,  et~al., 2013, \mn@doi [\mnras] {10.1093/mnras/stt601}, \href
  {http://adsabs.harvard.edu/abs/2013MNRAS.432.2433H} {432, 2433}

\bibitem[\protect\citeauthoryear{{Hikage}, {Takada}, {Hamana}  \&
  {Spergel}}{{Hikage} et~al.}{2011}]{Hikage11}
{Hikage} C.,  {Takada} M.,  {Hamana} T.,   {Spergel} D.,  2011, \mn@doi
  [\mnras] {10.1111/j.1365-2966.2010.17886.x}, \href
  {http://adsabs.harvard.edu/abs/2011MNRAS.412...65H} {412, 65}

\bibitem[\protect\citeauthoryear{{Hildebrandt} et~al.,}{{Hildebrandt}
  et~al.}{2017}]{Hildebrandt17}
{Hildebrandt} H.,  et~al., 2017, \mn@doi [\mnras] {10.1093/mnras/stw2805},
  \href {http://adsabs.harvard.edu/abs/2017MNRAS.465.1454H} {465, 1454}

\bibitem[\protect\citeauthoryear{{Hinshaw} et~al.,}{{Hinshaw}
  et~al.}{2013}]{Hinshaw13}
{Hinshaw} G.,  et~al., 2013, \mn@doi [\apjs] {10.1088/0067-0049/208/2/19},
  \href {http://adsabs.harvard.edu/abs/2013ApJS..208...19H} {208, 19}

\bibitem[\protect\citeauthoryear{{Hirata} \& {Seljak}}{{Hirata} \&
  {Seljak}}{2004}]{Hirata04}
{Hirata} C.~M.,  {Seljak} U.,  2004, \mn@doi [\prd]
  {10.1103/PhysRevD.70.063526}, \href
  {http://adsabs.harvard.edu/abs/2004PhRvD..70f3526H} {70, 063526}

\bibitem[\protect\citeauthoryear{{Hirata} \& {Seljak}}{{Hirata} \&
  {Seljak}}{2010}]{Hirata10}
{Hirata} C.~M.,  {Seljak} U.,  2010, \mn@doi [\prd]
  {10.1103/PhysRevD.82.049901}, \href
  {http://adsabs.harvard.edu/abs/2010PhRvD..82d9901H} {82, 049901}

\bibitem[\protect\citeauthoryear{{Hirata}, {Mandelbaum}, {Ishak}, {Seljak},
  {Nichol}, {Pimbblet}, {Ross}  \& {Wake}}{{Hirata} et~al.}{2007}]{Hirata07}
{Hirata} C.~M.,  {Mandelbaum} R.,  {Ishak} M.,  {Seljak} U.,  {Nichol} R.,
  {Pimbblet} K.~A.,  {Ross} N.~P.,   {Wake} D.,  2007, \mn@doi [\mnras]
  {10.1111/j.1365-2966.2007.12312.x}, \href
  {http://adsabs.harvard.edu/abs/2007MNRAS.381.1197H} {381, 1197}

\bibitem[\protect\citeauthoryear{{Hoekstra}, {van Waerbeke}, {Gladders},
  {Mellier}  \& {Yee}}{{Hoekstra} et~al.}{2002}]{Hoekstra02}
{Hoekstra} H.,  {van Waerbeke} L.,  {Gladders} M.~D.,  {Mellier} Y.,   {Yee}
  H.~K.~C.,  2002, \mn@doi [\apj] {10.1086/342228}, \href
  {http://adsabs.harvard.edu/abs/2002ApJ...577..604H} {577, 604}

\bibitem[\protect\citeauthoryear{{Hu} \& {White}}{{Hu} \& {White}}{2001}]{Hu01}
{Hu} W.,  {White} M.,  2001, \mn@doi [\apj] {10.1086/321380}, \href
  {http://adsabs.harvard.edu/abs/2001ApJ...554...67H} {554, 67}

\bibitem[\protect\citeauthoryear{{Jain} \& {Seljak}}{{Jain} \&
  {Seljak}}{1997}]{Jain97}
{Jain} B.,  {Seljak} U.,  1997, \mn@doi [\apj] {10.1086/304372}, \href
  {http://adsabs.harvard.edu/abs/1997ApJ...484..560J} {484, 560}

\bibitem[\protect\citeauthoryear{{Jarvis}, {Bernstein}  \& {Jain}}{{Jarvis}
  et~al.}{2004}]{Jarvis04}
{Jarvis} M.,  {Bernstein} G.,   {Jain} B.,  2004, \mn@doi [\mnras]
  {10.1111/j.1365-2966.2004.07926.x}, \href
  {http://adsabs.harvard.edu/abs/2004MNRAS.352..338J} {352, 338}

\bibitem[\protect\citeauthoryear{{Jee}, {Tyson}, {Schneider}, {Wittman},
  {Schmidt}  \& {Hilbert}}{{Jee} et~al.}{2013}]{Jee13}
{Jee} M.~J.,  {Tyson} J.~A.,  {Schneider} M.~D.,  {Wittman} D.,  {Schmidt} S.,
   {Hilbert} S.,  2013, \mn@doi [\apj] {10.1088/0004-637X/765/1/74}, \href
  {http://adsabs.harvard.edu/abs/2013ApJ...765...74J} {765, 74}

\bibitem[\protect\citeauthoryear{{Jee}, {Tyson}, {Hilbert}, {Schneider},
  {Schmidt}  \& {Wittman}}{{Jee} et~al.}{2016}]{Jee16}
{Jee} M.~J.,  {Tyson} J.~A.,  {Hilbert} S.,  {Schneider} M.~D.,  {Schmidt} S.,
   {Wittman} D.,  2016, \mn@doi [\apj] {10.3847/0004-637X/824/2/77}, \href
  {http://adsabs.harvard.edu/abs/2016ApJ...824...77J} {824, 77}

\bibitem[\protect\citeauthoryear{{Joachimi} \& {Bridle}}{{Joachimi} \&
  {Bridle}}{2010}]{Joachimi10}
{Joachimi} B.,  {Bridle} S.~L.,  2010, \mn@doi [\aap]
  {10.1051/0004-6361/200913657}, \href
  {http://adsabs.harvard.edu/abs/2010A%26A...523A...1J} {523, A1}

\bibitem[\protect\citeauthoryear{{Joachimi}, {Schneider}  \&
  {Eifler}}{{Joachimi} et~al.}{2008}]{Joachimi08}
{Joachimi} B.,  {Schneider} P.,   {Eifler} T.,  2008, \mn@doi [\aap]
  {10.1051/0004-6361:20078400}, \href
  {http://adsabs.harvard.edu/abs/2008A%26A...477...43J} {477, 43}

\bibitem[\protect\citeauthoryear{{Joachimi}, {Mandelbaum}, {Abdalla}  \&
  {Bridle}}{{Joachimi} et~al.}{2011}]{Joachimi11}
{Joachimi} B.,  {Mandelbaum} R.,  {Abdalla} F.~B.,   {Bridle} S.~L.,  2011,
  \mn@doi [\aap] {10.1051/0004-6361/201015621}, \href
  {http://adsabs.harvard.edu/abs/2011A%26A...527A..26J} {527, A26}

\bibitem[\protect\citeauthoryear{{Johnson} et~al.,}{{Johnson}
  et~al.}{2017}]{Johnson17}
{Johnson} A.,  et~al., 2017, \mn@doi [\mnras] {10.1093/mnras/stw3033}, \href
  {http://adsabs.harvard.edu/abs/2017MNRAS.465.4118J} {465, 4118}

\bibitem[\protect\citeauthoryear{{Joudaki} et~al.,}{{Joudaki}
  et~al.}{2016}]{Joudaki17kids}
{Joudaki} S.,  et~al., 2016, preprint, \href
  {http://adsabs.harvard.edu/abs/2016arXiv161004606J} {} (\mn@eprint {arXiv}
  {1610.04606})

\bibitem[\protect\citeauthoryear{{Joudaki} et~al.,}{{Joudaki}
  et~al.}{2017a}]{Joudaki17kids2df}
{Joudaki} S.,  et~al., 2017a, preprint, \href
  {http://adsabs.harvard.edu/abs/2017arXiv170706627J} {} (\mn@eprint {arXiv}
  {1707.06627})

\bibitem[\protect\citeauthoryear{{Joudaki} et~al.,}{{Joudaki}
  et~al.}{2017b}]{Joudaki17}
{Joudaki} S.,  et~al., 2017b, \mn@doi [\mnras] {10.1093/mnras/stw2665}, \href
  {http://adsabs.harvard.edu/abs/2017MNRAS.465.2033J} {465, 2033}

\bibitem[\protect\citeauthoryear{{Jullo} et~al.,}{{Jullo}
  et~al.}{2012}]{Jullo12}
{Jullo} E.,  et~al., 2012, \mn@doi [\apj] {10.1088/0004-637X/750/1/37}, \href
  {http://adsabs.harvard.edu/abs/2012ApJ...750...37J} {750, 37}

\bibitem[\protect\citeauthoryear{{Kilbinger}}{{Kilbinger}}{2015}]{Kilbinger15}
{Kilbinger} M.,  2015, \mn@doi [Reports on Progress in Physics]
  {10.1088/0034-4885/78/8/086901}, \href
  {http://adsabs.harvard.edu/abs/2015RPPh...78h6901K} {78, 086901}

\bibitem[\protect\citeauthoryear{{Kilbinger} et~al.,}{{Kilbinger}
  et~al.}{2017}]{Kilbinger17}
{Kilbinger} M.,  et~al., 2017, preprint, \href
  {http://adsabs.harvard.edu/abs/2017arXiv170205301K} {} (\mn@eprint {arXiv}
  {1702.05301})

\bibitem[\protect\citeauthoryear{{Kitching}, {Heavens}, {Taylor}, {Brown},
  {Meisenheimer}, {Wolf}, {Gray}  \& {Bacon}}{{Kitching}
  et~al.}{2007}]{Kitching07}
{Kitching} T.~D.,  {Heavens} A.~F.,  {Taylor} A.~N.,  {Brown} M.~L.,
  {Meisenheimer} K.,  {Wolf} C.,  {Gray} M.~E.,   {Bacon} D.~J.,  2007, \mn@doi
  [\mnras] {10.1111/j.1365-2966.2007.11473.x}, \href
  {http://adsabs.harvard.edu/abs/2007MNRAS.376..771K} {376, 771}

\bibitem[\protect\citeauthoryear{{Kitching} et~al.,}{{Kitching}
  et~al.}{2014}]{Kitching14}
{Kitching} T.~D.,  et~al., 2014, \mn@doi [\mnras] {10.1093/mnras/stu934}, \href
  {http://adsabs.harvard.edu/abs/2014MNRAS.442.1326K} {442, 1326}

\bibitem[\protect\citeauthoryear{{Kitching}, {Verde}, {Heavens}  \&
  {Jimenez}}{{Kitching} et~al.}{2016}]{Kitching16b}
{Kitching} T.~D.,  {Verde} L.,  {Heavens} A.~F.,   {Jimenez} R.,  2016, \mn@doi
  [\mnras] {10.1093/mnras/stw707}, \href
  {http://adsabs.harvard.edu/abs/2016MNRAS.459..971K} {459, 971}

\bibitem[\protect\citeauthoryear{{Kitching}, {Alsing}, {Heavens}, {Jimenez},
  {McEwen}  \& {Verde}}{{Kitching} et~al.}{2017}]{Kitching16}
{Kitching} T.~D.,  {Alsing} J.,  {Heavens} A.~F.,  {Jimenez} R.,  {McEwen}
  J.~D.,   {Verde} L.,  2017, \mn@doi [\mnras] {10.1093/mnras/stx1039}, 469,
  2737

\bibitem[\protect\citeauthoryear{{K{\"o}hlinger}, {Viola}, {Valkenburg},
  {Joachimi}, {Hoekstra}  \& {Kuijken}}{{K{\"o}hlinger}
  et~al.}{2016}]{Kohlinger16}
{K{\"o}hlinger} F.,  {Viola} M.,  {Valkenburg} W.,  {Joachimi} B.,  {Hoekstra}
  H.,   {Kuijken} K.,  2016, \mn@doi [\mnras] {10.1093/mnras/stv2762}, \href
  {http://adsabs.harvard.edu/abs/2016MNRAS.456.1508K} {456, 1508}

\bibitem[\protect\citeauthoryear{{K{\"o}hlinger} et~al.,}{{K{\"o}hlinger}
  et~al.}{2017}]{Kohlinger17}
{K{\"o}hlinger} F.,  et~al., 2017, preprint, \href
  {http://adsabs.harvard.edu/abs/2017arXiv170602892K} {} (\mn@eprint {arXiv}
  {1706.02892})

\bibitem[\protect\citeauthoryear{{Krause} \& {Eifler}}{{Krause} \&
  {Eifler}}{2016}]{Krause16}
{Krause} E.,  {Eifler} T.,  2016, preprint, \href
  {http://adsabs.harvard.edu/abs/2016arXiv160105779K} {} (\mn@eprint {arXiv}
  {1601.05779})

\bibitem[\protect\citeauthoryear{{Kwan} et~al.,}{{Kwan} et~al.}{2017}]{Kwan17}
{Kwan} J.,  et~al., 2017, \mn@doi [\mnras] {10.1093/mnras/stw2464}, \href
  {http://adsabs.harvard.edu/abs/2017MNRAS.464.4045K} {464, 4045}

\bibitem[\protect\citeauthoryear{{LSST Science Collaboration} et~al.,}{{LSST
  Science Collaboration} et~al.}{2009}]{LSST09}
{LSST Science Collaboration} et~al., 2009, preprint, \href
  {http://adsabs.harvard.edu/abs/2009arXiv0912.0201L} {} (\mn@eprint {arXiv}
  {0912.0201})

\bibitem[\protect\citeauthoryear{{Lacasa}, {Lima}  \& {Aguena}}{{Lacasa}
  et~al.}{2016}]{Lacasa16}
{Lacasa} F.,  {Lima} M.,   {Aguena} M.,  2016, preprint, \href
  {http://adsabs.harvard.edu/abs/2016arXiv161205958L} {} (\mn@eprint {arXiv}
  {1612.05958})

\bibitem[\protect\citeauthoryear{{Landy} \& {Szalay}}{{Landy} \&
  {Szalay}}{1993}]{Landy93}
{Landy} S.~D.,  {Szalay} A.~S.,  1993, \mn@doi [\apj] {10.1086/172900}, \href
  {http://adsabs.harvard.edu/abs/1993ApJ...412...64L} {412, 64}

\bibitem[\protect\citeauthoryear{{Laureijs} et~al.,}{{Laureijs}
  et~al.}{2011}]{Laureijs11}
{Laureijs} R.,  et~al., 2011, preprint, \href
  {http://adsabs.harvard.edu/abs/2011arXiv1110.3193L} {} (\mn@eprint {arXiv}
  {1110.3193})

\bibitem[\protect\citeauthoryear{{Lemos}, {Challinor}  \& {Efstathiou}}{{Lemos}
  et~al.}{2017}]{Lemos17}
{Lemos} P.,  {Challinor} A.,   {Efstathiou} G.,  2017, \mn@doi [\jcap]
  {10.1088/1475-7516/2017/05/014}, \href
  {http://adsabs.harvard.edu/abs/2017JCAP...05..014L} {5, 014}

\bibitem[\protect\citeauthoryear{{Lewis} \& {Bridle}}{{Lewis} \&
  {Bridle}}{2002}]{Lewis02}
{Lewis} A.,  {Bridle} S.,  2002, \mn@doi [\prd] {10.1103/PhysRevD.66.103511},
  \href {http://adsabs.harvard.edu/abs/2002PhRvD..66j3511L} {66, 103511}

\bibitem[\protect\citeauthoryear{{Lima}, {Cunha}, {Oyaizu}, {Frieman}, {Lin}
  \& {Sheldon}}{{Lima} et~al.}{2008}]{Lima08}
{Lima} M.,  {Cunha} C.~E.,  {Oyaizu} H.,  {Frieman} J.,  {Lin} H.,   {Sheldon}
  E.~S.,  2008, \mn@doi [\mnras] {10.1111/j.1365-2966.2008.13510.x}, \href
  {http://adsabs.harvard.edu/abs/2008MNRAS.390..118L} {390, 118}

\bibitem[\protect\citeauthoryear{{Lin} et~al.,}{{Lin} et~al.}{2012}]{Lin12}
{Lin} H.,  et~al., 2012, \mn@doi [\apj] {10.1088/0004-637X/761/1/15}, \href
  {http://adsabs.harvard.edu/abs/2012ApJ...761...15L} {761, 15}

\bibitem[\protect\citeauthoryear{{Liske} et~al.,}{{Liske}
  et~al.}{2015}]{Liske15}
{Liske} J.,  et~al., 2015, \mn@doi [\mnras] {10.1093/mnras/stv1436}, \href
  {http://adsabs.harvard.edu/abs/2015MNRAS.452.2087L} {452, 2087}

\bibitem[\protect\citeauthoryear{{Loverde} \& {Afshordi}}{{Loverde} \&
  {Afshordi}}{2008}]{Loverde08}
{Loverde} M.,  {Afshordi} N.,  2008, \mn@doi [\prd]
  {10.1103/PhysRevD.78.123506}, \href
  {http://adsabs.harvard.edu/abs/2008PhRvD..78l3506L} {78, 123506}

\bibitem[\protect\citeauthoryear{{MacCrann}, {Zuntz}, {Bridle}, {Jain}  \&
  {Becker}}{{MacCrann} et~al.}{2015}]{MacCrann15}
{MacCrann} N.,  {Zuntz} J.,  {Bridle} S.,  {Jain} B.,   {Becker} M.~R.,  2015,
  \mn@doi [\mnras] {10.1093/mnras/stv1154}, \href
  {http://adsabs.harvard.edu/abs/2015MNRAS.451.2877M} {451, 2877}

\bibitem[\protect\citeauthoryear{{Mandelbaum} et~al.,}{{Mandelbaum}
  et~al.}{2005}]{Mandelbaum05}
{Mandelbaum} R.,  et~al., 2005, \mn@doi [\mnras]
  {10.1111/j.1365-2966.2005.09282.x}, \href
  {http://adsabs.harvard.edu/abs/2005MNRAS.361.1287M} {361, 1287}

\bibitem[\protect\citeauthoryear{{Mandelbaum}, {Hirata}, {Ishak}, {Seljak}  \&
  {Brinkmann}}{{Mandelbaum} et~al.}{2006}]{Mandelbaum06}
{Mandelbaum} R.,  {Hirata} C.~M.,  {Ishak} M.,  {Seljak} U.,   {Brinkmann} J.,
  2006, \mn@doi [\mnras] {10.1111/j.1365-2966.2005.09946.x}, \href
  {http://adsabs.harvard.edu/abs/2006MNRAS.367..611M} {367, 611}

\bibitem[\protect\citeauthoryear{{Mandelbaum} et~al.,}{{Mandelbaum}
  et~al.}{2011}]{Mandelbaum11}
{Mandelbaum} R.,  et~al., 2011, \mn@doi [\mnras]
  {10.1111/j.1365-2966.2010.17485.x}, \href
  {http://adsabs.harvard.edu/abs/2011MNRAS.410..844M} {410, 844}

\bibitem[\protect\citeauthoryear{{Mandelbaum}, {Slosar}, {Baldauf}, {Seljak},
  {Hirata}, {Nakajima}, {Reyes}  \& {Smith}}{{Mandelbaum}
  et~al.}{2013}]{Mandelbaum13}
{Mandelbaum} R.,  {Slosar} A.,  {Baldauf} T.,  {Seljak} U.,  {Hirata} C.~M.,
  {Nakajima} R.,  {Reyes} R.,   {Smith} R.~E.,  2013, \mn@doi [\mnras]
  {10.1093/mnras/stt572}, \href
  {http://adsabs.harvard.edu/abs/2013MNRAS.432.1544M} {432, 1544}

\bibitem[\protect\citeauthoryear{{Mead}, {Peacock}, {Heymans}, {Joudaki}  \&
  {Heavens}}{{Mead} et~al.}{2015}]{Mead15}
{Mead} A.~J.,  {Peacock} J.~A.,  {Heymans} C.,  {Joudaki} S.,   {Heavens}
  A.~F.,  2015, \mn@doi [\mnras] {10.1093/mnras/stv2036}, \href
  {http://adsabs.harvard.edu/abs/2015MNRAS.454.1958M} {454, 1958}

\bibitem[\protect\citeauthoryear{{M{\'e}nard}, {Scranton}, {Schmidt},
  {Morrison}, {Jeong}, {Budavari}  \& {Rahman}}{{M{\'e}nard}
  et~al.}{2013}]{Menard13}
{M{\'e}nard} B.,  {Scranton} R.,  {Schmidt} S.,  {Morrison} C.,  {Jeong} D.,
  {Budavari} T.,   {Rahman} M.,  2013, preprint, \href
  {http://adsabs.harvard.edu/abs/2013arXiv1303.4722M} {} (\mn@eprint {arXiv}
  {1303.4722})

\bibitem[\protect\citeauthoryear{{Miller} et~al.,}{{Miller}
  et~al.}{2013}]{Miller13}
{Miller} L.,  et~al., 2013, \mn@doi [\mnras] {10.1093/mnras/sts454}, \href
  {http://adsabs.harvard.edu/abs/2013MNRAS.429.2858M} {429, 2858}

\bibitem[\protect\citeauthoryear{{More}, {Miyatake}, {Mandelbaum}, {Takada},
  {Spergel}, {Brownstein}  \& {Schneider}}{{More} et~al.}{2015}]{More15}
{More} S.,  {Miyatake} H.,  {Mandelbaum} R.,  {Takada} M.,  {Spergel} D.~N.,
  {Brownstein} J.~R.,   {Schneider} D.~P.,  2015, \mn@doi [\apj]
  {10.1088/0004-637X/806/1/2}, \href
  {http://adsabs.harvard.edu/abs/2015ApJ...806....2M} {806, 2}

\bibitem[\protect\citeauthoryear{{Morrison}, {Hildebrandt}, {Schmidt},
  {Baldry}, {Bilicki}, {Choi}, {Erben}  \& {Schneider}}{{Morrison}
  et~al.}{2017}]{Morrison17}
{Morrison} C.~B.,  {Hildebrandt} H.,  {Schmidt} S.~J.,  {Baldry} I.~K.,
  {Bilicki} M.,  {Choi} A.,  {Erben} T.,   {Schneider} P.,  2017, \mn@doi
  [\mnras] {10.1093/mnras/stx342}, \href
  {http://adsabs.harvard.edu/abs/2017MNRAS.467.3576M} {467, 3576}

\bibitem[\protect\citeauthoryear{{Nicola}, {Refregier}  \& {Amara}}{{Nicola}
  et~al.}{2016}]{Nicola16}
{Nicola} A.,  {Refregier} A.,   {Amara} A.,  2016, \mn@doi [\prd]
  {10.1103/PhysRevD.94.083517}, \href
  {http://adsabs.harvard.edu/abs/2016PhRvD..94h3517N} {94, 083517}

\bibitem[\protect\citeauthoryear{{Nicola}, {Refregier}  \& {Amara}}{{Nicola}
  et~al.}{2017}]{Nicola17}
{Nicola} A.,  {Refregier} A.,   {Amara} A.,  2017, \mn@doi [\prd]
  {10.1103/PhysRevD.95.083523}, \href
  {http://adsabs.harvard.edu/abs/2017PhRvD..95h3523N} {95, 083523}

\bibitem[\protect\citeauthoryear{{Peebles}}{{Peebles}}{1973}]{Peebles73}
{Peebles} P.~J.~E.,  1973, \mn@doi [\apj] {10.1086/152431}, \href
  {http://adsabs.harvard.edu/abs/1973ApJ...185..413P} {185, 413}

\bibitem[\protect\citeauthoryear{{Pen}}{{Pen}}{1998}]{Pen98}
{Pen} U.-L.,  1998, \mn@doi [\apj] {10.1086/306098}, \href
  {http://adsabs.harvard.edu/abs/1998ApJ...504..601P} {504, 601}

\bibitem[\protect\citeauthoryear{{Planck Collaboration} et~al.,}{{Planck
  Collaboration} et~al.}{2016}]{Planck15}
{Planck Collaboration} et~al., 2016, \mn@doi [\aap]
  {10.1051/0004-6361/201525830}, \href
  {http://adsabs.harvard.edu/abs/2016A%26A...594A..13P} {594, A13}

\bibitem[\protect\citeauthoryear{{Schmidt}, {M{\'e}nard}, {Scranton},
  {Morrison}  \& {McBride}}{{Schmidt} et~al.}{2013}]{Schmidt13}
{Schmidt} S.~J.,  {M{\'e}nard} B.,  {Scranton} R.,  {Morrison} C.,   {McBride}
  C.~K.,  2013, \mn@doi [\mnras] {10.1093/mnras/stt410}, \href
  {http://adsabs.harvard.edu/abs/2013MNRAS.431.3307S} {431, 3307}

\bibitem[\protect\citeauthoryear{{Schneider}, {van Waerbeke}, {Kilbinger}  \&
  {Mellier}}{{Schneider} et~al.}{2002}]{Schneider02}
{Schneider} P.,  {van Waerbeke} L.,  {Kilbinger} M.,   {Mellier} Y.,  2002,
  \mn@doi [\aap] {10.1051/0004-6361:20021341}, \href
  {http://adsabs.harvard.edu/abs/2002A%26A...396....1S} {396, 1}

\bibitem[\protect\citeauthoryear{{Schneider}, {Eifler}  \&
  {Krause}}{{Schneider} et~al.}{2010}]{Schneider10}
{Schneider} P.,  {Eifler} T.,   {Krause} E.,  2010, \mn@doi [\aap]
  {10.1051/0004-6361/201014235}, \href
  {http://adsabs.harvard.edu/abs/2010A%26A...520A.116S} {520, A116}

\bibitem[\protect\citeauthoryear{{Seljak} et~al.,}{{Seljak}
  et~al.}{2005}]{Seljak05}
{Seljak} U.,  et~al., 2005, \mn@doi [\prd] {10.1103/PhysRevD.71.043511}, \href
  {http://adsabs.harvard.edu/abs/2005PhRvD..71d3511S} {71, 043511}

\bibitem[\protect\citeauthoryear{{Sif{\'o}n}, {Hoekstra}, {Cacciato}, {Viola},
  {K{\"o}hlinger}, {van der Burg}, {Sand}  \& {Graham}}{{Sif{\'o}n}
  et~al.}{2015}]{Sifon15}
{Sif{\'o}n} C.,  {Hoekstra} H.,  {Cacciato} M.,  {Viola} M.,  {K{\"o}hlinger}
  F.,  {van der Burg} R.~F.~J.,  {Sand} D.~J.,   {Graham} M.~L.,  2015, \mn@doi
  [\aap] {10.1051/0004-6361/201424435}, \href
  {http://adsabs.harvard.edu/abs/2015A%26A...575A..48S} {575, A48}

\bibitem[\protect\citeauthoryear{{Simon}, {Hetterscheidt}, {Schirmer}, {Erben},
  {Schneider}, {Wolf}  \& {Meisenheimer}}{{Simon} et~al.}{2007}]{Simon07}
{Simon} P.,  {Hetterscheidt} M.,  {Schirmer} M.,  {Erben} T.,  {Schneider} P.,
  {Wolf} C.,   {Meisenheimer} K.,  2007, \mn@doi [\aap]
  {10.1051/0004-6361:20065904}, \href
  {http://adsabs.harvard.edu/abs/2007A%26A...461..861S} {461, 861}

\bibitem[\protect\citeauthoryear{{Singh}, {Mandelbaum}  \& {More}}{{Singh}
  et~al.}{2015}]{Singh15}
{Singh} S.,  {Mandelbaum} R.,   {More} S.,  2015, \mn@doi [\mnras]
  {10.1093/mnras/stv778}, \href
  {http://adsabs.harvard.edu/abs/2015MNRAS.450.2195S} {450, 2195}

\bibitem[\protect\citeauthoryear{{Singh}, {Mandelbaum}, {Seljak}, {Slosar}  \&
  {Vazquez Gonzalez}}{{Singh} et~al.}{2016}]{Singh16}
{Singh} S.,  {Mandelbaum} R.,  {Seljak} U.,  {Slosar} A.,   {Vazquez Gonzalez}
  J.,  2016, preprint, \href
  {http://adsabs.harvard.edu/abs/2016arXiv161100752S} {} (\mn@eprint {arXiv}
  {1611.00752})

\bibitem[\protect\citeauthoryear{{Smith} et~al.,}{{Smith}
  et~al.}{2003}]{Smith03}
{Smith} R.~E.,  et~al., 2003, \mn@doi [\mnras]
  {10.1046/j.1365-8711.2003.06503.x}, \href
  {http://adsabs.harvard.edu/abs/2003MNRAS.341.1311S} {341, 1311}

\bibitem[\protect\citeauthoryear{{Spergel} et~al.,}{{Spergel}
  et~al.}{2015}]{Spergel15}
{Spergel} D.,  et~al., 2015, preprint, \href
  {http://adsabs.harvard.edu/abs/2015arXiv150303757S} {} (\mn@eprint {arXiv}
  {1503.03757})

\bibitem[\protect\citeauthoryear{{Takada} \& {Hu}}{{Takada} \&
  {Hu}}{2013}]{Takada13}
{Takada} M.,  {Hu} W.,  2013, \mn@doi [\prd] {10.1103/PhysRevD.87.123504},
  \href {http://adsabs.harvard.edu/abs/2013PhRvD..87l3504T} {87, 123504}

\bibitem[\protect\citeauthoryear{{Takahashi}, {Sato}, {Nishimichi}, {Taruya}
  \& {Oguri}}{{Takahashi} et~al.}{2012}]{Takahashi12}
{Takahashi} R.,  {Sato} M.,  {Nishimichi} T.,  {Taruya} A.,   {Oguri} M.,
  2012, \mn@doi [\apj] {10.1088/0004-637X/761/2/152}, \href
  {http://adsabs.harvard.edu/abs/2012ApJ...761..152T} {761, 152}

\bibitem[\protect\citeauthoryear{{Tr{\"o}ster} et~al.,}{{Tr{\"o}ster}
  et~al.}{2017}]{Troster17}
{Tr{\"o}ster} T.,  et~al., 2017, \mn@doi [\mnras] {10.1093/mnras/stx365}, \href
  {http://adsabs.harvard.edu/abs/2017MNRAS.467.2706T} {467, 2706}

\bibitem[\protect\citeauthoryear{{Viola} et~al.,}{{Viola}
  et~al.}{2015}]{Viola15}
{Viola} M.,  et~al., 2015, \mn@doi [\mnras] {10.1093/mnras/stv1447}, \href
  {http://adsabs.harvard.edu/abs/2015MNRAS.452.3529V} {452, 3529}

\bibitem[\protect\citeauthoryear{{Zehavi} et~al.,}{{Zehavi}
  et~al.}{2011}]{Zehavi11}
{Zehavi} I.,  et~al., 2011, \mn@doi [\apj] {10.1088/0004-637X/736/1/59}, \href
  {http://adsabs.harvard.edu/abs/2011ApJ...736...59Z} {736, 59}

\bibitem[\protect\citeauthoryear{{Zu} \& {Mandelbaum}}{{Zu} \&
  {Mandelbaum}}{2015}]{Ying15}
{Zu} Y.,  {Mandelbaum} R.,  2015, \mn@doi [\mnras] {10.1093/mnras/stv2062},
  \href {http://adsabs.harvard.edu/abs/2015MNRAS.454.1161Z} {454, 1161}

\bibitem[\protect\citeauthoryear{{de Jong} et~al.,}{{de Jong}
  et~al.}{2013}]{DeJong13}
{de Jong} J.~T.~A.,  et~al., 2013, The Messenger, \href
  {http://adsabs.harvard.edu/abs/2013Msngr.154...44J} {154, 44}

\bibitem[\protect\citeauthoryear{{de Jong} et~al.,}{{de Jong}
  et~al.}{2017}]{DeJong17}
{de Jong} J.~T.~A.,  et~al., 2017, preprint, \href
  {http://adsabs.harvard.edu/abs/2017arXiv170302991D} {} (\mn@eprint {arXiv}
  {1703.02991})

\makeatother
\end{thebibliography}


\begin{appendix}

\section{Accuracy of the power spectrum estimators}\label{app_val_ps}
\begin{figure*}
   \includegraphics[width=1.\linewidth]{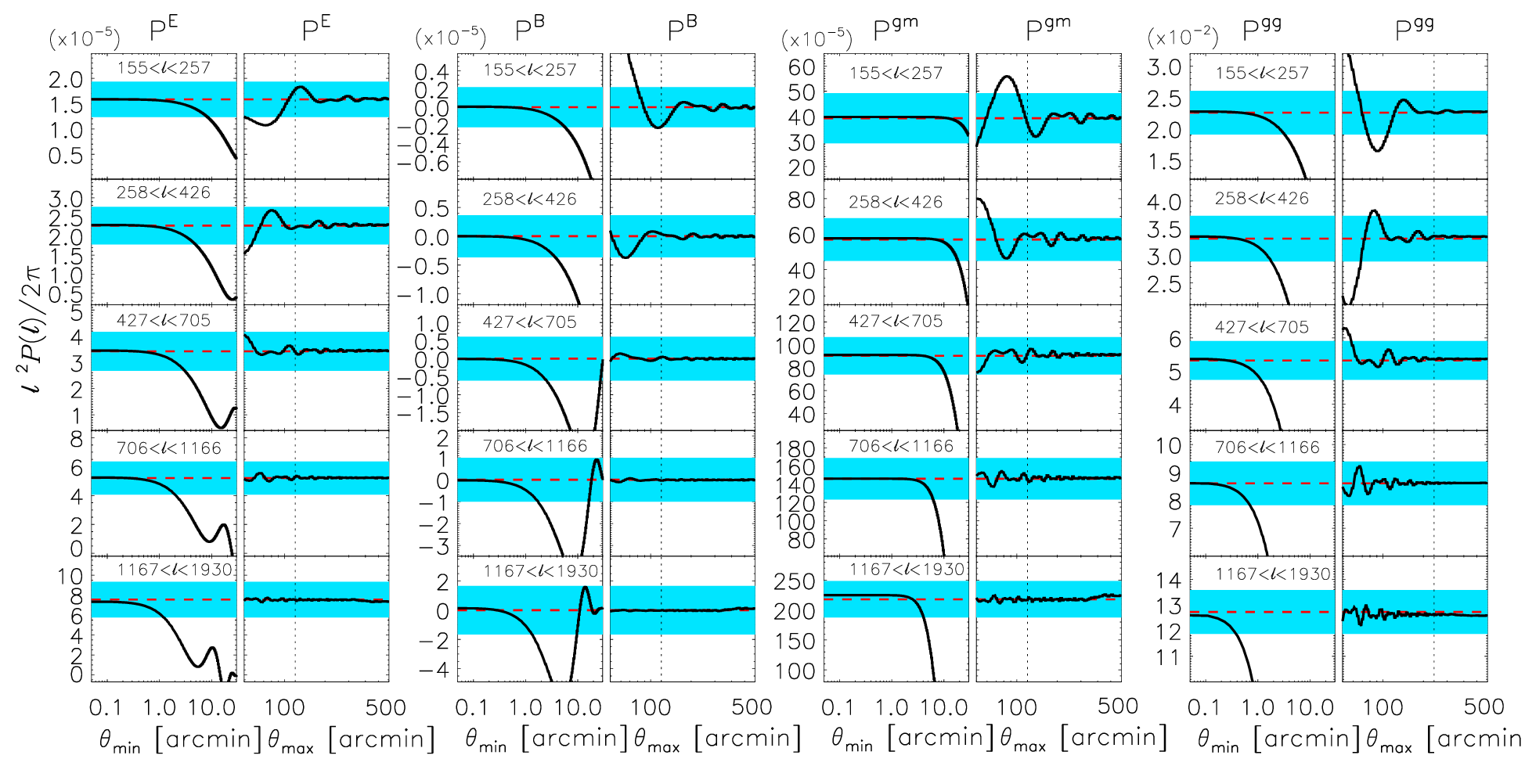}
   \caption{Recovered band powers from analytically computed real-space correlation functions, obtained by varying the lower limit of the conversion integral whilst keeping the upper limit fixed to 600 arcmin (the first, third, fifth and seventh column), or by varying the upper limit whilst keeping the lower limit fixed to 0.06 arcmin (the second, fourth, sixth and eighth column). The black line indicates the recovered power spectrum, the red line the input power spectrum at the logarithmic mean of the $\ell$ bin. The five rows correspond to five different $\ell$-ranges, indicated in the panels. The first two columns are for $P^{\rm E}$, column 3--4 for $P^{\rm B}$, column 5--6 for $P^{\rm gm}$ and column 7--8 for $P^{\rm gg}$. We use the lens and source redshift distributions from the data, and select those bins that have the highest signal-to-noise. For $P^{\rm E/B}$, that is the S4-S4 bin, for $P^{\rm gm}$ it is F2-S4, and for $P^{\rm gg}$ it is F2-F2. Other bins show similar trends. The cyan regions indicate the analytical error for those bins in our data. The thin dotted vertical lines show the maximum scales available in the data (2 degrees for $P^{\rm E}$ and $P^{\rm gm}$, 4 degrees for $P^{\rm gg}$), showing that the lowest $\ell$ bin may suffer from a small bias, which we correct for. }
   \label{plot_P_scaletst}
\end{figure*}
By definition, our power spectrum estimators from Eqs. (\ref{eq_pe},\ref{eq_pgm2},\ref{eq_pgg2}) are unbiased when we integrate the corresponding real space correlation functions from zero to infinity. However, even if the correlation functions are measured over a finite range, we can extract practically unbiased band powers over a considerable range in $\ell$. Outside this range, the measured power spectra can be corrected for this integral bias. We show that this correction is robust, but recommend only applying it when the bias correction is smaller than the statistical errors, as is the case here.   \\
\indent For this test, we compute the 3-D matter power spectrum with the non-linear corrections from \citet{Takahashi12} and the transfer function fit by \citet{Eisenstein98}, adopting cosmological parameters that correspond to the best-fitting values of \citet{Planck15}. We convert this into the convergence power spectrum using the redshift distribution of our fourth tomographic bin, which has the highest signal-to-noise and is therefore the most conservative test case. This is then converted into the shear correlation functions $\xi_{+/-}$ using Eq. (\ref{eq_xipm}), adopting the same $\theta$ binning as on the data. These are treated as the observed correlation functions, and inserted in Eq. (\ref{eq_pe}) and (\ref{eq_pb}) to estimate $P^{\rm E}$ and $P^{\rm B}$, respectively, using the same $\ell$-ranges that we adopted in the data. \\
\indent In the first column on Fig. \ref{plot_P_scaletst}, we show how the recovered band powers vary when we increase the lower limit of the integrals of Eq. (\ref{eq_pe}). Each row corresponds to a given $\ell$-range, as indicated in each panel. The solid black line is the recovered band power, while the dashed red line corresponds to the theoretical power spectrum at the logarithmic mean of the $\ell$-bin. The band powers become biased when the lower limit of the integral becomes larger than 1 arcmin. The higher $\ell$-bins are more strongly affected, as expected, as they retrieve more information from small angular scales. The cyan area indicates the analytical error on $P^{\rm E}$ for the fourth tomographic bin of KiDS-450. Since we can measure the shear correlation functions to much smaller scales than 1 arcmin, we conclude that our band powers are not biased from shifting the minimum scale from zero to 0.06 arcmin. \\
\indent In the second column, we repeat the exercise, but now changing the upper limit of the integral. The highest $\ell$ bins are completely unaffected. The band powers in the lower $\ell$ bin become biased if the upper limit becomes less than 200 arcmin. We measure the shear correlation functions in the data up to 120 arcmin. For that scale, there is a small bias for the lowest $\ell$ bin, but it is smaller than our statistical error. We apply an integral bias correction (IBC) by adding the difference between the observed and input power spectrum, to the one measured on the data. A similar correction scheme was implemented for a power spectrum estimator in \citet{Troster17} in the context of a cross-correlation study of gamma-ray maps with weak lensing data. \\
\indent The third and fourth columns show the corresponding B-modes. The trends are similar as for $P^{\rm E}$: we recover B-modes in the range where the E-modes are biased. However, if there are significant B-modes in the data, it does not automatically mean that they are caused by leakage in our band power estimators. If leakage is present, it is most likely to affect the first and the last $\ell$ bin. \\
\indent The fifth and sixth column show the results for the galaxy-matter power spectrum. The analytical power spectra are determined with Eq. (\ref{eq_pgm}), where we used the redshift distribution of the F2 foreground sample from GAMA, adopted an effective bias of unity, and used the redshift distribution of the fourth tomographic shape sample from KiDS-450 to estimate the lensing efficiencies. We recover a similar result. Given the $\ell$-ranges, our band powers are unbiased as long as we use a minimum lower limit of 2 arcmin or less, which is trivially met. Furthermore, for a maximum scale of 120 arcmin (the maximum scale available in the data), there is a small bias in the lowest $\ell$ bin, which we correct for. \\
\indent The final two columns repeat this test for the angular power spectrum. The angular power spectrum is more sensitive to the lower limit of the integral, and the highest $\ell$ bin becomes significantly biased if the minimum scale is less than 0.2 arcmin; we measure the angular correlation function up to 0.06 arcmin in the data, hence the highest $\ell$ bin is not affected by the integral bias. The lowest $\ell$ bin is recovered without bias as long as the maximum scale is 200 arcmin or more. Since we trust the angular correlation function up to four degrees, also this bin should not be biased. The larger sensitivity to the lower limit of the integral comes from the mixing between $\theta$ and $\ell$ scales through $J_0$. $P^{\rm E}$, for example, uses information from both $\xi_+$ and $\xi_-$, which yields a different mixing between $\theta$ and $\ell$ scales through the combination of $J_0$ (for $\xi_+$) and $J_4$ (for $\xi_-$). If we adopt $K_+=1$ to compute $P^{\rm E}$ (i.e. only using information from $\xi_+$), the bias increases much faster when the lower limit of the integral increases, similar to what is observed for $P^{\rm gg}$.  \\
\indent Note that in Fig. \ref{plot_P_scaletst}, we chose the combinations of foreground sample and shape sample bins that resulted in the highest signal-to-noise, which are the most affected by the integral bias. We have checked that the relative impact of the integral bias is smaller for the other power spectra, relative to their statistical powers. \\
\indent We computed the IBC for all power spectra and applied it to the data. We note that applying this correction has a negligible effect on our results; the constraints on $S_8$ shift by less than 0.2$\sigma$ if we do not apply the correction. The error on the IBC correction will therefore cause a shift of $S_8$ that is much smaller than  0.2$\sigma$. \\
\indent To test the dependence of the IBC on cosmology, we repeated the test for a Planck cosmology where we either increased or decreased the value of $\sigma_8$ by ten percent. As a result, the curves shown in Fig. \ref{plot_P_scaletst} shifted vertically, meaning that the error on the IBC linearly depends on the error of the amplitude of the theoretical power spectrum used to compute the correction. It is unlikely that the amplitude of the theoretical power spectrum is off by more than 10\%, which puts an upper limit on the bias of the IBC of $\sim$10\% as well. As long as the IBC is smaller than the measurement errors, such an error is subdominant and will not impact the results. Second order effects, such as small differences in the shape of the power spectrum, are expected to lead to even smaller biases. We note that the issue of a cosmology dependence of the IBC can be circumvented by computing a new correction for every cosmological model evaluated in the MCMC and applying that to the data before determining the likelihood. \\

\section{Validation on mocks}\label{app_val}
Next, we determine the accuracy of our power spectrum estimators on numerical simulations, which enables us to test whether we measure the correlation functions correctly, and whether the redshift distributions are properly accounted for. The main purpose of these numerical simulations, however, is to compute the covariance matrix from different realisations in order to validate the analytical covariance matrix, and secondly, to test our cosmological inference pipeline by fitting the mock power spectra (using the analytical covariance matrix, hence mimicking the observational procedure), to verify whether we recover the input cosmology. \\ 
\indent The mock catalogues are based on the Scinet Light Cone Simulations \citep[SLICS;][]{HarnoisDeraps15}, a series of about 1000 $N$-body simulations tailored for weak lensing surveys. Each realization follows the evolution of 1536$^3$ dark matter particles in a cube that measures 505 $\mbox{Mpc}/h$ on a side, which are projected on 18 redshift mass planes in the range $0 < z < 3$.  Light cones are then extracted from these planes on 7745$^2$ pixel grids and turned into convergence and shear maps by ray-tracing. The cosmology is fixed to \emph{WMAP}9 + SN + BAO\footnote{https://lambda.gsfc.nasa.gov}, that is $\Omega_{\rm m} = 0.2905$, $\Omega_{\Lambda} = 0.7095$, $\Omega_{\rm b} = 0.0473$, $h = 0.6898$, $\sigma_8 = 0.826$ and $n_{\rm s} = 0.969$. \\
\indent These mocks have been tailored specifically for the KiDS-450 analyses and were first presented in \citet{Hildebrandt17}, although the version we use here has a larger opening angle (100 deg$^2$ instead of 60 deg$^2$). Source galaxies are placed at random positions in the mocks and are assigned both a true redshift \citep[enforcing the tomographic $n(z)$ estimated with the DIR method, see][]{Hildebrandt17} and a photometric redshift, $z_B$, based on the joint probability distribution of these two quantities. This enables the selection of tomographic redshift bins whose true redshift distributions exactly match those of KiDS. The gravitational shears are interpolated from the simulated shear maps at the galaxy positions,  while the intrinsic ellipticities are from a Gaussian with a width equalling the intrinsic ellipticity dispersion measured for KiDS galaxies. \\  
\indent To simulate a foreground galaxy sample, which we need to measure the galaxy-matter power spectrum and the angular power spectrum, we use the simulation boxes that are at a mean redshift of $z=0.221$ and span a redshift range of 0.1747 to 0.2680. Lens positions are drawn from the projected mass maps with a probability that is proportional to the density, ensuring that the bias of the lens sample is constant at $b=1$ and does not depend on scale. Having a lens sample with a known bias is advantageous as we can check whether we recover it. The disadvantage is that it does not enable us to test whether the effective galaxy bias is scale-independent in the simulations. Although our mock foreground sample does not reproduce the galaxy sample from GAMA, it allows us to test our pipelines by replacing the GAMA clustering and redshift distribution by that of these mock foreground galaxies.\\
\indent  For simplicity, we only use one lens sample and two source samples, selected with $0.1<z_{\rm B}\leq0.5$ and $0.5<z_{\rm B}\leq0.9$, respectively. The various real-space correlation functions are measured with the same pipelines as those applied to the data, and converted into power spectra using the same angular scales.

 
\subsection{Validating the power spectrum estimator  \label{app_val_signal}}
\begin{figure}
   \centering
   \includegraphics[width=0.95\linewidth,angle=270]{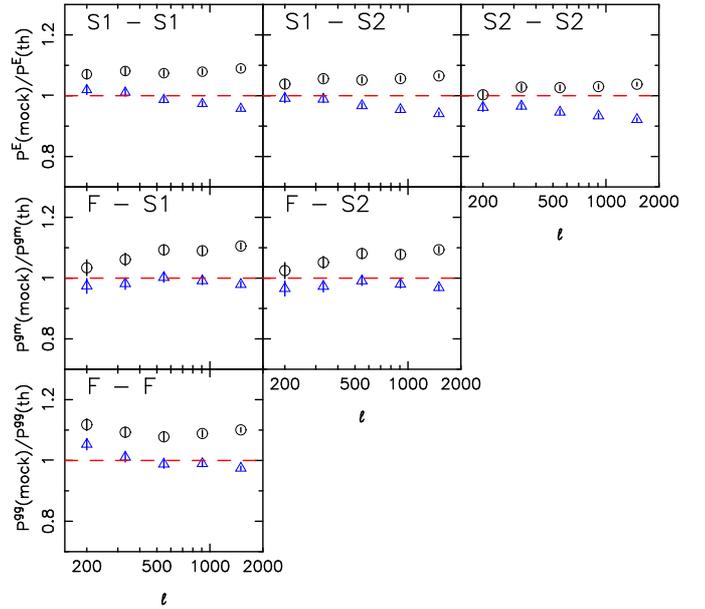}
   \caption{Ratio of power spectra measured on the numerical simulations and the theoretical power spectra. The top row shows the auto-correlation and cross-correlation of the cosmic shear power spectra for the two shape samples, the middle row shows the galaxy-matter power spectrum for one foreground sample and two shape samples, and the bottom row shows the angular power spectrum of the foreground sample. Open black circles (blue triangles) correspond to theoretical power spectra computed using the non-linear corrections of \citet{Smith03} (\citet{Takahashi12}). Error bars show the error on the mean of the different mock realisations.}
   \label{plot_p_mocksignal}
\end{figure}
\begin{figure*}
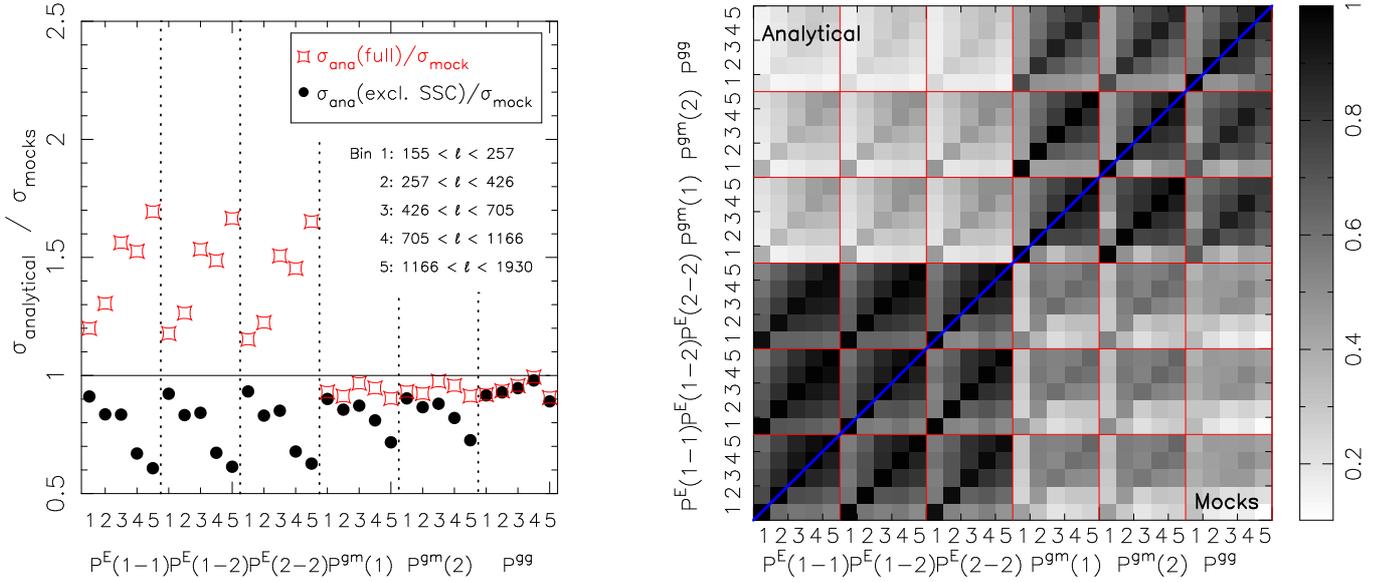

  \begin{minipage}[t]{0.43\linewidth}
      \includegraphics[width=1.\linewidth,angle=-90]{plot_covarDiag_noShapeNoise.ps}
  \end{minipage}
  \hspace{10mm}
  \begin{minipage}[t]{0.43\linewidth}
      \includegraphics[width=1.\linewidth,angle=-90]{plot_corrMat_noShapeNoise.ps}
  \end{minipage}
   \caption{Comparison of the analytical and numerical simulation covariance matrix, for the case without shape noise. The left-hand side shows the ratio of the square root of the diagonals, separately for the six different power spectra. Numbers on the horizontal axis correspond to the different $\ell$ ranges, as indicated in the figure. Red open squares correspond to the full analytical covariance matrix, black circles to the analytical predictions excluding SSC. These two extremes bracket the mock covariance, which has some SSC, but not the full effect. The right-hand side shows the correlation matrix, with the analytical covariance including SSC in the top left and the numerical simulation covariance matrix in the lower right. The off-diagonal terms agree reasonably well. }
  \label{plot_errcomp}
\end{figure*}

\begin{figure*}
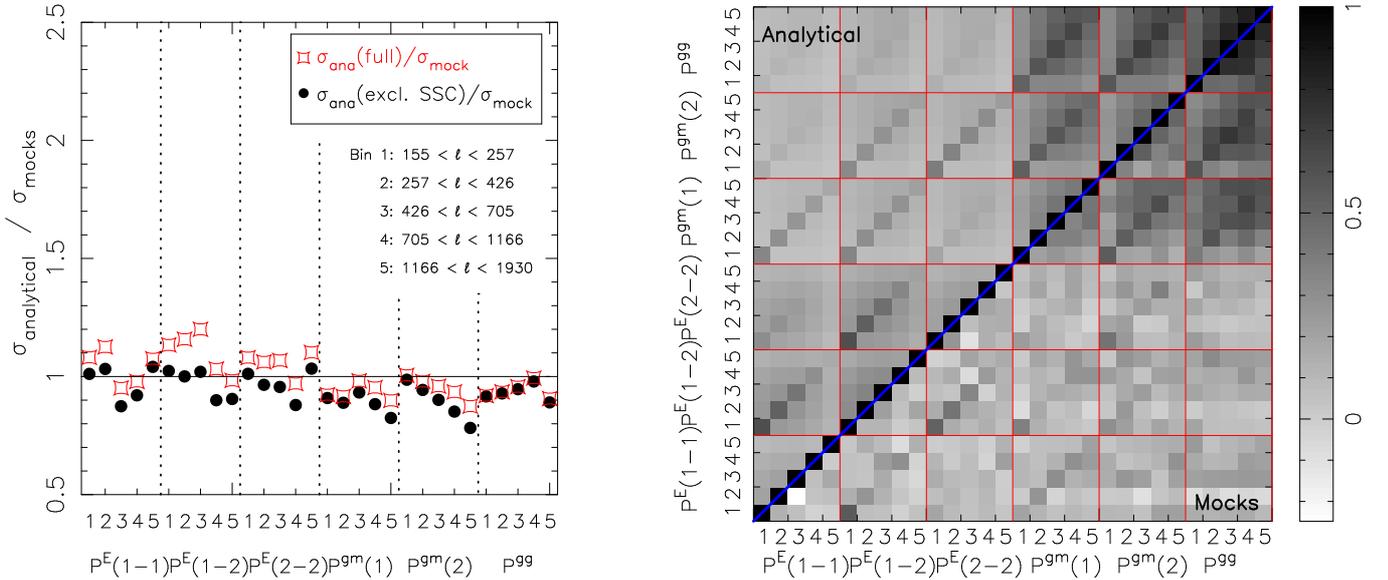

  \begin{minipage}[t]{0.43\linewidth}
      \includegraphics[width=1.\linewidth,angle=-90]{plot_covarDiag_ShapeNoise.ps}
  \end{minipage}
  \hspace{10mm}
  \begin{minipage}[t]{0.43\linewidth}
      \includegraphics[width=1.\linewidth,angle=-90]{plot_corrMat_ShapeNoise.ps}
  \end{minipage}
   \caption{Similar to Fig. \ref{plot_errcomp}, but with the inclusion of shape noise.}
     \label{plot_errcompshape}
\end{figure*}
We measured the cosmic shear correlation functions of our two source samples (two auto-correlations and one cross-correlation), the two tangential shear signals around the foreground sample, and the clustering signal of the foreground sample, and converted those into their respective power spectra, using the same angular ranges that we adopted in the data. We compare that to theoretical predictions, computed for the same cosmological parameters that were used in the mocks and using the non-linear corrections of either \citet{Smith03} or \citet{Takahashi12}. These theoretical power spectra bracket the power spectrum measured directly from the simulations \citep[see][]{HarnoisDeraps15}. \\
\indent The ratio of the power spectra measured on the mocks and the theoretical power spectra is shown in Fig. \ref{plot_p_mocksignal}. The theoretical power spectra computed using \citet{Smith03} are up to 10\% smaller than the power spectra measured on the mocks, while the theoretical predictions using \citet{Takahashi12} agree very well, and only overestimate the signal by $\sim$5\% of the cosmic shear S2--S2 power spectrum. If the mock power spectra lie right in the middle of the \citet{Smith03} and \citet{Takahashi12} predictions \citep[as was found in][]{HarnoisDeraps15}, we conclude that our power spectra are accurate to better than $\sim$5\%. This is the relative precision with respect to the simulations, not the absolute precision of our estimator. As demonstrated in Appendix \ref{app_val_ps}, the absolute error of our estimator is much smaller than the statistical errors and can therefore be safely ignored in our error budget. \\
\indent Note that the tangential shear signal around the foreground sample steeply drops on scales $<2$ arcmin, which is a sign that their positions do not exactly coincide with the centre of their dark matter haloes. This may be due to the finite resolution of the simulations. However, as Fig. \ref{plot_P_scaletst} shows, this does not affect $P^{\rm gm}$, as the band power in the highest $\ell$ bin only becomes biased if the lower limit of the integral is larger than 3 arcmin.


\subsection{Validating the covariance matrix}\label{app_val_err}

We computed the covariance matrix of the numerical power spectra using 136 different realisations of the mocks. We do this separately using galaxy shapes with and without intrinsic shape noise. We consider the case without shape noise as it enables us to assess whether the non-Gaussian in-survey term and the super-sample covariance (SSC) term, which are subdominant in the presence of shape noise, are correctly modelled.  \\
\indent We computed the analytical covariance matrix using the specifics of the mocks, that is with the same cosmological parameters, the same mock foreground and source redshift distributions, the same mock intrinsic shear dispersion, and the same mock survey coverage (a total survey area of 100 deg$^2$, with complete overlap between the three probes). \\
\indent A comparison of the covariance matrix for the mocks without shape noise is shown in Fig. \ref{plot_errcomp}. The left-hand panel shows the ratio of the square root of the diagonals, while the right-hand panel compares the off-diagonal terms. In the left-hand panel, we compare to analytical covariance matrices with and without a SSC contribution. The mocks have some SSC from the regions in the original simulation boxes that were outside the light cones, but not the full effect, hence we expect the two analytical predictions to bracket the result from the numerical simulations, which they do for the cosmic shear power spectra. For $P^{\rm gm}$ and $P^{\rm gg}$, the analytical error is $\sim$10\% smaller than the error estimated from the mocks. \\
\indent In the right-hand panel, we compare the off-diagonal terms of the mock covariance to the analytical one including SSC. We find a good overall agreement, suggesting that the cross-correlation of the different power spectra is correctly modelled with our analytical covariance matrix. The mean difference of the off-diagonal elements of the analytical and mock correlation matrix is only $\sim0.06$.\\
\indent In Figure \ref{plot_errcompshape}, we repeat the test for the case with shape noise. The left-hand panel shows that the analytical covariance with and without SSC is quite similar, because the covariance matrix is dominated by shape noise. We find a good overall agreement between the diagonals, with typical differences of the order 10\% for the three power spectra. Given the small residual differences in the power spectra measurements, this level of agreement is expected. Also the off-diagonal terms match well. The covariance between the different power spectra is much smaller, and there is little correlation left between $P^{\rm E}$ and $P^{\rm gm}$ or $P^{\rm gg}$.


\subsection{Validating the cosmological inference \label{app_val_cosm}}

\begin{figure}
   \centering
   \includegraphics[width=\linewidth]{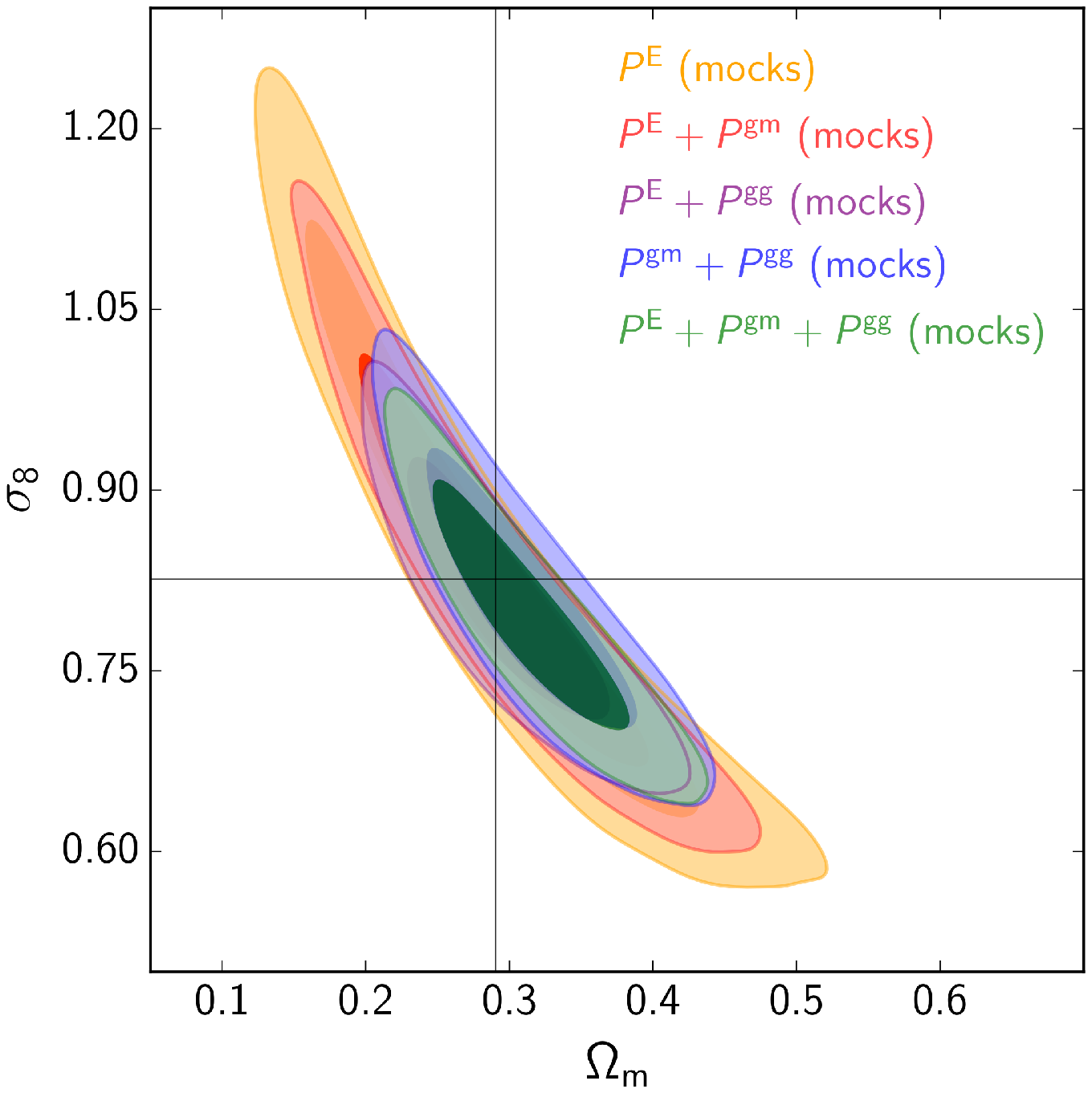}
   \caption{Constraints on $\Omega_{\rm m}$ - $\sigma_8$ for different combinations of mock power spectra, measured on the numerical simulations, assuming a total effective survey area of 500 deg$^2$. The cross hair indicates the input cosmology, which is comfortably recovered by our fitting pipeline.}
   \label{plot_cosm_mock}
\end{figure}
\begin{figure}
   \centering
   \includegraphics[width=\linewidth]{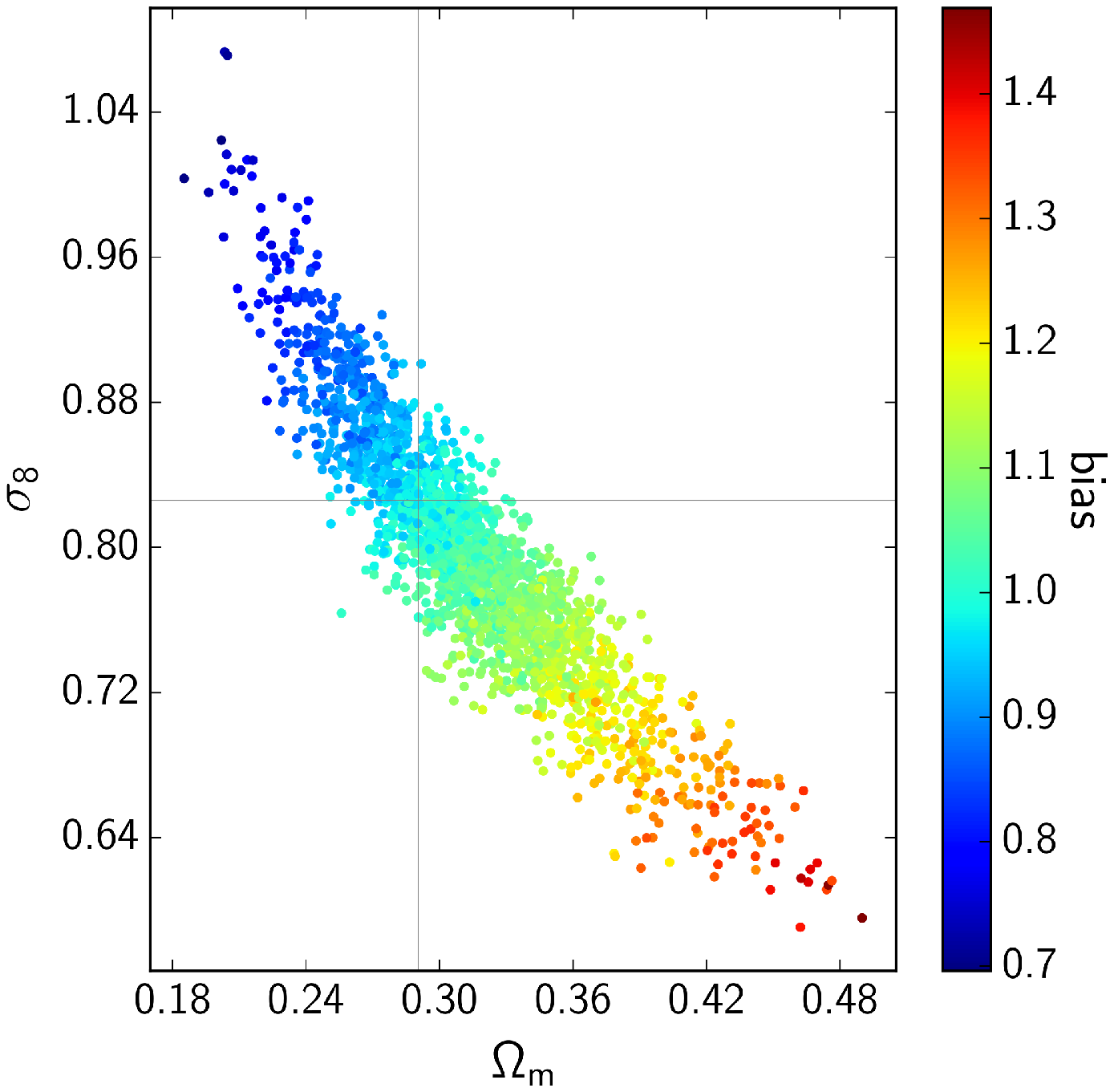}
   \caption{Constraints on $\Omega_{\rm m}$ - $\sigma_8$ obtained by fitting all mock power spectra simultaneously, colour-coded by the value of the bias. It illustrates the degeneracy between the bias and the $\Omega_{\rm m}$ - $\sigma_8$ degeneracy. The cross hair indicates the input cosmology, at which point the bias has a value of unity.}
   \label{plot_cosm_mock_bias}
\end{figure}
The final purpose of the mocks is to test our cosmological inference pipeline. We fit different combinations of mock power spectra, using the analytical covariance matrix that includes SSC. The mocks have an area of 100 deg$^2$, which is smaller than the survey area of the data. To ensure that the cosmological inference is unbiased given our current statistical precision on the data, we simply rescale the analytical covariance matrix such that it corresponds to an area of 500 deg$^2$. Consequently, the SSC contribution to the analytical covariance matrix is not completely correctly modelled, but since its impact is relatively small it is not expected to affect our results much. \\
\indent We use the extended version of {\sc cosmoMC} using the same settings as on the data and with the same convergence criterion. The resulting constraints on $\Omega_{\rm m} -\sigma_8$ are shown in Fig. \ref{plot_cosm_mock}. The input cosmology is indicated with the grey cross-hair and falls comfortably within the 1$\sigma$ contours for all combinations of power spectra that we fit. Hence our cosmological inference pipeline correctly retrieves the input parameters of the mocks, and any remaining systematic bias is smaller than our statistical precision. The constraints from the various power spectrum combinations show little scatter, since we fit the mean signal of the different realisations of the mocks. \\
\indent In all fits, the bias of the foreground sample is consistent with unity, the input value (except when we only fit $P^{\rm E}$, which leaves $b$ unconstrained), although for $P^{\rm E}+P^{\rm gg}+P^{\rm gm}$ it is shifted by 0.3$\sigma$ towards lower values. This explains why the cross-hair in Fig. \ref{plot_cosm_mock} is not exactly centred in the middle of the contours for this power spectra combination. If we fix the bias to unity, the contours shrink dramatically and centre on the cross hair. The degeneracy between the bias and the degeneracy direction of the posterior in the $\Omega_{\rm m} -\sigma_8$ plane is further illustrated in Fig. \ref{plot_cosm_mock_bias}, where we colour coded the posterior of the $P^{\rm E}+P^{\rm gg}+P^{\rm gm}$ fit using the average values of the bias. \\
\indent Ideally, one would like to know whether our estimators are unbiased for a much larger survey (e.g. ten times larger than KiDS). However, we already noted in Sect. \ref{app_val_signal} that we do not exactly recover the input power spectra, but that there are differences left of up to $\sim$5\%. It is unclear whether these differences are caused by limitations of our mocks, of our analytical predictions or of our power spectra estimators, although we note that the tests on the analytical power spectrum -- correlation function pairs in Appendix \ref{app_val_ps}, as well as tests we have performed on Gaussian random fields (not reported here), suggest that the accuracy of our power spectrum estimator is much better than 5\%. For future surveys with improved statistical power, these tests need to be revisited to a higher level of precision.

\section{Comparison to quadratic estimator}\label{app_qe}
\begin{figure*}
   \centering
   \includegraphics[width=\linewidth]{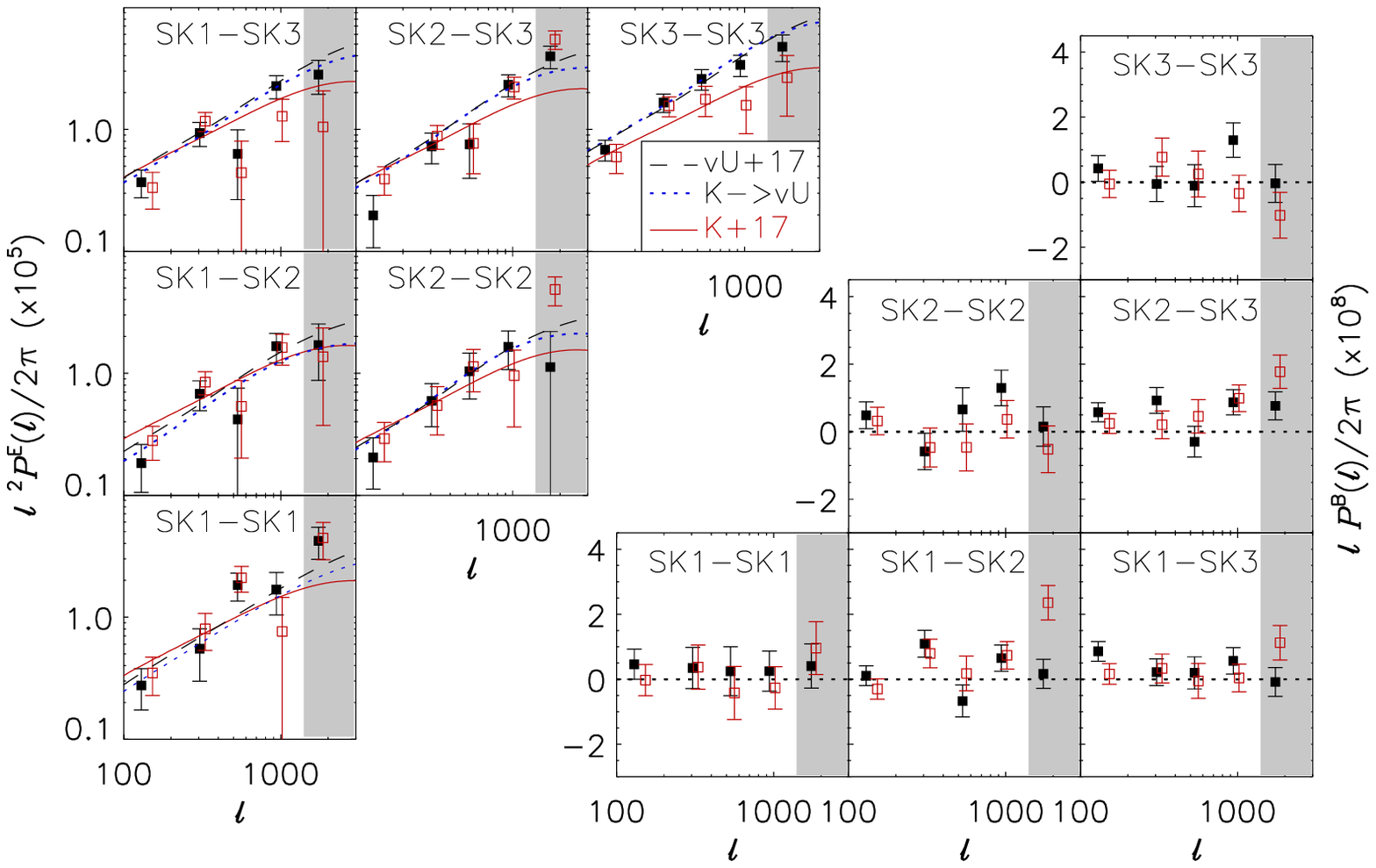}
   \caption{Comparison of $P^{\rm E}$ (left-hand panels) and $P^{\rm B}$ (right-hand panels) for the various tomographic bin combinations, measured with our estimator (black, solid squares) and the quadratic estimator from \citet{Kohlinger17} (red, open squares). The black dashed lines indicate the best-fitting model from $P^{\rm E}$ alone from this work, the dotted blue line indicates the best-fitting model to $P^{\rm E}$ using the cosmological inference code from \citet{Kohlinger17}, and the solid red lines indicate the best-fitting model from \citet{Kohlinger17}. The measurements in the grey area were excluded from the cosmological inferences with the \citet{Kohlinger17} model. The only noticeable difference between the estimators is observed for the highest tomographic bin of $P^{\rm E}$, where our estimator returns higher band powers at large $\ell$ than the quadratic estimator.}
   \label{plot_P_comparison}
\end{figure*}
We compare our band powers to the quadratic estimator from \citet{Kohlinger17}. To do that, we adopt the redshift binning from that work, that is three tomographic source redshift bins selected with ${\rm SK}1=0.1<z_{\rm B}\le0.3$, ${\rm SK}2=0.3<z_{\rm B}\le0.6$ and ${\rm SK}3=0.6<z_{\rm B}\le0.9$. We measure the shear correlation functions with our standard angular binning, apply the shear calibration bias corrections, and estimate our power spectra for the same $\ell$ binning as \citet{Kohlinger17}. Note that we use the full KiDS-450 area to measure the shear correlation functions, while \citet{Kohlinger17} exclude 36 deg$^2$ from disconnected patches. We have tested that excluding those fields does not affect our measurements. Furthermore, we do not apply an IBC to our band powers, which means that our first $\ell$ bin is biased low. These limitations should not affect the comparison much. The results are shown in Fig. \ref{plot_P_comparison}. We plot the full statistical error bars in both cases. The differences between the power estimates should be substantially smaller than these errors as the input data is practically identical, although residual fluctuations will occur because the measurements are weighted differently in the quadratic and band power estimates. \\
\indent We find a fair agreement for $P^{\rm E}$ for the first and second tomographic bin. For the third tomographic bin, our band powers at high $\ell$ are higher than the quadratic estimator. This is most apparent in the SK3--SK3 bin. The B-modes also appear to show fair agreement, except for the highest $\ell$-bin in the SK3-SK3 combination and in the SK2-SK3 combination, where again the band powers are higher than the power determined from the quadratic estimator. We note that both estimators were tested on very similar (B-mode free) Gaussian random field simulations and found to faithfully reproduce the input power, which implies that the differences seen are due to some systematic trend in the data that was not included in the mocks. A corresponding trend can be seen in Fig. D11 of \citet{Hildebrandt17}. In their auto-correlation of the third tomographic bin (which has substantial overlap with SK3), $\xi_{\rm B}$ displays a positive signal below $\sim 4\,$arcmin. While our band power estimator is sensitive to these scales in the correlation function, the quadratic estimator is expected to be immune to systematics at a few arcminutes or less. Note that we find similar reduced $\chi^2$ values of the null-hypothesis as for our fiducial four redshift bin analysis, while \citet{Kohlinger17} report that their B-modes are consistent with zero. \\
\indent To test for the potential impact of B-modes on the cosmological inference, \citet{Hildebrandt17} applied a correction under the assumption that the underlying systematics contribute equally to the E- and B-modes, which led to a downward shift in $S_8$ by $\sim 1\,\sigma$, in the direction of the \citet{Kohlinger17} results. We repeated this test by creating a new data vector where we substracted the B-modes from the E-modes. We updated the covariance matrix by adding the analytical B-mode covariance matrix to the E-mode one. The cosmological inference resulted in $S_8=0.787\pm0.034$, which is within 0.5$\sigma$ of our fiducial result. As in \citet{Hildebrandt17}, the shift is towards the results from \citet{Kohlinger17}, but does not close the gap. As the source of the low-level B-mode contamination is currently unknown \citep[see the discussion in][]{Hildebrandt17}, we do not know how it affects the E-modes, hence we do not attempt to remove it in our fiducial analysis but defer this to forthcoming work. It is interesting to note that the B-mode signature appears to vary with the choice of redshift binning (cf. Figs. C1 and 2). The link between cosmic shear B-modes and binning in terms of a photometric redshift point estimate will be explored in more detail in Asgari et al. (in prep.). \\
\indent Figure \ref{plot_P_comparison} also shows the best-fitting model from \citet{Kohlinger17}, our best-fitting model from fitting $P^{\rm E}$ only on the default KiDS-450 four-bin data (but shown for the current three tomographic redshift bins), as well as the best-fitting model obtained by applying the cosmological inference method of \citet{Kohlinger17} to our band powers. The differences between the latter two are small, which suggests that differences in modelling choices are not driving the shift of $S_8$; these include the approaches to the non-linear matter power spectrum, baryon feedback and massive neutrinos, which might contribute at a lower level. The lower quadratic estimator band powers at high $\ell$ for the SK3--SK3 bin is accommodated by the model with a negative intrinsic alignment amplitude, $A_{\rm IA}=-1.72$ (weighted median), albeit with large error bars. The fit of the \citet{Kohlinger17} to our band powers results in an intrinsic alignment amplitude of  $A_{\rm IA}=1.53$. Since $A_{\rm IA}$ is correlated with $S_8$ (see Fig. \ref{plot_zbias}), this lowers the constraints on $S_8$ from \citet{Kohlinger17} relative to our results. Such a large, negative intrinsic alignment amplitude is not expected from physical models of the effect. Given the findings of Sect. \ref{sec_zshift}, the low value of $A_{\rm IA}$ could point to inconsistencies in the relative strengths of the tomographic power spectra caused by biases in the redshift distributions. \\

\section{Iterated analytical covariance matrix}\label{app_iter}

In our first run on the data, we used an analytical covariance matrix computed using the best-fit parameters of \citet{Planck15}, and effective galaxy biases of unity for the two foreground samples. If the parameter values we adopted for the covariance model were far from the high-probability region in the resulting posterior, our inferred cosmological model would be inconsistent with the error model used in the likelihood. For example, if we underestimated the effective galaxy biases, the error bars on $P^{\rm gm}$ and $P^{\rm gg}$ would be too small and these power spectra would get too much weight in the cosmological inference. Therefore, we updated the analytical covariance matrix using the best-fit parameters from our initial fit to the data, and repeated the cosmological inference. We repeated this procedure a second time to make sure that this iterative approach is stable and converging. Ideally, one would like to update the analytical covariance matrix in each step in the chain during the cosmological inference, but the computational demands make this approach currently infeasible. Note that in these runs, we did not marginalize over the uncertainty of the source redshift distributions, as we found that to be less stable due to the increased noise. \\
\begin{figure}
   \centering
   \includegraphics[width=\linewidth]{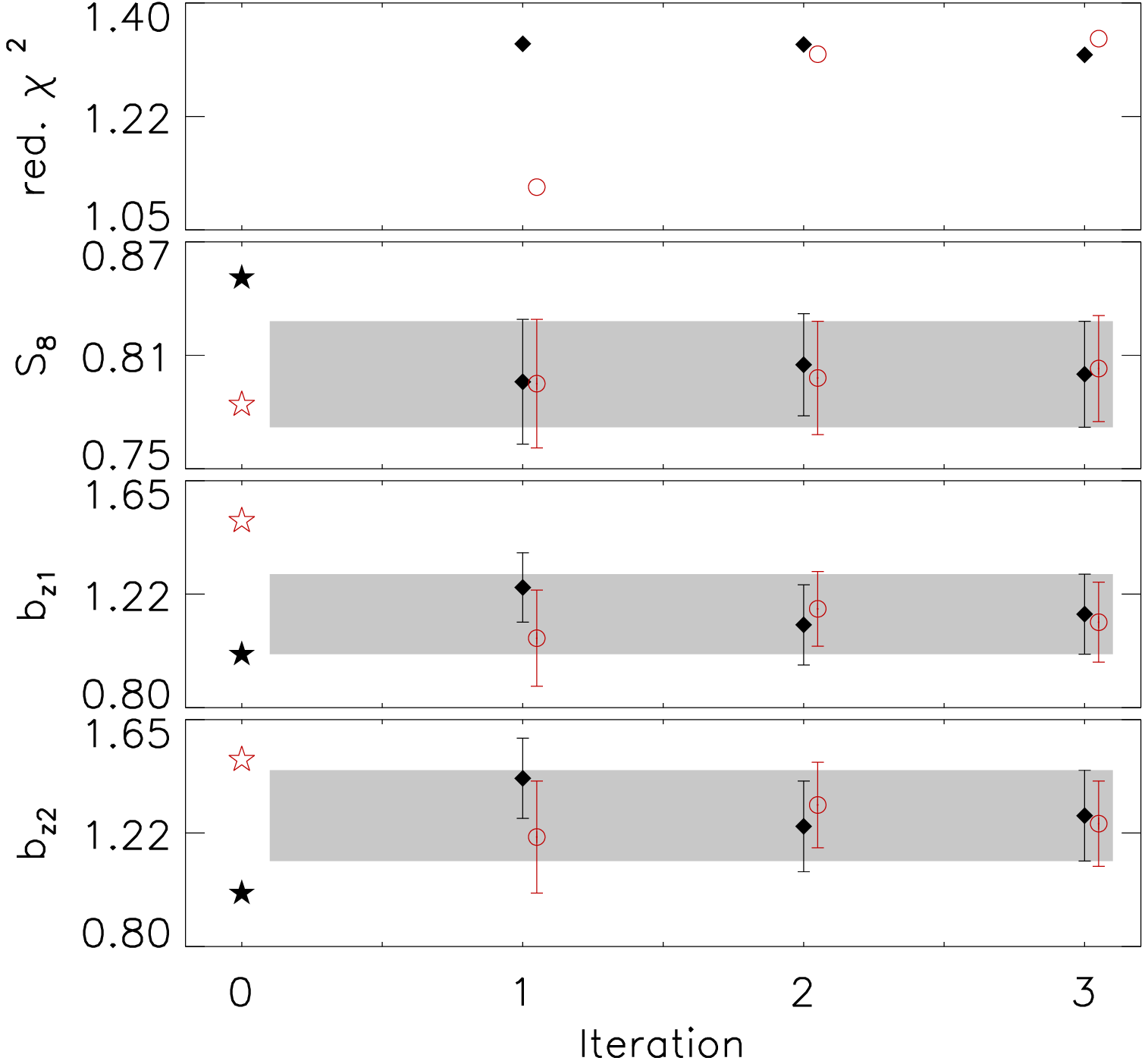}
   \caption{Constraints on the reduced $\chi^2$ value of the best-fitting model and the fit parameters $S_8$, $b_{z1}$ and $b_{z2}$, where we iteratively update the analytical covariance matrix using the best-fitting parameter values from the previous run, and starting with a fiducial \citet{Planck15} cosmology and effective galaxy biases of unity (black stars) or starting with a fiducial WMAP9 cosmology and effective galaxy biases of 1.5 (red stars). The subsequent constraints are shown by black diamonds and red circles for the Planck and WMAP9 runs, respectively. The constraints on the galaxy biases shift significantly after the first iteration, but do not change after the second iteration.}
   \label{plot_iter}
\end{figure}
\indent In Fig. \ref{plot_iter}, we show the reduced $\chi^2$ of the best-fitting model, the constraints on $S_8$ and on the nuisance parameters $b_{z1}$ and $b_{z2}$ in each step of the iteration. The reduced $\chi^2$ of the initial fit is 1.34 and remains constant. The constraints on $S_8$ does not change much either and is therefore not sensitive to small changes in the parameters of the analytical covariance matrix. The effective biases, however, change significantly. The constraints from the initial fit are $b_{\rm z1}=1.25\pm0.13$ and $b_{\rm z2}=1.43\pm0.15$, while the constraints after the second iteration are $b_{\rm z1}=1.11\pm0.15$ and $b_{\rm z2}=1.25\pm0.17$. Hence the posteriors shift not only by about 1$\sigma$, but also the 68\% confidence intervals increase. The cosmological parameters that are degenerate with the galaxy biases, in particular $\Omega_{\rm m}$ and $\sigma_8$, shifted by similar amounts. All cosmological inferences on data in this paper use the analytical covariance matrix based on the best-fitting parameters of the second iteration, as the parameters do not change significantly after another iteration. \\
\indent To test the stability of this iterative procedure, we also computed the analytical covariance matrix using a WMAP9 cosmology \citep{Hinshaw13} and effective galaxy biases of 1.5 and used that as the starting point. The resulting parameter constraints are also shown in Fig. \ref{plot_iter}. The results converge after the first iteration and are thus not very sensitive to the exact starting point of the iterative procedure.

\section{Full posterior}\label{app_fullpost}

\begin{figure*}
   \centering
   \includegraphics[width=\linewidth]{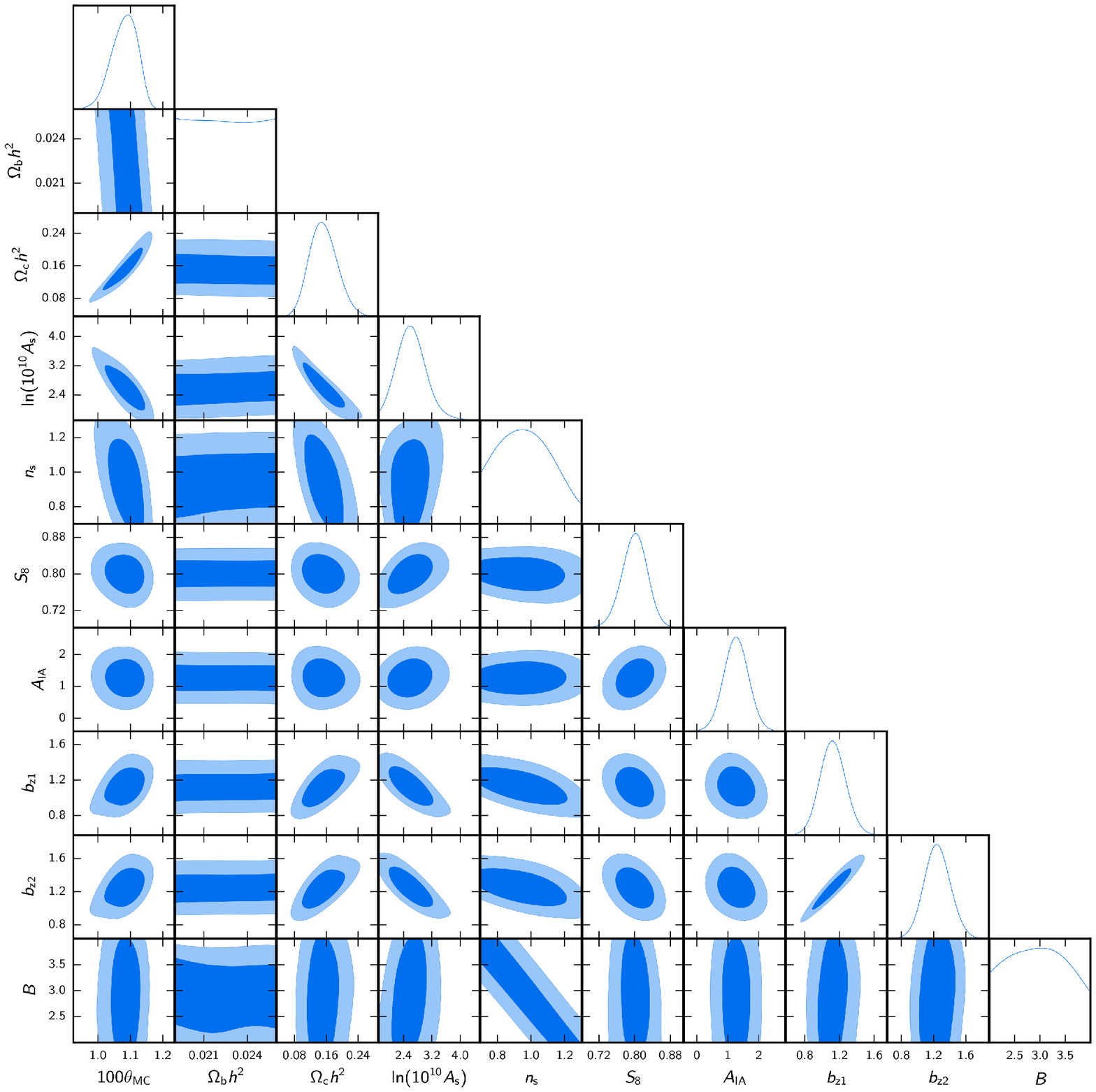}
   \caption{Posterior of combinations of all fit parameters, obtained by marginalizing over all other parameters. The contours indicate the 1$\sigma$ and 2$\sigma$ regimes.}
   \label{plot_postdiag}
\end{figure*}

\begin{table*}
  \centering
  \renewcommand{\tabcolsep}{0.06cm}
  \caption{Mean and 68\% confidence interval of the fit parameters (row 1--10) and derived parameters (row 11-13), $\chi^2$ of the best-fitting model (row 14) and the number of degrees of freedom (row 15). Also shown are the results from our conservative runs where we excluded the lowest $\ell$ bin of all power spectra (`cons-1') and the highest $\ell$ bin of $P^{\rm gm}$ and $P^{\rm gg}$ (`cons-2') in the fit. }
  \begin{tabular}{c c c c c c c c c} 
  \hline
 & $P^{\rm E}+P^{\rm gm}+P^{\rm gg}$ & $P^{\rm E}$ & $P^{\rm E}+P^{\rm gm}$ & $P^{\rm E}+P^{\rm gg}$ & $P^{\rm gm}+P^{\rm gg}$ & $P^{\rm E}+P^{\rm gm}+P^{\rm gg}$ & $P^{\rm E}+P^{\rm gm}+P^{\rm gg}$ & $P^{\rm gm}+P^{\rm gg}$  \\
Parameter & & & & & & cons-1 & cons-2 & cons-2 \\
 & & & & & & & &\\
$100\theta_{\rm MC}$ & $1.082_{-0.034}^{+0.045}$ & $1.064_{-0.038}^{+0.056}$  & $1.063_{-0.039}^{+0.050}$ & $1.093_{-0.032}^{+0.040}$ & $1.087_{-0.035}^{+0.046}$ & $1.064_{-0.039}^{+0.049}$ & $1.094_{-0.032}^{+0.040}$ & $1.095_{-0.033}^{+0.043}$ \\
$\Omega_{\rm c}h^2$  &  $0.153_{-0.038}^{+0.032}$ & $0.134_{-0.046}^{+0.044}$ & $0.132_{-0.042}^{+0.033}$ & $0.169_{-0.038}^{+0.034}$ &  $0.159_{-0.041}^{+0.034}$ & $0.133_{-0.041}^{+0.031}$ & $0.169_{-0.038}^{+0.032}$ & $0.173_{-0.042}^{+0.037}$\\
$\Omega_{\rm b}h^2$ ($\times 10^{-2}$) & $2.25\pm0.35$ & $2.24_{-0.34}^{+0.36}$ & $2.24_{-0.34}^{+0.36}$ & $2.25\pm0.35$ & $2.25\pm0.35$ & $2.25\pm0.35$ & $2.25\pm0.35$ &  $2.25\pm0.35$\\
$\ln(10^{10}A_{\rm s})$ & $2.63_{-0.44}^{+0.37}$ & $2.75_{-1.05}^{+0.29}$ & $2.80_{-0.77}^{+0.52}$ & $2.30_{-0.54}^{+0.20}$ & $2.67_{-0.45}^{+0.37}$ & $2.92_{-0.58}^{+0.45}$ &  $2.33_{-0.58}^{+0.22}$ & $2.43_{-0.58}^{+0.31}$ \\
$n_{\rm s}$ & $0.97_{-0.19}^{+0.14}$ & $1.11_{-0.05}^{+0.19}$ & $1.08_{-0.06}^{+0.22}$ & $0.97_{-0.18}^{+0.14}$ & $0.93_{-0.23}^{+0.07}$ & $1.03_{-0.13}^{+0.21}$ & $0.99_{-0.18}^{+0.15}$ & $0.91_{-0.21}^{+0.06}$ \\
$h$ & $0.73_{-0.05}^{+0.06}$ & $0.74_{-0.05}^{+0.06}$ & $0.74_{-0.05}^{+0.06}$ & $0.73\pm0.06$ & $0.73_{-0.05}^{+0.06}$ & $0.74_{-0.05}^{+0.06}$ & $0.73\pm0.06$ & $0.73\pm0.06$ \\
$A_{\rm IA}$ & $1.27\pm0.39$ & $0.92_{-0.60}^{+0.76}$ & $1.46\pm0.41$ & $0.88_{-0.52}^{+0.69}$ & $1.38\pm0.47$ & $1.27\pm0.46$ & $1.20\pm0.41$ & $1.42\pm0.55$\\
$B$ & $2.97_{-0.69}^{+0.56}$ & $3.26_{-0.22}^{+0.74}$ & $3.28_{-0.22}^{+0.72}$ & $3.08_{-0.41}^{+0.80}$ & $2.86_{-0.86}^{+0.27}$ & $3.03_{-0.55}^{+0.70}$ & $2.99\pm0.62$ & $2.82_{-0.82}^{+0.25}$ \\
$b_{\rm z1}$ & $1.12\pm0.15$ & - & $0.78_{-0.18}^{+0.14}$ & $1.21\pm0.14$ & $1.13\pm0.15$ & $1.00\pm0.17$ & $1.23_{-0.17}^{+0.16}$ & $1.24\pm0.18$ \\
$b_{\rm z2}$ & $1.25\pm0.16$ & - & $1.45_{-0.32}^{+0.27}$ & $1.45_\pm0.18$ & $1.23\pm0.17$ & $1.11\pm0.19$ & $1.37\pm0.17$ & $1.36\pm0.19$ \\
$\Omega_{\rm m}$ & $0.33_{-0.06}^{+0.05}$ & $0.29_{-0.09}^{+0.08}$ & $0.29_{-0.07}^{+0.06}$ & $0.36\pm0.06$ & $0.34_{-0.06}^{+0.05}$ & $0.29_{-0.06}^{+0.05}$ & $0.36\pm0.06$ & $0.37_{-0.07}^{+0.06}$\\
$\sigma_8$ & $0.78_{-0.08}^{+0.06}$ & $0.80_{-0.17}^{+0.10}$ & $0.81_{-0.13}^{+0.10}$ & $0.70_{-0.08}^{+0.05}$ &$0.80_{-0.09}^{+0.07}$ & $0.84_{-0.11}^{+0.08}$ & $0.72_{-0.08}^{+0.06}$ & $0.74_{-0.09}^{+0.07}$ \\
$S_8$ & $0.800_{-0.027}^{+0.029}$ & $0.761\pm0.038$ & $0.769_{-0.033}^{+0.036}$ & $0.759_{-0.032}^{+0.036}$ & $0.835\pm0.037$ & $0.808_{-0.028}^{+0.030}$ & $0.778\pm0.033$ & $0.805_{-0.043}^{+0.044}$ \\
$\chi^2$ & 115.9 & 61.5 & 97.0 & 67.3 & 45.4 & 84.5 & 101.6 & 32.5\\
d.o.f. & 90 & 42 & 80 & 50 & 40 & 70 & 80 & 30\\
  \hline\hline
  \end{tabular}
  \label{tab_posterior}
\end{table*}

We show the full posterior of all fit parameters in our fiducial run in Fig. \ref{plot_postdiag}, which highlights the degeneracies between parameters. The mean and 68\% confidence interval of the fit parameters for the different combinations of power spectra are listed in Table \ref{tab_posterior}. \\

\section{Validity of flat-sky approximation} \label{app_flat}

\begin{figure}
\centering
\includegraphics[height=0.95\columnwidth,angle=270,trim={0 0 4.5cm 0},clip]{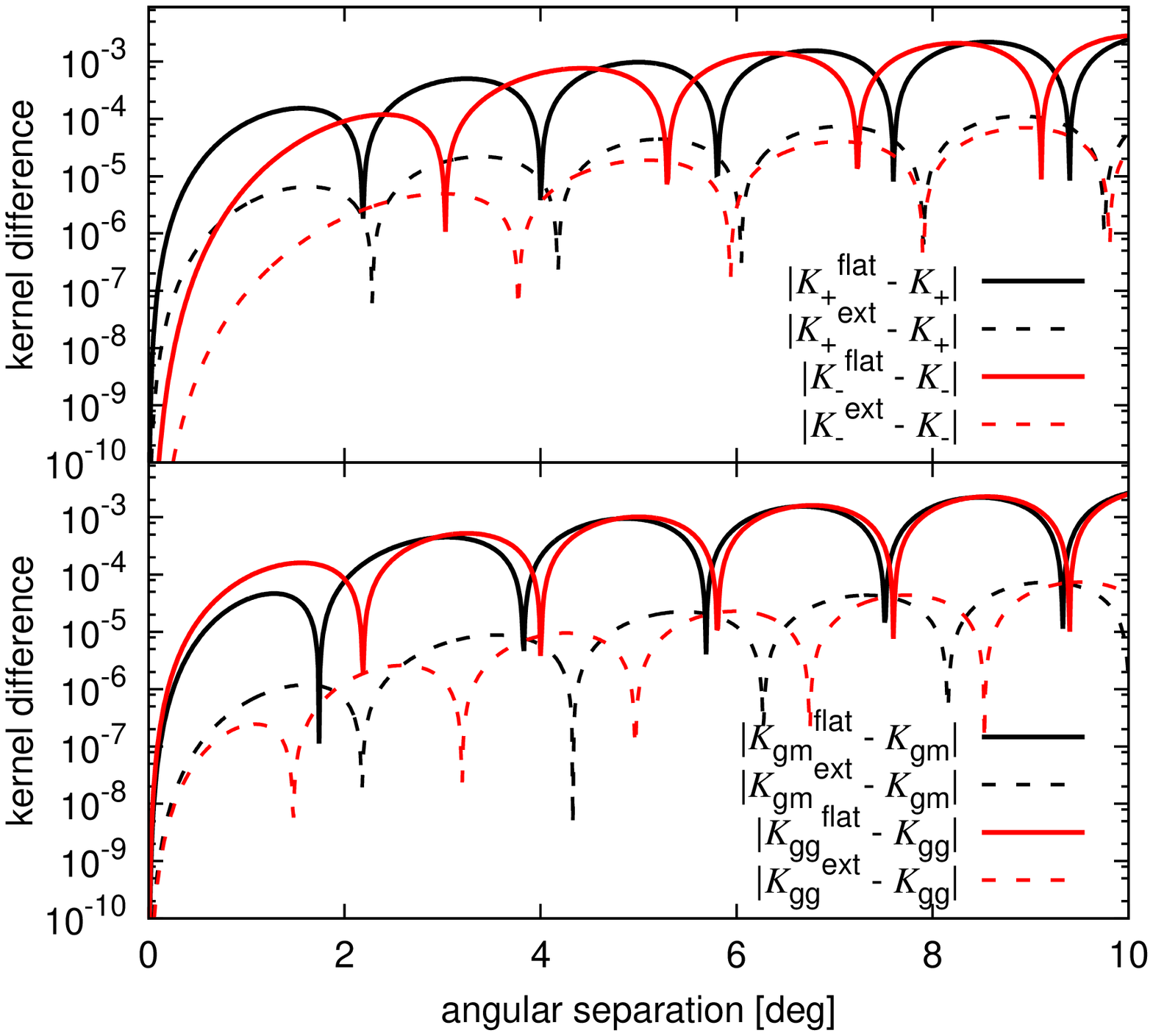}
\caption{Comparison between kernels in the conversion from correlation functions to power spectra. Shown is the difference between the standard flat-sky (${\cal K}^{\rm flat}$, solid lines), as well as the extended flat-sky (${\cal K}^{\rm ext}$, dashed lines), approximation and the exact full-sky kernel as a function of angular separation for fixed $\ell=100$. The top panel shows the cosmic shear kernels relating to $\xi_\pm$; the bottom panel those for $P^{\rm gm}$ and $P^{\rm gg}$.}
\label{fig:kernel_comp}
\end{figure}
We check if the flat-sky approximation that underlies our power spectrum estimators has any impact on the accuracy of our measurements. Before taking band-power averages, the estimators can be expressed as
\eqa{
&&\hspace*{-8mm} \widehat{P^{\rm E}_\kappa}(\ell) = \pi \int_0^\infty \dd \theta \; \bc {{\cal K}_+ (\ell, \theta)\; \widehat{\xi_+}(\theta) + {\cal K}_- (\ell, \theta)\; \widehat{\xi_-}(\theta) };  \hspace{4mm} \\
&&\hspace*{-8mm} \widehat{P^{\rm gm}}(\ell) = 2 \pi \int_0^\infty \dd \theta \; {\cal K}_{\rm gm} (\ell, \theta)\; \widehat{\gamma_{\rm t}} (\theta)\;;  \\
&&\hspace*{-8mm} \widehat{P^{\rm gg}}(\ell) = 2 \pi \int_0^\infty \dd \theta \; {\cal K}_{\rm gg} (\ell, \theta)\; \widehat{w} (\theta)\;.
}
The full-sky expressions for the kernels ${\cal K}_\pm$ were derived in \citet{Chon04}, while their result for the $TE$ spectrum can be adopted for our galaxy-galaxy lensing estimator. Together with the well-known full-sky relation between scalar power spectrum and correlation function \citep{Peebles73}, one obtains
\eqa{
{\cal K}_+ (\ell, \theta) &=& d^{\ell}_{22} (\theta)\, \sin \theta\;; \\
{\cal K}_- (\ell, \theta) &=& d^{\ell}_{2-2} (\theta)\, \sin \theta\;; \\
{\cal K}_{\rm gm} (\ell, \theta) &=& d^{\ell}_{20} (\theta)\, \sin \theta\;; \\
{\cal K}_{\rm gg} (\ell, \theta) &=& d^{\ell}_{00} (\theta)\, \sin \theta\;,
}
where $d^{\ell}_{m m'}$ is the Wigner small-d matrix, and $d^{\ell}_{00}(\theta) = P_\ell(\cos \theta)$ is a Legendre polynomial. The flat-sky (large $\ell$, small $\theta$) approximations for these kernels, denoted by ${\cal K}^{\rm flat}$, can be read off Eqs. (\ref{eq_Pkappa}, \ref{eq_pgm}, \ref{eq_pgg}); see \citet{Kitching16,Kilbinger17,Lemos17} for more detailed discussions. \\
\indent We additionally propose an extended flat-sky approximation with the following kernels,
\eqa{
\label{eq:kernel_ext_start}
{\cal K}^{\rm ext}_+ (\ell, \theta) &=& J_0\bb{(\ell+1/2)\theta} \, \theta\;; \\
{\cal K}^{\rm ext}_- (\ell, \theta) &=& J_4\bb{(\ell+1/2)\theta} \, \theta\;; \\
{\cal K}^{\rm ext}_{\rm gm} (\ell, \theta) &=& J_2\bb{(\ell+1/2)\theta} \, \theta\;; \\
\label{eq:kernel_ext_end}
{\cal K}^{\rm ext}_{\rm gg} (\ell, \theta) &=& J_0\bb{(\ell+1/2)\theta} \, \theta\;,
}
obtained from the standard flat-sky expression by replacing $\ell \rightarrow \ell+1/2$. In Fig.$\,$\ref{fig:kernel_comp} we show the difference between the standard and extended flat-sky approximated kernels and the full-sky expressions at $\ell=100$, which is slightly beyond the largest scales that we consider. The extended approximation can readily be implemented in our current estimators, with the integral corresponding to the band power average now performed numerically. \\
\indent We compare the power spectrum estimates under the assumptions of the standard and extended flat-sky approximations in the case of galaxy clustering, which extends to the largest angular scales and should thus be most affected. As expected, we find the largest discrepancy for the lowest $\ell$-bin, which amounts to a relative difference of $3.3 \times 10^{-5}$ in F1 and less than $10^{-6}$ in F2. Since the extended flat-sky approximation is typically two orders of magnitude more accurate over the scales used in this analysis (see Fig.$\,$\ref{fig:kernel_comp}, bottom panel), this implies that the discrepancy that we have measured is also the one between standard flat-sky approximation and full expression to within $1\,\%$. It is reasonable to expect this discrepancy to be very small because we choose to altogether ignore large scales which are inaccessible from our data, rather than merely misrepresenting their geometry. We have demonstrated in Appendix \ref{app_val_ps} that the IBC, resulting from cutting in particular the large scales from the integrals in our estimators, is much larger but still well controlled in our measurements.

\onecolumn

\section{Cross-survey covariance}\label{app_covarmat}

In this analysis we combine probes from two surveys with substantially different areas, where the smaller survey is fully contained in the larger's area coverage. While the covariances of individual probes can be modelled analytically with the respective survey footprint over which they are measured, there may be significant cross-variances between probes due to the shared sky that need to be included in the joint likelihood analysis. \\
\indent We model the cross-variances under the same assumptions applied to previous analytic approaches \citep[e.g.][]{Takada13}, neglecting the explicit effect of the survey mask on modes well within the footprint, and only taking it into account for the coupling between in-survey and super-survey modes (the super-sample covariance). Under these assumptions, the spin-2 nature of gravitational shear does not impact on the covariance, so it is sufficient to work with the scalar weak lensing convergence and the galaxy number density. Moreover, shot/shape noise does not contribute to the cross-variances we consider. Therefore, we consider the simple, idealised power spectrum estimator \citep[see][for the analogous full calculation for gravitational shear estimates]{Joachimi08} for survey A
\eqa{
\label{eq:psestimate}
\widehat{P_{\rm A}}(\ell) \equiv \frac{1}{\Omega_{\rm A}\, \Omega_{\rm R}(\ell)} \int_{\Omega_{\rm R}(\ell)} \dd^2 \ell_1 | \tilde{x}_{\rm A}^{\rm ob}(\vec{\ell}_1)|^2 \;,
}
with
\eqa{
\tilde{x}_{\rm A}^{\rm ob}(\vec{\ell}) = \int \dd^2 \ell'\; \tilde{x}(\vec{\ell'})\, \Delta_{\rm A}(\vec{\ell}-\vec{\ell'})\;,
}
where $\Omega_{\rm A}$ is the (effective) survey area, $\Omega_{\rm R}(\ell)$ is the area of an annulus centred on $\ell$, and $\Delta_{\rm A}$ is the Fourier transform of the survey aperture, divided by $4 \pi^2$ \citep[see eq. 12 of][]{Joachimi08}. For ease of notation, we consider a single random field $x$, but the following calculations hold equally for any combination of the weak lensing convergence and galaxy number density fields. The Fourier-transformed field is denoted by a tilde.

This power spectrum estimator is unbiased under certain assumptions about the survey footprint, which can be seen by taking the expectation value,
\eqa{
\ba{ \widehat{P_{\rm A}}(\ell) } &=& \frac{1}{\Omega_{\rm A}\, \Omega_{\rm R}(\ell)} \int_{\Omega_{\rm R}(\ell)} \dd^2 \ell_1 \int \dd^2 \ell'\; \int \dd^2 \ell''\; \ba{  \tilde{x}(\vec{\ell'})\, \tilde{x}(\vec{\ell''}) }  \Delta_{\rm A}(\vec{\ell}_1 - \vec{\ell'})\,  \Delta_{\rm A}(-\vec{\ell}_1 - \vec{\ell''})\; \nn \\
&=& \frac{(2 \pi)^2}{\Omega_{\rm A}\, \Omega_{\rm R}(\ell)} \int_{\Omega_{\rm R}(\ell)} \dd^2 \ell_1 \int \dd^2 \ell'\; P_x(\ell')\; \Delta_{\rm A}^2(\vec{\ell}_1 - \vec{\ell'})\;,
}
where in the second equality the definition of the power spectrum of the random field $x$ was used,
\eqa{
\ba{  \tilde{x}(\vec{\ell})\, \tilde{x}(\vec{\ell'}) } = (2 \pi)^2\, \delta_{\rm D}(\vec{\ell} + \vec{\ell'})\; P_x(\ell)\;,
}
with $\delta_{\rm D}$ the Dirac delta distribution. The larger the survey area, the closer $\Delta_{\rm A}$ will be to a Dirac delta distribution. It is therefore appropriate to approximate $\Delta_{\rm A}^2(\vec{\ell}) \approx \delta_{\rm D}(\vec{\ell})\; \Omega_{\rm A}/(2 \pi)^2$, as long as we consider modes well within the survey footprint. Inserting this expression, we find that Eq. (\ref{eq:psestimate}) is an unbiased estimate of the annular average of the power spectrum of $x$,
\eqa{
\label{eq:unbiasedp}
\ba{ \widehat{P_{\rm A}}(\ell) } = \frac{1}{\Omega_{\rm R}(\ell)} \int_{\Omega_{\rm R}(\ell)} \dd^2 \ell_1\; P_x(\ell_1)\;.
} 

We use this estimator to calculate the cross-variance between surveys A and B,
\eqa{
\label{eq:crosscov}
\ba{ \Delta P_{\rm A}(\ell)\; \Delta P_{\rm B}(\ell') } &=& \frac{1}{\Omega_{\rm A}\, \Omega_{\rm B}\, \Omega_{\rm R}(\ell)\, \Omega_{\rm R}(\ell')} \int_{\Omega_{\rm R}(\ell)} \dd^2 \ell_1 \int_{\Omega_{\rm R}(\ell')} \dd^2 \ell_2 \\ \nn 
&& \hspace*{-4cm} \times \Big\{ \ba{ \tilde{x}_{\rm A}^{\rm ob}(\vec{\ell}_1) \tilde{x}_{\rm B}^{\rm ob}(\vec{\ell}_2) }  \ba{ \tilde{x}_{\rm A}^{\rm ob}(-\vec{\ell}_1) \tilde{x}_{\rm B}^{\rm ob}(-\vec{\ell}_2) } + \ba{ \tilde{x}_{\rm A}^{\rm ob}(\vec{\ell}_1) \tilde{x}_{\rm B}^{\rm ob}(-\vec{\ell}_2) }  \ba{ \tilde{x}_{\rm A}^{\rm ob}(-\vec{\ell}_1) \tilde{x}_{\rm B}^{\rm ob}(\vec{\ell}_2) } +  \ba{ \tilde{x}_{\rm A}^{\rm ob}(\vec{\ell}_1) \tilde{x}_{\rm B}^{\rm ob}(-\vec{\ell}_1) \tilde{x}_{\rm A}^{\rm ob}(\vec{\ell}_2) \tilde{x}_{\rm B}^{\rm ob}(-\vec{\ell}_2) }_{\rm c} \Big\}\;,
}
where the subscript $c$ denotes the connected correlator that encapsulates the non-Gaussian cosmic variance contributions. Here, $\Delta P$ denotes the fluctuation of the power spectrum estimator around its expectation. We continue with the first term of Eq. (\ref{eq:crosscov}), assuming without loss of generality that survey A is the larger of the two,
\eqa{
&& \hspace*{-7mm}  \ba{ \tilde{x}_{\rm A}^{\rm ob}(\vec{\ell}_1) \tilde{x}_{\rm B}^{\rm ob}(\vec{\ell}_2) }  \ba{ \tilde{x}_{\rm A}^{\rm ob}(-\vec{\ell}_1) \tilde{x}_{\rm B}^{\rm ob}(-\vec{\ell}_2) }\\ \nn
&=& \int \dd^2 \ell_a\; \int \dd^2 \ell_b\; \int \dd^2 \ell_c\; \int \dd^2 \ell_d\; \ba{ \tilde{x}(\vec{\ell}_a) \tilde{x}(\vec{\ell}_b) }  \ba{ \tilde{x}(\vec{\ell}_c) \tilde{x}(\vec{\ell}_d) } \Delta_{\rm A}(\vec{\ell}_1-\vec{\ell}_a)\; \Delta_{\rm B}(-\vec{\ell}_2-\vec{\ell}_b)\; \Delta_{\rm A}(-\vec{\ell}_1-\vec{\ell}_c)\; \Delta_{\rm B}(\vec{\ell}_2-\vec{\ell}_d)\; \\ \nn
&=& (2 \pi )^4  \int \dd^2 \ell_a\; \int \dd^2 \ell_c\; P_x(\ell_a)\, P_x(\ell_c)\, \Delta_{\rm A}(\vec{\ell}_1-\vec{\ell}_a)\; \Delta_{\rm A}(-\vec{\ell}_1-\vec{\ell}_c)\; \Delta_{\rm B}(-\vec{\ell}_2+\vec{\ell}_a)\; \Delta_{\rm B}(\vec{\ell}_2+\vec{\ell}_c)\; \\ \nn
&\approx&  (2 \pi )^4\;  P_x^2(\ell_1)\,  \Delta_{\rm B}^2(\vec{\ell}_1-\vec{\ell_2})\; \\ \nn
&\approx&  (2 \pi )^4\;  \Omega_{\rm B}\, P_x^2(\ell_1)\, \delta_{\rm D}(\vec{\ell}_1-\vec{\ell_2})\;.
}
To arrive at the third equality, we approximated the $\Delta_{\rm A}$ of the larger survey A by the Dirac distribution. This assumption is fair as long as $\Omega_{\rm A} \gg \Omega_{\rm B}$; however, if the two surveys approached similar coverage, the standard expression for a single survey could be used to good accuracy anyway. The final equality results from the same approximation that was made in the derivation of Eq. (\ref{eq:unbiasedp}). \\
\indent The remaining terms in Eq. (\ref{eq:crosscov}) are processed in full analogy to yield the following expression for the in-survey contributions to the cross-variance,
\eqa{
\ba{ \Delta P_{\rm A}(\ell)\; \Delta P_{\rm B}(\ell') }_{\rm in-survey} \approx \frac{8 \pi^2}{\Omega_{\rm max}\, \Omega_{\rm R}(\ell)}\; \delta_{\ell \ell'}  P_x^2(\ell) + \frac{1}{\Omega_{\rm max}\, \Omega_{\rm R}(\ell)\, \Omega_{\rm R}(\ell')}  \int_{\Omega_{\rm R}(\ell)} \dd^2 \ell_1 \int_{\Omega_{\rm R}(\ell')} \dd^2 \ell_2\; T_x(\vec{\ell}_1,\vec{\ell}_2,-\vec{\ell}_1,-\vec{\ell}_2)\;,
}
where $\delta_{\ell \ell'}$ is the Kronecker delta acting on the angular frequency bins at which the covariance is evaluated, and where $T_x$ is the trispectrum of the field $x$. We have defined $\Omega_{\rm max} \equiv {\rm max}( \Omega_{\rm A}, \Omega_{\rm B})$. This result reduces to the standard expression for $\Omega_{\rm A} \rightarrow \Omega_{\rm B}$ and is straightforward to generalise to the multi-field and tomographic case \citep[see e.g.][]{Joachimi10,Krause16}. \\
\indent It remains to be shown how the SSC term is modified for the cross-variance across different surveys. We follow \citet{Takada13} closely and refer to their paper for detailed calculations. They showed that in the single-survey case the SSC of angular power spectra (again for a single field) is given by
\eqa{
\label{eq:ssc}
\ba{ \Delta P_{\rm A}(\ell)\; \Delta P_{\rm A}(\ell') }_{\rm SSC} = \int_0^{\chi_{\rm hor}} \dd \chi\; \frac{K^4(\chi)}{f_{\rm K}(\chi)^6}\; \frac{\partial P_\delta \bb{\ell/f_{\rm K}(\chi)}}{\partial \delta_{\rm b}}\; \frac{\partial P_\delta \bb{\ell'/f_{\rm K}(\chi)}}{\partial \delta_{\rm b}}\; \sigma_{\rm A}^2(\chi)\;,
}
where $\chi$ is comoving distance, $f_{\rm K}(\chi)$ comoving angular diameter distance, and $K(\chi)$ the line-of-sight kernel of the signal under consideration. The expression above is based on the Limber approximation, which we expect to hold similarly well as for the signals we model, due to the broad line-of-sight distributions entering our signals. The derivatives of the matter power spectrum $P_\delta$ with respect to a fluctuation in the background density $\delta_{\rm b}$ provide the response of the measurement to super-survey modes, while
\eqa{
\sigma_{\rm A}^2(\chi) = \frac{1}{\Omega_{\rm A}^2} \int \frac{\dd^2 \ell_1}{(2 \pi)^2}\; P_\delta^{\rm lin} \br{\frac{\ell_1}{f_{\rm K}(\chi)}} |\Delta_{\rm A}(\vec{\ell}_1)|^2
}
is the variance of this background density field within the mask of survey A. The linear matter power spectrum is employed in this expression. Repeating the derivation of \citet{Takada13}, but now with the last term of Eq. (\ref{eq:crosscov}) as the starting point, which accounts for different survey footprints, we obtain
\eqa{
\sigma_{\rm AB}^2(\chi) = \frac{1}{\Omega_{\rm A}\, \Omega_{\rm B}} \int \frac{\dd^2 \ell_1}{(2 \pi)^2}\; P_\delta^{\rm lin} \br{\frac{\ell_1}{f_{\rm K}(\chi)}} \Delta_{\rm A}(\vec{\ell}_1)\;  \Delta_{\rm B}(-\vec{\ell}_1)\;.
}
Note that $\sigma_{\rm AB}^2$ remains real because the imaginary parts of the Fourier transforms of the survey masks are anti-symmetric and will thus vanish after the area integration over $\ell_1$. Our implementation of the SSC contribution uses Eq. (\ref{eq:ssc}) with $\sigma_{\rm AB}^2$ determined from the explicit GAMA and KiDS survey footprints provided in the form of Healpix maps. A more detailed study of the impact of survey geometry on covariance contributions will be presented in a forthcoming publication; see also \citet{Lacasa16} for a detailed discussion of effects pertaining to the SSC term.

\end{appendix}

\end{document}